\documentclass[a4paper,12pt,twoside,titlepage]{article}

\RequirePackage[OT1]{fontenc}
\RequirePackage{amsthm,amsmath,amssymb}

\usepackage[comma,authoryear,round]{natbib}
\usepackage{graphicx}
\usepackage{color}
\usepackage[colorlinks,citecolor=blue,urlcolor=blue,linkcolor=blue]{hyperref}

\usepackage{dsfont}                 
\usepackage{multirow}  
\usepackage{makecell}
\usepackage{mathrsfs}
\usepackage{url}

\usepackage{multirow}

\usepackage{textcomp}  

\newcommand{\bZ}{\mathbf{Z}}
\newcommand{\Z}{\mathrm{Z}}

\newcommand{\Mu}{\boldsymbol{\mu}}
\newcommand{\bSigma}{\boldsymbol{\Sigma}}
\newcommand{\true}{^\text{true}}
\newcommand{\perla}{\textsc{perla}}

\newcommand{\virgolette}[1]{``#1''}
\newcommand{\review}[1]{{\color{black}#1}}
\newcommand{\reviewm}[1]{{\color{black}#1}}

\usepackage{fancyhdr}
\title{\bf Bayesian Mapping of Mortality Clusters\vspace{2cm}}

\author{{A. Sottosanti, E. Bovo, P. Belloni, G. Boccuzzo} \\
	{University of Padova, Department of Statistical Sciences}\vspace{2cm}
}

\date{\today \\ \vspace{3cm}
	Address for correspondence: \texttt{andrea.sottosanti@unipd.it}}

\begin{document}
	
	\maketitle
	
	\begin{abstract}
		Disease mapping analyses the distribution of several disease outcomes within a territory. Primary goals include identifying areas with unexpected changes in mortality rates, studying the relation among multiple diseases, and dividing the analysed territory into clusters based on the observed levels of disease incidence or mortality. In this work, we focus on detecting spatial mortality clusters, that occur when neighbouring areas within a territory exhibit similar mortality levels due to one or more diseases. When multiple causes of death are examined together, it is relevant to identify not only the spatial boundaries of the clusters but also the diseases that lead to their formation. However, existing methods in literature struggle to address this dual problem effectively and simultaneously. To overcome these limitations, we introduce \perla, a multivariate Bayesian model that clusters areas in a territory according to the observed mortality rates of multiple causes of death, also exploiting the information of external covariates. Our model incorporates the spatial structure of data directly into the clustering probabilities by leveraging the stick-breaking formulation of the multinomial distribution. Additionally, it exploits suitable global-local shrinkage priors to ensure that \review{the detection of clusters depends on diseases showing concrete increases or decreases in mortality levels, while excluding uninformative diseases.}
We propose an MCMC algorithm for posterior inference that consists of closed-form Gibbs sampling moves for nearly every model parameter, without requiring complex tuning operations. This work is primarily motivated by a case study on the territory of a local unit within the Italian public healthcare system, known as ULSS6 \textit{Euganea}. To demonstrate the flexibility and effectiveness of our methodology, we also validate \perla~with a series of simulation experiments and an extensive case study on mortality levels in U.S. counties.\\[3cm]
	\end{abstract}

\section{Introduction}
\label{sec:introduction}

\subsection{Motivation and case studies}
\label{subsec:motivation}



In the analysis of environmental and epidemiological phenomena, researchers often gather geolocalised data in the form of multiple outcomes. For example, studies on the effects of long-term exposure to air pollution and high temperatures on public health generally collect and analyse both multiple environmental and health indicators \citep{Kunzli_etal.2000, Wong_etal.2008, Lee_etal.2009, Taylor_etal.2015}. In epidemiological studies, several outcomes are used to determine the frailty characteristics of specific groups of patients  \citep{Li_etal.2021, Qian_etal.2023}. 
Therefore, analysing the joint distribution of multiple outcomes and their interactions is a crucial step to uncover patterns in space and relationships that may not be evident when analysing each outcome on its own. \review{This approach provides a more comprehensive understanding of the underlying environmental and epidemiological processes in a territory}.

Disease mapping, in particular, aims at studying the spatial distribution of health phenomena to identify significant changes of diseases incidence or mortality. It therefore represents the first step for understanding geographical disparities in health outcomes and developing strategies for disease prevention and control. 
Disease mapping techniques have evolved significantly over the last decades, moving from simple representations of a disease occurrence to more \review{advanced} statistical analyses and models \citep{waller.2004, lawson.2013, Banerjee_etal.2015}. The use of \review{sophisticated modelling techniques} yields a more detailed representation of the spatial distribution of phenomena in space and of the dependence across outcomes, resulting in more accurate explanatory and  predictive performances \citep{Aungkulanon_etal.2017, Coker_etal.2023, griffith.2023, tesema.2023}.
One of the main goals of disease mapping is discovering and locating \review{ groups of spatially proximate areas that show similar levels of the outcomes considered. We refer to these groups as \textit{spatial clusters}.} 
In public health, discovering spatial clusters of disease incidence or mortality represents a  step forward in the identification of possible underlying risk factors in specific areas of the territory \citep{kulldorff.1997, Kulldorff_etal.2005, Robertson_etal.2010}. 

\begin{figure}%
    \centering
    \includegraphics[width=0.44\linewidth]{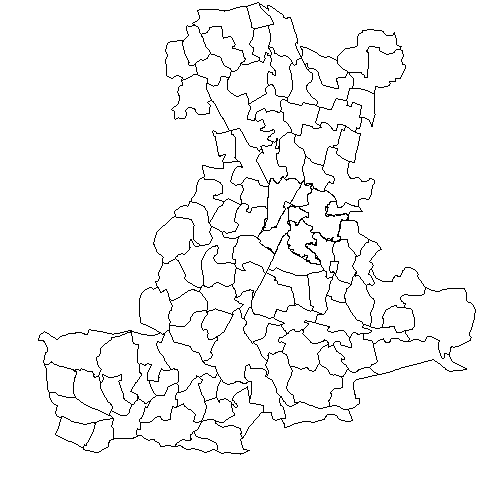}
    \includegraphics[width=0.5\linewidth]{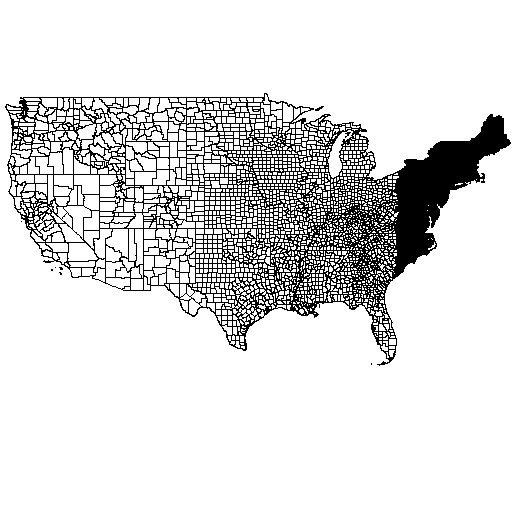}
    \caption{Left: map of the Padua province territory, which covers about 2,144~km$^2$ with a population of approximately 936,000 residents. \reviewm{The geographical position of the Padua province in Italy is shown in the left panel of Supplementary Figure 1.} 
    Right: map of the U.S. divided into counties. We highlight in black the 388 counties on the U.S. East coast that we analyse in Section \ref{subsec:USA}. Maps are not shown on the same scale.}
    \label{fig:maps_Padua_US}
\end{figure}

As an example, we consider a study of mortality levels in the Padua province, located in northeastern Italy, with data provided by the local healthcare unit ULSS6 \textit{Euganea}.
This institution oversees health services across the province, whose territory is divided into 106 administrative sub-areas \reviewm{(see the left panel of Figure \ref{fig:maps_Padua_US})}. The city of Padua is not considered as a stand-alone municipality but is divided into six districts of size and population comparable to other municipalities in the province. 
The goal of ULSS6 is to optimise logistical and financial resources by identifying significant variations in mortality levels across either individual areas or clusters of areas.
Their interest is particularly focused on the mortality due to three major death causes: \textit{diseases of the circulatory system}, \textit{diseases of the respiratory system}, and \textit{malignant neoplasms}.
According to \cite{fedeli_etal.2021}, these three categories collectively represent 70\% of the overall mortality rate recorded in the province. Mortality data for these death causes are available from 2017 to 2019, divided by gender. 
\cite{Bovo_etal.2023} made a first attempt to detect mortality clusters in this province by analysing each cause of death separately. Nonetheless, \reviewm{this approach has some important limitations:} treating the outcome variables independently \reviewm{does not exploit the relationships across causes of death and} can lead to the identification of clusters that vary substantially for each cause. In contrast, analysing all causes of death simultaneously enables the identification of territorial clusters with specific mortality profiles, \textit{i.e.}, estimated risk levels for each cause.


The case study of ULSS6 mortality data is relevant due to its depiction of a setting where areas have similar territorial characteristics, comparable population sizes and socio-economic conditions.
In other contexts, the territory under consideration may be highly heterogeneous in one or more aspects.
For instance, in the United States of America (U.S.) the administrative division of the territory into counties provides a far different scenario from the province of Padua. The distribution of population in space is heterogeneous, as well as population density, urbanisation and wealth; in addition, different U.S. states may have different health care systems, and this leads to another source of variability.
The distribution of U.S. counties in space is displayed in the right panel of~Figure~\ref{fig:maps_Padua_US}.
The Centres for Disease Control and Prevention (CDC) collected data on mortality levels by counties in the U.S. from 2016 to 2019, available through the WONDER Online Databases~\citep{WONDER}. 
In such heterogeneous contexts, it is crucial not only to quantify different risk levels across areas, but also to determine the existence of possible geographically concentrated clusters with specific epidemiological relevance. 

Both the discussed examples reveal the need for a statistical method that can: \emph{i.)} perform multivariate clustering of a territory based on the joint analyses of multiple outcomes, exploiting the spatial structure of data and the dependence across diseases; \emph{ii.)} leverage information provided by exogenous variables that influence the outcomes; and \emph{iii.)} identify which outcome variables lead to the formation of spatial clusters.
This last point becomes more and more relevant as the number of outcomes increases or the size of the territory grows, making it essential to identify clusters of disease prevalence or excess mortality that are geographically concentrated in small areas while avoiding spurious signals.

\subsection{Spatial clustering of mortality levels}
\label{sec:spatial_mortality_clusters}

In this article, we consider the problem of clustering areas of a territory based on the observed outcomes of multiple diseases. We assume that the territory is divided into polygons defined by administrative boundaries, and that the outcome variables are aggregates of measures taken within the boundaries of each area.
In both the case studies presented in Section \ref{subsec:motivation}, the outcome variables are measurements of mortality from multiple causes of death. The areas in the Padua province correspond to municipalities or administrative districts, while in the U.S. dataset  they represent the U.S. counties.

A common approach for identifying groups of contiguous areas with excess mortality is through \textit{spatial scan statistics} (SSS, \citealp{kulldorff.1997}). These techniques detect spatial clusters with unexpected changes in outcome levels compared to the rest of the territory using suitable test statistics \citep{Cucala.2016, Liu_etal.2018, ahmed_genin.2020}. Over the past twenty years, these methods have been extended to the multivariate framework \citep{Cucala_etal.2017, Cucala_etal.2019} and to fit complex data structures, such as multivariate functional data \citep{Frevent_etal.2023} and spatio-temporal data, in both frequentist \citep{Kulldorff_etal.2005,Robertson_etal.2010,Amini_etal.2023} and Bayesian paradigms \citep{Li_etal.2012}. For a detailed review of applications of multivariate SSS to different contexts, readers can refer to the Introduction section of \cite{Frevent_etal.2023}.
Despite their extensive use, SSS may fail to detect significant changes in mortality within spatial clusters that exhibit highly irregular shapes \citep{tango.2021}. Moreover, they are not suitable for partitioning the entire map into clusters.

Another established approach for detecting spatial clusters is through hidden Potts models (HPM), an extension of the Gaussian mixture models (GMM) that account for the spatial dependence across adjacent neighbours \citep{Potts.1952}.
Readers can refer to \cite{Moores_etal.2018} for a comprehensive review of HPM estimation methods. 
Despite their popularity, \reviewm{the use of the Potts model} presents some relevant drawbacks, particularly related to the estimation of the inverse temperature parameter.
Recently, \cite{moores_etal.2020} conducted a detailed survey of existing approaches to address such estimation challenges, with particular attention to Approximate Bayesian Computation algorithms, and proposed their own scalable solution. However, the current statistical literature does not adequately address the extension of HPM to the joint analysis of multiple outcomes. Moreover, to the best of our knowledge, no software implementation is currently available for fitting multivariate HPM, for instance in the \texttt{R} programming language \citep{Rlibrary}.

To address the need to identify mortality clusters jointly considering multiple causes of death, we propose \perla~(PEnalised Regression with Localities Aggregation), a multivariate Bayesian model that aggregates areas of a territory into spatially informed clusters based on different levels of outcome variables, while accounting for the presence of exogenous variables and for the marginal correlation across responses. 
\reviewm{This is done by embedding} the data spatial proximity directly into the clustering probabilities using the \textit{multinomial stick-breaking} strategy \citep{Linderman_etal.2015}, which
 consists of rewriting a multinomial model as a sequence of binomial random variables. \reviewm{This solution allows} the use of the well-established  P\'olya-gamma data augmentation \reviewm{even with} multinomial outcomes, \reviewm{and makes it possible to derive the conditional posterior distributions of the model parameters in closed form}.
\reviewm{It also eliminates the need} of the Potts model to account for the neighbourhood structure and consequently avoiding the known inferential problems associated with it. Alternative approaches to link a probability vector defined on a simplex to a series of Gaussian random variables include, for example, the logistic normal distribution \citep{Lafferty_Blei.2005, Russo_etal.2022}. %
In addition, \perla~regulates the clusters intercepts through global-local shrinkage priors \citep{polson_scott.2011}, allowing to affirm or inhibit the role of specific diseases in the formation of clusters.
The shrinkage priors play also a regularising role in the selection of the number of clusters, as they help suppressing the effect of excess clusters.
The reader can refer to \cite{Bhadra_etal.2019} and \cite{vanErp_etal.2019} for two reviews of global-local shrinkage priors. 

This work is strongly motivated by the two epidemiological case studies presented in Section \ref{subsec:motivation} and can be of great usefulness in the analysis of several other social and epidemiological phenomena, including the distribution of the level of education, the incidence of diseases, and the occurrence of multiple kind of crimes.
Nonetheless, \perla~can be applied to any kind of multivariate areal data.
Two examples of the numerous possible applications outside the epidemiological context are image segmentation \citep{Moores_etal.2015} and spatial transcriptomics \citep{zhao_etal.2021}.
For these reasons, we implement our methodological solution in the software library \texttt{perla} for the $\texttt{R}$ programming language.

The remainder of this article is structured as follows.
Section \ref{sec:statistical_model} introduces the \perla~model, highlighting the formulation of the prior distributions and their role in the detection of spatial clusters.
Section \ref{sec:posterior_inference} delineates an MCMC algorithm to perform posterior inference and explains how to operate model selection and post-processing using the posterior samples. 
Section \ref{sec:simulation_experiments} presents \review{three} simulation experiments. The goal of the first experiment is to demonstrate the significant improvements in \review{interpretability of the results} and inferential conclusions provided by our global-local shrinkage priors compared to standard, non-informative priors. In addition, it performs a comparison of the clustering accuracy with other competing methods. 
The second experiment investigates whether different a priori choices on the parameters that handle the spatial correlation of the clustering probabilities have an impact on the final clustering performance, in particular when spatial clusters present different levels of spatial association. \review{The third experiment evaluates the performance of the proposed methodology against competing methods using data simulated from a different generating mechanism that diverges from \perla~in several key aspects.}
Section \ref{sec:application} presents the application of \perla~on the two cases studies described in Section \ref{subsec:motivation}, showing the usefulness of our methodology in responding to specific epidemiological questions. 
Finally, Section \ref{section:discussion} takes some concluding remarks and outlines future perspectives.

\section{Model formulation}
\label{sec:statistical_model}

\subsection{The statistical model}\label{subsec2}
\label{subsec:the_statistical_model}

We assume that data are collected in the form of an $n\times d$ matrix $\mathbf{Y}$, where $y_{ij}\in \mathbb{R}$ denotes the mortality level due to the $j$-th disease ($j = 1,\dots,d$) measured at the $i$-th area of a map ($i = 1,\dots,n$). Data are structured such that $y_{ij} > 0$ represents a mortality excess, and $y_{ij} < 0$ represents a mortality deficit. 
In addition, let $\mathbf{X}$ be an $n\times p$ matrix collecting the values of $p$ exogenous variables for the $n$ areas. We assume that the $n$ areas can be divided into $K$ territorial clusters, implying that the average mortality levels in the $i$-th spatial area due to the $d$ diseases considered vary based on the cluster to which the $i$-th area belong. We collect the information about the unknown clusters into an $n\times K$ matrix $\bZ$, where $\Z_{ik} = 1$ if the $i$-th area belongs to the $k$-th cluster (with $k = 1,\dots, K$), and 0 otherwise. \reviewm{As each area can belong to only one cluster, every row of $\mathbf{Z}$ contains exactly one non-zero element.} 
To access the rows and the columns of a generic matrix of parameters $\boldsymbol{\theta}$, we  use the notation $\boldsymbol{\theta}_{i.}$ and $\boldsymbol{\theta}_{.j}$ to refer to its $i$-th row and its $j$-th column, respectively. The same notation is used also for the data matrices $\mathbf{Y}$ and $\mathbf{X}$, with the only exception that their row and column vectors are denoted with bold lowercase letters.

We assume that regions within a spatial cluster have similar mortality levels for each disease considered, whereas significant variations in mortality levels occur across clusters. In addition, we assume that exogenous variables influence the outcomes but do not regulate the formation of clusters \citep{Coker_etal.2023}. 
Based on these considerations, we formulate the multivariate regression model
\begin{equation}
    \label{formula:model}
    \mathbf{y}_{i.} = \bZ^T_{i.} \Mu + \mathbf{x}^T_{i.}\boldsymbol{\beta}+\boldsymbol{\epsilon}_{i.},\hspace{1cm}\boldsymbol{\epsilon}_{i.}|\bSigma\sim \mathcal{N}_d(\mathbf{0},\bSigma),
\end{equation}
where $\bZ_{i.}$ and $\mathbf{x}_{i.}$ are the $i$-th row of $\bZ$ and $\mathbf{X}$, $\Mu=\{\mu_{kj}\}_{k = 1,\dots, K;j = 1,\dots,d}$ is a matrix containing the cluster-~and disease-specific intercepts, $\boldsymbol{\beta}=\{\beta_{\ell j}\}_{\ell = 1,\dots, p;j = 1,\dots,d}$ is the matrix of regression coefficients, and $\boldsymbol{\epsilon}_{i.}$ is the vector of error terms. $\bSigma$ aims at capturing the correlation across the $d$ diseases.
The formulation of the regression model in \eqref{formula:model} is equivalent to assuming a matrix variate Gaussian distribution on the outcome matrix $\mathbf{Y}$, with mean matrix $\mathbf{Z\Mu}+\mathbf{X\boldsymbol{\beta}}$, the identity matrix $\mathbf{I}_{n}$ as covariance matrix of the rows, and $\bSigma$ as covariance matrix of the columns.

By definition of $\mathbf{Z}$, we define a cluster as a set of labelled  areas of the map with the same average levels of the $d$ responses, net of the effects of the covariates $\mathbf{X}$.
Since the clustering matrix $\bZ$ is unknown, we assume that $\bZ_{i.}$ distributes according to a multinomial distribution of size 1 and vector of probabilities $\boldsymbol{\pi}_i = (\pi_{i1},\dots,\pi_{iK})$. To let the clustering draw information by the spatial structure of data, we first express the multinomial model as a series of binomial probability functions using the multinomial stick-breaking representation of \cite{Linderman_etal.2015}
\begin{equation}
\label{formula:stick_breaking}
\Pr(\bZ_{i.} |\tilde{\boldsymbol{\pi}}_i) = \prod_{k = 1}^{K-1}\left\{\tilde{\pi}_{ik}^{\mathrm{Z}_{ik}}(1-\tilde\pi_{ik})^{1-\mathrm{Z}_{ik}}\right\}^{N_{ik}},
\end{equation}
where $N_{i1} = 1$ (there is only one element \reviewm{in $\mathbf{Z}_{i.}$} that is equal to one), $\tilde{\pi}_{i1}={\pi}_{i1}$, and, for $k>1$,
$$
\hspace{.7cm}N_{ik} = 1-\sum_{k'<k}\mathrm{Z}_{ik'},\hspace{.5cm}\tilde{\pi}_{ik} = \dfrac{\pi_{ik}}{1-\sum_{k'<k}\pi_{ik'}}.$$ 
By construction, $\tilde{\pi}_{iK} = 1$. Then, for $k = 1,\dots,K-1$, we reparameterise the probabilities $\tilde{\pi}_{ik}$ using the logistic function, thus writing $\tilde{\pi}_{ik} = \exp(\psi_{ik})/\{1+\exp(\psi_{ik})\}$.  The vector $\boldsymbol{\psi}_{.k} = (\psi_{1k},\dots,\psi_{nk})$ contains $n$ realisations in space of the transformed conditional probabilities of belonging to cluster $k$, given that the observations were not assigned to the previous $k-1$ clusters $(\boldsymbol{\psi}_{.1}$ simply represents the transformed probabilities of belonging to cluster 1). 
To introduce spatial dependence across the transformed clustering probabilities, we assume that each vector $\boldsymbol{\psi}_{.k}$ distributes according to a conditionally autoregressive (CAR) model of the form 
\begin{equation}
\label{formula:CAR}
\boldsymbol{\psi}_{.k}|\tau_k,\rho_k\sim\mathcal{N}_n\{\boldsymbol{0},\tau_k(\mathbf{D}-\rho_k\mathbf{W})^{-1}\},    
\end{equation}
where $\mathbf{W}$ has null diagonal and $\mathrm{W}_{ii'} = 1$ if regions $i$ and $i'$ are neighbours, and 0 otherwise, $\mathbf{D}$ is a diagonal matrix with $\mathrm{D}_{ii}$ equals to the number of neighbours of $i$, $\tau_k >0$ and $\rho_k\in[0,1)$.  

Although the model in \eqref{formula:CAR} does not force the areas within the same cluster to be connected by a path, therefore allowing even spatially distant areas to belong to the same cluster, modelling the \reviewm{transformed probabilities $\{\boldsymbol{\psi}_{.k}\}_{k = 1,\dots,K-1}$} with CAR distributions as in \eqref{formula:CAR} encourages the identification of spatially separated clusters. The principal role of $\rho_k$ is ensuring that $\mathbf{D}-\rho_k\mathbf{W}$ is positive semidefinite. \reviewm{Within the CAR parametrisation considered by \cite{Assuncao_Krainski.2009} (which is not the one adopted in this work), the positive definiteness is guaranteed with a more restrictive condition that imposes $\rho_k$ to lie between the reciprocals of the maximum and minimum eigenvalues of  $\mathbf{D}^{-1}\mathbf{W}$.} As discussed in Section 4.3.1 of \cite{Banerjee_etal.2015},  $\rho_k$ cannot be interpreted as a measure of spatial correlation of the data, similarly for example to what is expressed by the \textit{Moran's I} statistic. The only exception is $\rho_k = 0$, which denotes the case of independence across spatial areas. In addition, \cite{Assuncao_Krainski.2009} highlighted that $\rho_k$ can assume also negative values, which however can still be associated to positive correlation values. These inconsistencies were further underlined also by \cite{VerHoef_etal.2018}, who stated that, in geostatistics, it is common to set $\rho_k$ to be positive, as negative values of spatial correlation are generally not considered. Therefore, following also \cite{Carlin_Banerjee.2003}, in this work we assume $\rho_k$ to be only positive.
To solve these oddities, \cite{Datta_etal.2019} proposed an alternative to the CAR formulation, named DAGAR, using a directed acyclic graph (DAG) to model the precision matrix of $\boldsymbol{\psi}_{.k}$. One of the key advantages of their approach lies in the interpretation of $\rho_k$, which now represents the degree of spatial correlation within the data. However, for this study, we have opted to retain the CAR formulation for mainly two  reasons. Firstly, the DAGAR model necessitates the ordering of map areas. While \cite{Datta_etal.2019} demonstrate that ordering does not heavily impact the final results, it is not clear how the ordering operation should be performed when the statistical units, \textit{i.e.}, the spatial areas, are divided into clusters. In addition, the concept of ordering spatial data may pose challenges for non-experts in the field. Given our intention for \perla~to be used in epidemiological studies, we prefer to maintain a standard framework that does not rely on ordering. Secondly, our decision is reinforced by the favourable performance of the CAR model compared to DAGAR, as evidenced by \cite{Datta_etal.2019}. Their study indicates that both models yield comparable results from various perspectives, with the interpretability of $\rho_k$ emerging as the primary advantage of DAGAR over CAR. In our application, we employ the spatial model solely to incorporate spatial correlation into the reparameterised clustering probabilities $\boldsymbol{\psi}_{.k}$. Thus, we are not using a spatial model for the raw data itself, but rather for a latent parameter. Therefore, while interpretable, we believe that the use of the DAGAR would offer a few epidemiological insights.

With the current formulation and assumptions, the stick-breaking construction offers a natural ordering of the discovered clusters. In fact, taking $\mathbb{E}(\boldsymbol{\psi}_{.k}|\tau_k,\rho_k) = \mathbf{0}$ is equivalent to assuming that the marginal expected value of $\tilde{\pi}_{ik}$ is $\mathbb{E}(\tilde{\pi}_{ik}|\tau_k,\rho_k) = 0.5$, for $i=1,\dots,n$ and $k = 1,\dots,K-1$. As a direct consequence, $\mathbb{E}({\pi}_{ik}|\tau_k,\rho_k) > \mathbb{E}({\pi}_{ik'}|\tau_{k'},\rho_{k'})$ when $k > k'$, therefore the average number of observations expected in the $k$-th cluster decreases as $k$ grows, and clusters are naturally interpreted from the most to the least predominant.

The model in Equation \eqref{formula:model} can be rephrased also in terms of a multivariate Gaussian mixture model. In fact, it corresponds to assuming $\mathbf{y}_{i.}|\Z_{ik}=1\sim\mathcal{N}_d(\Mu_{k.}+\mathbf{x}^T_{i.}\boldsymbol{\beta}, \bSigma)$, where $\Pr(\Z_{ik} = 1|\pi_{ik}) = \pi_{ik}$. Although, at a data level, \perla~and HPM are practically equivalent, they adopt distinct strategies for embedding the spatial structure of data into the clustering labels. In fact, HPM works directly with the conditional distributions $\Z_{ik}|\mathbf{Z}_{\mathscr{N}(i),k}$, where $\mathscr{N}(i)$ denotes the spatial neighbours of region $i$.  
Instead, \perla~assumes that the clustering probabilities are spatially correlated. In Section \ref{sec:posterior_inference}, we show how this strategy offers practical advantages in terms of Bayesian posterior inference.

\subsection{Prior specification}
\label{subsec:priors}

In the Bayesian framework, the prior distributions of the model parameters are used either for conveying previous knowledge about the analysed phenomena or for specific modelling needs, like bounding the parameter space or performing variable selection. In this section, we discuss the a priori assumptions that we make to help our model retrieving the clustering structure of the data, while inferring the underlying epidemiological phenomena.

\perla~is designed to disentangle spatial clusters based on the variations of disease-specific mortality in space. The elements in the \review{Formula \eqref{formula:model}} that differentiate the clusters are the intercepts $\Mu = \{\mu_{kj}\}_{k = 1,\dots,K;j = 1,\dots,d}$.  For instance, it is possible that not all considered diseases contribute to the formation of spatial clusters, or that certain diseases contribute to the formation of clusters only in specific areas of the map. \review{In practice, we state that the $j$-th disease contributes to the formation of the $k$-th cluster if $\mu_{kj}$ deviates from zero, either positively or negatively. A positive deviation indicates a mortality excess in the cluster due to disease $j$, while a negative deviation suggests a mortality deficit.} Therefore, we \review{design an a priori setting for} $\Mu$ that promotes the clustering \review{only} when it detects significant deviations from zero of the mean levels of the response variables across different areas of the map. 
For instance, suppose that in an epidemiological study comparing the mortality due to two diseases, 
\review{the mean levels related to the first disease ($j = 1$) are indicative of the presence of two spatial clusters, while the mean levels related to the second disease ($j = 2$) are close to zero.} 
We aim for a prior distribution on \review{the} $\mu_{kj}$ that \review{leverages the} diversity between $\{\mu_{k1}\}_{k = 1,2}$ \review{to delineate the clusters, }while shrinking the disparities between $\{\mu_{k2}\}_{k = 1,2}$ toward zero\review{, as they do not inform the clustering structure of the territory}. We translate these concepts into the following global-local shrinkage prior:
\begin{equation}
    \label{formula:prior_mu}
    \mu_{kj}| \phi, \zeta_j,\gamma_{kj}\sim \mathcal{N}\left(0,{\phi\zeta_j\gamma_{kj}}\right),
\end{equation}
where 
$$
\sqrt{\phi}\sim\mathcal{C}_+(0,1),\hspace{.5cm}\sqrt{\zeta_j}\sim\mathcal{C}_+(0,1),\hspace{.5cm}\sqrt{\gamma_{kj}}\sim\mathcal{C}_+(0,1),
$$
and $\mathcal{C}_+$ denotes the half-Cauchy distribution. While $\phi$ represents a global shrinkage factor, shared among all the $\mu_{kj}$, $\zeta_j$ is a disease-specific parameter that is in common with the cluster intercepts $\{\mu_{kj}\}_{k=1,\dots,K}$. A large value of $\zeta_j$ emphasises the contribution of disease $j$ to the clustering process, and a small value of  $\zeta_j$ diminishes its contribution. Lastly, the local factor $\gamma_{kj}$ acts specifically on the $kj$-th mean component, highlighting or shrinking the contribution of disease $j$ to the identification of cluster $k$. We collect the shrinkage parameters appearing in \eqref{formula:prior_mu} into the vectors $\boldsymbol{\zeta} = \{\zeta_j\}_{j = 1,\dots,d}$ and $\boldsymbol{\gamma} = \{\gamma_{kj}\}_{k = 1,\dots,K;j = 1,\dots,d}$.

By introducing progressively more local factors in Formula \eqref{formula:prior_mu}, we model the influence of each disease on the overall clustering structure of the data through the $\boldsymbol{\zeta}$, and then we investigate the contribution of each disease to the formation of individual mortality clusters through the $\boldsymbol{\gamma}$.
The local factors $\boldsymbol{\gamma}$ become particularly relevant when some diseases are informative only of a few, specific clusters. For instance, suppose that a disease $j$ is not informative of any cluster but the $k$-th. We expect the model to \reviewm{impose a large shrinkage on disease $j$} through a small value of $\zeta_j$, and to highlight the contribution of \reviewm{the same disease} in identifying the $k$-th cluster through a large value of  $\gamma_{kj}$. The total number of shrinkage parameters introduced with the prior model in \eqref{formula:prior_mu} is $Kd+d+1$ . Without $\zeta_j$, our prior distribution reduces to the well-know \textit{horseshoe prior} of  \cite{Carvalho_etal.2010}. 

 In alternative to \eqref{formula:prior_mu}, we consider also the prior distribution
\begin{equation}
    \label{formula:prior_mu2}
    \mu_{kj}| \phi, \zeta_j,\delta_{k}\sim \mathcal{N}\left(0,{\phi\delta_{k}\zeta_j}\right).
\end{equation}
The parameter $\sqrt{\delta_k}\sim\mathcal{C}_+(0,1)$ highlights or shrinks toward zero the differences across diseases within the whole $k$-th cluster, \review{following a similar rationale of \cite{Denti_etal.2023}}. We collect all the shrinkage parameters introduced through Formula \eqref{formula:prior_mu2} into the vector $\boldsymbol{\delta}=\{\delta_k\}_{k = 1,\dots,K}$. Not only does the prior in \eqref{formula:prior_mu2} simplify \eqref{formula:prior_mu}, but it is also more parsimonious than the horseshoe prior when $K, d \geq 2$, with at least one of them strictly greater than 2. This is because it requires only $K + d + 1$ shrinkage parameters, compared to $Kd + 1$ in the horseshoe prior. The assumption of separability between cluster and disease effects allows to quantify the whole contribution of disease $j$ in the formation of clusters through $\zeta_j$, and the amount of variability across diseases in cluster $k$ through $\delta_k$.   
In addition, despite not operating directly as a model selection method, $\delta_k$  shrinks all the coefficients in $\Mu_{k.}$ toward zero if the $k$-th cluster finds poor evidence from the data, allowing for a clearer identification of the number of clusters in the post-processing phase. Finally, \review{we can also obtain simpler priors taking} only a subset of the shrinkage parameters in Formulas \eqref{formula:prior_mu}-\eqref{formula:prior_mu2}. For example, a prior having only the elements $\phi$ and $\boldsymbol{\zeta}$ applies a global shrinkage and a local disease-specific shrinkage.

The use of the half-Cauchy distribution could be potentially extended by allowing its scale parameter to differ from 1, thereby enabling the inferential process to incorporate some a priori knowledge about the analysed phenomena \citep{Piironen_Vehtari.2017}. This extension can, for example, be used to specify which diseases are more likely to contribute to the clustering. However, we opt for the $\mathcal{C}_+(0,1)$ distribution because, a priori, there is often little or no knowledge of how many diseases are responsible of the formation of clusters. This is because, in mortality mapping, the primary question is whether clusters of mortality excess or deficit exist at all, with the number of clusters investigated subsequently (\citealt{Bivand_etal.2013}, Section 10.6). This differs fundamentally from other unsupervised analyses: for example, in transcriptomics,  cells are clustered based on their gene expression levels and there is typically a priori knowledge of both the number of cell types present in a tissue and of the genes known to be primarily linked to its biological processes \citep{zhao_etal.2021}.

The parameters $\tau_k$ and $\rho_k$ calibrate the spatial distribution of $\boldsymbol{\psi}_{.k}$. In Supplementary Section 1, we discuss the roles of these two parameters on the clustering probabilities. In summary, it emerges that  $\tau_k$ has no impact on the mean of the values in $\boldsymbol{\psi}_{.k}$, but it has a large impact on their variability: the larger $\tau_k$, the greater the variability associated with each $\psi_{ik}$ (see for instance Supplementary Figure 4).
The inverse gamma distribution is conjugate with the Gaussian model; therefore, a natural choice of a prior model for $\tau_k$ would be $\mathcal{IG}(\alpha_\tau,\beta_\tau)$. However, \review{one could also avoid inferring an extra parameter and instead fixing $\tau_k$ sufficiently large to guarantee an adequate level of uncertainty in the clustering probabilities. In the simulation experiment proposed in Section \ref{subsec:simulation2}, we compare different fixed values of $\tau_k$ with the use of its conjugate prior.} 

\begin{figure}[t]
    \centering
    \includegraphics[width=0.5\linewidth]{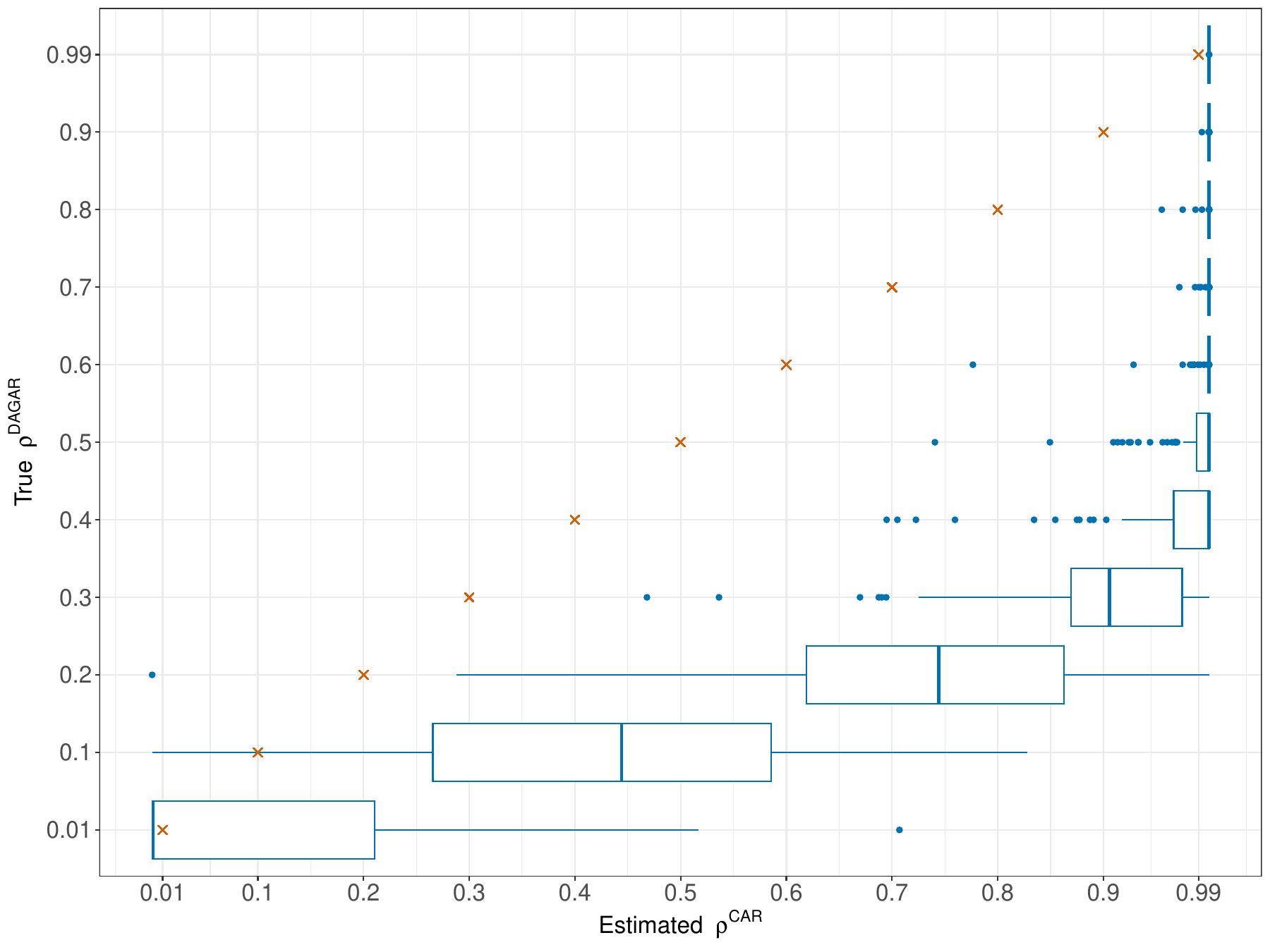}
    \caption{Distribution of the maximum likelihood estimates of $\rho^{\mathrm{CAR}}$ obtained under different realisations of a DAGAR spatial process. \reviewm{Data are simulated using} different values of spatial correlation $\rho^{\mathrm{DAGAR}}$ that appear of the y-axis. For every $\rho^{\mathrm{DAGAR}}$, we simulate {200} spatial realisations and we estimate the CAR. Red crosses have been inserted to help visualising the discrepancy between the true true $\rho^{\mathrm{DAGAR}}$ and the estimated $\rho^{\mathrm{CAR}}$. \reviewm{Simulations are based on the map displayed in Supplementary Figure 2.}}
    \label{fig:DAGARvsCAR}
\end{figure}

\review{A second relevant aspect that emerges from the discussion in Supplementary Section 1 is that only values of $\rho_k$ close to 1 allow the CAR prior to handle spatially correlated data. 
The limit case $\rho_k = 1$ is not admittable because $\mathbf{D}-\mathbf{W}$ is not a full-rank precision matrix, implying that the integral of the joint prior density of $\boldsymbol{\psi}_{.k}$ is not finite. 
Nonetheless, when the CAR model is used as a prior, \cite{Banerjee_etal.2015} encourage using the kernel of a CAR with $\rho_k = 1$}, trusting that the posterior distribution will be proper due to the contribution of the likelihood function. \review{This solution is known in the literature as intrinsic CAR.} However, \review{in \perla, the CAR serves not as a prior for a likelihood parameter, but for a parameter that defines the distribution of $\mathbf{Z}$, which is itself a random variable. Therefore, being in a second level of hierarchy in the model structure, we believe that the usage of an intrinsic CAR prior should be properly investigated and compared with a proper CAR prior that takes for instance $\rho_k = 0.99$ for every $k$.} Alternatively, we propose also inferring $\rho_k$ from the data. \cite{Carlin_Banerjee.2003} suggest using $\rho_k\sim\mathcal{B}(18,2)$, where $\mathcal{B}$ denotes the beta distribution, to give large probability to values of $\rho_k$ close to 1. It is worth noting that the density of this beta distribution peaks at 0.944 and decreases rapidly as values approach 1, thus discouraging the limit case $\rho_k = 1$. \review{To construct a prior distribution for $\rho_k$, first it could be worth evaluating how the estimate of $\rho_k$ changes with respect to different levels of spatial correlation in the data. To do so,} we simulated a series of spatial datasets using a DAGAR model, based on a map of $n = 259$ U.S. counties on the West side of the U.S. map (see Supplementary Figure 2). \review{The choice of the DAGAR as a data generative mechanism allows us to control the level of spatial correlation (called $\rho^{\mathrm{DAGAR}}$) in the simulations.} For every level of $\rho^{\mathrm{DAGAR}}$ used as input, we produced 200 realisations from the spatial process, and we estimated the CAR model. Figure \ref{fig:DAGARvsCAR} reports the distribution of the maximum likelihood estimates of $\rho^{\mathrm{CAR}}$ under different $\rho^{\mathrm{DAGAR}}$. We notice that a clear correspondence between the  \review{estimate of $\rho^{\mathrm{CAR}}$ and the level of spatial correlation $\rho^{\mathrm{DAGAR}}$} does not emerge. In fact, when $\rho^{\mathrm{DAGAR}}\geq 0.4$, the distributions of the estimates of $\rho^{\mathrm{CAR}}$ concentrate above 0.9, while, when $0.1 \leq \rho^{\mathrm{DAGAR}}\leq0.3$, the estimates of $\rho^{\mathrm{CAR}}$ span over the whole $(0,1)$ interval. Lastly, when $\rho^{\mathrm{DAGAR}} = 0.01$, the most likely value of $\rho^{\mathrm{CAR}}$ appears to be 0.01, although some uncertainty is also evident. \review{Based on these considerations, a prior distribution for $\rho_k$ that favours values near 0 or 1 would be appropriate to capture a wide range of spatial correlation scenarios.}  
We thus propose the following mixture prior:
\begin{equation}
\label{formula:prior_rho_SS}
\rho_k\sim 0.5\mathcal{B}(2,18)+0.5\mathcal{B}(18,2).
\end{equation}
We choose a symmetric structure to maintain non-informativeness about the value of $\rho_k$, giving equal probability to very small or very high values. The mixture component $\mathcal{B}(2,18)$ reflects the idea of the $\mathcal{B}(18,2)$ component, assigning high probability mass to small values of $\rho_k$, while discouraging the case of complete independence $\rho_k = 0$, which is unusual in spatial data modelling. By placing most of the probability mass toward values close to zero or one - reflecting the almost absence or large presence of spatial correlation - the distribution in \eqref{formula:prior_rho_SS} retraces the idea of a continuous spike-and-slab prior \citep{rockova.2018}. Notice also that \eqref{formula:prior_rho_SS} represents a mixture model that integrates out the mixing probability $\theta_k$, whose prior distribution is assumed to be $\theta_k\sim\mathcal{B}(c_\theta,c_\theta)$, for any $c_\theta > 0$.  One of the goals of the simulation experiments proposed in Section \ref{subsec:simulation2} is comparing the two prior proposals for $\rho_k$, investigating whether they make any difference in terms of clustering accuracy and model fitting.

We complete the model specification with the prior distributions on $\boldsymbol{\beta}$ and $\bSigma$. We generically assume the conjugate and poorly informative priors ${\beta}_{\ell j}\sim\mathcal{N}(0,10)$ and $\bSigma\sim\mathcal{IW}(d,\mathbf{I}_d)$, where $\mathcal{IW}$ denotes the inverse Wishart distribution.

\begin{figure}[t]
    \centering
    \includegraphics[width=0.5\linewidth]{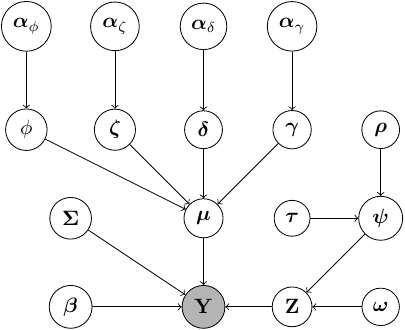}
    \caption{
    Directed acyclic graph (DAG) illustrating the structure of PERLA, including also the augmented variables. Grey circles represent observed data, while white circles represent random variables. Notice that this graph contains the parameters from both the prior formulations \eqref{formula:prior_mu} and \eqref{formula:prior_mu2} considered for $\Mu$. Hyperparameters are omitted as they have been pre-set. 
    }
    \label{fig:DAG}
\end{figure}

Figure \ref{fig:DAG} illustrates the hierarchical structure of \perla~through a DAG. Note that this scheme also includes the augmented variables $\boldsymbol{\omega}$, $\boldsymbol{\alpha}_\phi$, $\boldsymbol{\alpha}_\zeta$, $\boldsymbol{\alpha}_\delta$ and $\boldsymbol{\alpha}_\gamma$, which will be introduced in the next section to allow the posterior simulation of some model parameters to be performed via Gibbs sampling.

\section{Posterior inference}
\label{sec:posterior_inference}

\subsection{The MCMC algorithm}
\label{subsec:mcmc_algorithm}

We carry out the inference on \reviewm{\perla~via posterior simulation}. We present a Markov chain Monte Carlo scheme that allows a direct sampling from the conditional posterior distributions of most of the model parameters, requiring a Metropolis-Hastings step only for a few of them. This translates into the practical advantage of requiring fewer tuning parameters that can be hard to set when either the number of clusters $K$ or the number of responses $d$ grow.  With respect to HPM, which requires particular attention for sampling the inverse temperature parameter, our method presents no critical step.

As we already mentioned, the stick-breaking representation in \eqref{formula:stick_breaking} allows to rewrite the multinomial model as a sequence of Bernoulli probability functions. To facilitate the posterior sampling, we introduce the latent variables $\omega_{ik}\sim \mathcal{PG}(N_{ik},0)$, for $i=1,\dots,n$ and $k = 1,\dots,K-1$, where $\mathcal{PG}(\cdot,\cdot)$ denotes the P\'olya-gamma random variable. Through this data augmentation strategy, we can update the transformed probabilities $\boldsymbol{\psi}_{.k}$ by drawing values from the conditional posterior distribution of $\boldsymbol{\psi}_{.k}|\omega_k,\mathbf{Z}_{.k},\rho_k$, that is
\begin{align*}
\pi(\boldsymbol{\psi}_{.k}|\omega_k,\mathbf{Z}_{.k},&\tau_k,\rho_k)\propto \left\{\prod_{i=1}^n\exp\left(\kappa_{ik}\psi_{ik}-\dfrac{\omega_{ik}\psi^2_{ik}}{2}\right)\right\}\mathcal{N}_n\{\boldsymbol{\psi}_{.k}|\mathbf{0},\tau_k(\mathbf{D}-\rho_k\mathbf{W})^{-1}\},
\end{align*}
where $\kappa_{ik} = \Z_{ik}-N_{ik}/2$ and $\mathcal{N}_q(\mathbf{a}|\mathbf{b},\mathbf{C})$ denotes the density of the $q$-variate normal distribution with mean vector $\mathbf{b}$ and covariance matrix $\mathbf{C}$, evaluated in $\mathbf{a}$. As the number of spatial areas $n$ grows, and with it the dimension of the neighbour matrix $\mathbf{W}$, it becomes more convenient to update the $\psi_{ik}$ iteratively, rather than generating the full vector $\boldsymbol{\psi}_{.k}$. This can be performed thanks to the \textit{Brook's lemma}, which allows to factorise the joint distribution in \eqref{formula:CAR} into a series of univariate conditional distributions derived from the neighbourhood structure imposed by $\mathbf{W}$ \citep{Besag.1974}. The conditional posterior of $\psi_{ik}$ is 
$$
\psi_{ik}|\boldsymbol{\psi}_{-i,k},\mathrm{Z}_{ik},\omega_{ik},\tau_k,\rho_k\sim \mathcal{N}\left\{
\varsigma^2_{ik}\left(\kappa_{ik}+\dfrac{\rho_k}{\tau_k}\mathbf{W}^T_{i,-i}\boldsymbol{\psi}_{-i,k}\right),\varsigma^2_{ik}\right\},
$$
where  $\varsigma^2_{ik} = {\tau_k}/\left({\tau_k\omega_{ik}+\mathrm{D}_{ii}}\right)$ \reviewm{and $\boldsymbol{\psi}_{-i,k}$ denotes the vector $\boldsymbol{\psi}_{.k}$ without the $i$-th element}. The augmented variable $\omega_{ik}$ is updated drawing from $\omega_{ik}|\mathbf{Z}_{i.}, \psi_{ik}\sim \mathcal{PG}(\kappa_{ik},\psi_{ik})$ if $N_{ik} = 1$, or setting $\omega_{ik} = 0$  if $N_{ik} = 0$.

The parameter $\rho_k$ under the prior model in \eqref{formula:prior_rho_SS} is the only case for which a closed-form updating scheme is not available. We implement a random walk Metropolis-Hastings on the transformed parameter $\mathrm{logit}(\rho_k)$, using a Gaussian distribution centred in the  value sampled at the previous iteration, and with variance $\tilde{\sigma}^2_\rho = 3$, in the role of proposal.

To update the clustering labels, it is now sufficient to transform the quantities $\{\boldsymbol{\psi}_{.k}\}_{k=1,\dots,K-1}$ into the clustering probabilities $\{\boldsymbol{\pi}_i\}_{i=1,\dots,n}$, then drawing a value from the multinomial distribution whose probabilities are defined as
$$
\Pr(\Z_{ik}=1|{\pi}_{ik},\mathbf{y}_{i.},\Mu_{k.},\boldsymbol{\beta},\bSigma)\propto \pi_{ik}\mathcal{N}_d(\mathbf{y}_{i.}|\Mu_{k.}+\mathbf{x}^T_{i.}\boldsymbol{\beta},\bSigma).
$$

Under the prior model in \eqref{formula:prior_mu}, we update the shrinkage parameters $\boldsymbol{\zeta}$,  $\boldsymbol{\gamma}$ and $\phi$ following the data augmentation strategy of \cite{Makalic_Schmidt.2016}, which exploits the scale mixture representation of the half-Cauchy distribution. We thus introduce a set of augmented scale parameters for each shrinkage parameter considered: $\boldsymbol{\alpha}_\zeta = \{\alpha_{\zeta_j}\}_{j = 1,\dots,d}$, $\boldsymbol{\alpha}_\gamma = \{\alpha_{\gamma_{kj}}\}_{k = 1,\dots,K;j = 1,\dots,d}$, and $\alpha_{\phi}$. Each of these augmented parameters is assumed to distribute according to an inverse gamma $\mathcal{IG}(1/2,1)$. Then, for $j = 1,\dots,d$, draw from
\begin{align}
\label{formula:Gibbs_updateZeta}
\alpha_{\zeta_j}|\zeta_j&\sim \mathcal{IG}\left(1,\dfrac{1}{\zeta_j}+1\right),\hspace{.7cm}\zeta_j|\boldsymbol{\mu}_{.j},\phi,\boldsymbol{\gamma}_{.j},\alpha_{\zeta_j}\sim \mathcal{IG}\left(\dfrac{K+1}{2},\dfrac{1}{2\phi}\sum_{k = 1}^K\dfrac{\mu_{kj}^2}{\gamma_{kj}}+\dfrac{1}{\alpha_{\zeta_j}}\right).
\end{align}
For $k = 1,\dots,K$ and $j = 1,\dots,d$, draw from
\begin{align}
\label{formula:Gibbs_updateGamma}
\alpha_{\gamma_{kj}}|\gamma_{kj}\sim \mathcal{IG}\left(1,\dfrac{1}{\gamma_{kj}}+1\right),\hspace{.7cm}\gamma_{kj}|{\mu}_{kj},\phi,{\zeta}_j,\alpha_{\gamma_{kj}}\sim \mathcal{IG}\left(1,\dfrac{\mu_{kj}^2}{2\phi\zeta_j}+\dfrac{1}{\alpha_{\gamma_{kj}}}\right).
\end{align}
Lastly, draw from
\begin{align}
\label{formula:Gibbs_updatePhi}
a_{\phi}|\phi\sim \mathcal{IG}\left(1,\dfrac{1}{\phi}+1\right),\hspace{.7cm}
\phi|\boldsymbol{\mu},\boldsymbol{\zeta},\boldsymbol{\gamma},\alpha_{\phi}\sim \mathcal{IG}\left(\dfrac{Kd+1}{2},\sum_{j = 1}^d\dfrac{1}{2\zeta_j}\sum_{k = 1}^K\dfrac{\mu_{kj}^2}{\gamma_{jk}}+\dfrac{1}{\alpha_{\phi}}\right).
\end{align}
Let $\mathcal{C}_k = \{i = 1,\dots,n: \mathrm{Z}_{ik} = 1\}$ be the set denoting which observations are assigned to cluster $k$, and $n_k = |\mathcal{C}_k|$. The cluster intercepts are updated drawing from
\begin{equation}
    \label{formula:Gibbs_updateMu}
\boldsymbol{\mu}_{k.}|\mathbf{Y}_{\mathcal{C}_k.}, \boldsymbol{\beta},\phi,\boldsymbol{\zeta},\boldsymbol{\gamma}_{k.},\bSigma\sim \mathcal{N}_d\left[\bSigma^*_{k}\left\{\bSigma^{-1}\left(\mathbf{Y}_{\mathcal{C}_k.}-\mathbf{X}_{\mathcal{C}_k.}\boldsymbol{\beta}\right)\right\},\bSigma^*_{k}\right],
\end{equation}
where $\bSigma^*_{k} = \left[\left\{\phi\;\mathrm{diag}(\zeta_1\gamma_{k1},\dots,\zeta_d\gamma_{kd})\right\}^{-1}+n_k\bSigma^{-1}\right]^{-1}$.

If instead the prior model in \eqref{formula:prior_mu2} is considered, we introduce the set of augmented scale parameters $\boldsymbol{\alpha}_\delta = \{\alpha_{\delta_k}\}_{k = 1,\dots, K}$, each of which distributes according to an inverse gamma $\mathcal{IG}(1/2,1)$. Then, for $k = 1,\dots,K$, draw from
\begin{align*}
\alpha_{\delta_k}|\delta_k\sim \mathcal{IG}\left(1,\dfrac{1}{\delta_k}+1\right),\hspace{.7cm}
\delta_k|\boldsymbol{\mu}_{k.},\phi,\boldsymbol{\zeta},\alpha_{\delta_k}\sim \mathcal{IG}\left(\dfrac{d+1}{2},{\dfrac{1}{2\phi}\sum_{j = 1}^d\dfrac{\mu_{kj}^2}{\zeta_j}}+\dfrac{1}{\alpha_{\delta_k}}\right).
\end{align*}
The rest of the updates can be performed as in Formulas \eqref{formula:Gibbs_updateZeta}, \eqref{formula:Gibbs_updatePhi} and \eqref{formula:Gibbs_updateMu}, just replacing $\gamma_{kj}$ with $\delta_k$.

Finally, the updates of $\boldsymbol{\tau}=\{\tau_k\}_{k=1,\dots,K-1}$, $\bSigma$ and $\boldsymbol{\beta}$ are straightforward to perform, thanks to the conjugacy between the likelihood and the selected priors. Details can be found for example in \cite{Marin_Robert.2007}.

\review{
In our software implementation, we initialise the MCMC algorithm as follows. The $n$ observations are randomly assigned to the $K$ clusters; each element in $\boldsymbol{\psi}$ and $\boldsymbol{\beta}$ is generated from $\mathcal{N}(0,1)$; the P\'olya-gamma parameters in $\boldsymbol{\omega}$ are drawn from a gamma random variable $\mathcal{G}(1,1)$; and all shrinkage parameters are drawn from $\mathcal{C}_+(0,1)$.  
The cluster centroids $\Mu$ and the covariance matrix $\bSigma$ are the only two parameters not randomly initialised. The former are taken equal to the ordinary least squares estimate $(\mathbf{Z}^T\mathbf{Z})^{-1}\mathbf{Z}^T\mathbf{Y}$, while the latter are initialised as $\sum_{i=1}^n (\mathbf{y}_{i.} - \overline{\mathbf{y}})^T (\mathbf{y}_{i.} - \overline{\mathbf{y}}) / (n-1)$, where $\overline{\mathbf{y}}$ denotes the vector of sample means for the $d$ outcome variables.
} 

To provide point estimates of the model parameters, we consider the posterior mode. With the notation $\mathcal{M}(\theta|\mathrm{data})$, we denote the posterior mode of a generic model parameter $\theta$, given the data. A relevant quantity in mortality mapping analyses is also the posterior probability of observing an excess mortality in a specific region, compared to the whole territory. Following the considerations of \cite{Richardson_etal.2004}, we state that cluster $k$ shows evidence of mortality excess due to disease $j$ if $\Pr(\mu_{kj}>0|\mathrm{data})\geq 0.95$, and a deficit of mortality if $\Pr(\mu_{kj}>0|\mathrm{data})\leq 0.05$. These quantities can be computed straightforwardly from the output of the MCMC algorithm.

\subsection{Post-processing and model selection}
\label{subsec:postprocessing_modelselection}

The presented MCMC scheme is susceptible to the common label-switching phenomenon, which arises due to the unidentifiability of the mixture components and leads the posterior distributions of $\Mu$ to be multimodal\reviewm{, even when the number of mixture components is correctly specified}. In practice, this makes inference on model parameters challenging. To disentangle the mixture components, we explore the ECR algorithm of \cite{Papastamoulis_Iliopoulos.2010}, considering the partition associated to the largest log-likelihood value as the pivotal one. In addition to this method, the \texttt{R} package \texttt{label.switching} \citep{Papastamoulis.2016} implements numerous alternative algorithms. Not only are these tools useful to relabel the posterior draws, but they are also designed to return an optimal partition of the data based on the MCMC output. This process can also be accomplished using other algorithms, such as the SALSO greedy search method proposed by \cite{Dahl_etal.2022}.


\perla~can be fitted with different setups, varying the number of clusters $K$ and the prior distributions of the model parameters. Therefore, a criterion that allows users to evaluate different model setups and determine which is preferable for the analysed data is necessary. 
This approach aligns with the philosophy of parsimonious clustering \citep{celeux_govaert.1995,scrucca_etal.2016}, which involves comparing the same type of model with different degrees of flexibility and determining the setup that best fits the data.

We conduct the model selection using the $\mathrm{DIC}_3$ information criterion of \cite{Celeux_etal.2006}, which is recommended for mixture models, and whose properties have been studied by \cite{Kim.2021}. \review{A detailed description on how to compute $\mathrm{DIC}_3$ for \perla~is given in Supplementary Section 2. Models that return smaller $\mathrm{DIC}_3$ values are preferable to those producing larger values.}

We already discussed in Section \ref{subsec:priors} that different setups for $\Mu$ imply different numbers of model parameters. Assuming that the global shrinkage parameter $\phi$ is always inserted into the model, we report in \review{Table \ref{tab:list_of_shrinkage_parameters}} the possible combinations of shrinkage factors that we consider for \perla, along with the corresponding code notation used in our \texttt{R} package \texttt{perla}. Given the model dimension, users can evaluate all the five combinations of shrinkage factors shown in the table, starting with simpler models and growing the complexity.

Lastly, it is crucial to underline that model selection cannot rely solely on an information criterion, but must be performed considering also other factors, such as the quality of model fitting, evaluated through the MCMC output, and the interpretability of the clusters.

\section{Simulation studies}
\label{sec:simulation_experiments}

We study the performance of \perla~with \review{three} simulation experiments that aim at recreating some possible real data scenarios and conditions. The map considered for the simulations consists of $n = 259$ counties of the U.S. map introduced in Section \ref{sec:introduction}, located West of the 110\textdegree~W longitude meridian. A representation of this territory is given in Supplementary Figure 2. 

Every version of \perla~considered is fitted running the MCMC algorithm four times separately, each for 10,000 iterations, discarding the first half of each run, and merging the remaining draws into a unique chain of length 20,000. 
Finally, the ECR algorithm is applied to resolve eventual label switching. 

\subsection{Simulation model}
\label{subsec:simulation_model}

 The data generating mechanism that we adopt follows the scheme of \perla~displayed in Figure \ref{fig:DAG}, \review{but using the} DAGAR prior in place of the CAR to generate the clustering probabilities.  This choice allows us to control the amount of spatial correlation simulated in each cluster. 
The use of the DAGAR requires the ordering of the areas. We thus order the counties from the most southern to the most northern. If $i = 1,\dots,259$ is the index denoting the counties, then $i = 1$ corresponds to the most southern county, and $i = 259$ to the most northern. Being $K\true$ the true number of clusters in the map, the simulation model generates the transformed clustering probabilities as
\begin{equation*}
    \label{formula:modelsimulation1}
    \psi\true_{ik}|\boldsymbol{\psi}\true_{\mathscr{N}(i),k}\sim \mathcal{N}\left(
\dfrac{\rho\true_k}{1+(n_{<i}-1){\rho\true_k}^2},
\dfrac{1-{\rho\true_k}^2}{1+(n_{<i}-1){\rho\true_k}^2}
    \right)
\end{equation*}
where $k = 1,\dots,K\true-1$, $\mathscr{N}(i)$ denotes the set of spatial areas that are neighbours of $i$ and precede $i$ according to the imposed ordering, and $n_{<i} = |\mathscr{N}(i)|$.  The quantity $\rho\true_k\in(0,1)$ denotes the amount of spatial correlation of the probabilities of belonging to cluster $k$. For further details on the DAGAR model, the reader can refer to \cite{Datta_etal.2019}.

By transforming the quantities $\{\psi\true_{ik}\}_{i=1,\dots,259;k = 1,\dots,K\true-1}$ into $\{\pi\true_{ik}\}_{i=1,\dots,259;k = 1,\dots,K\true}$ through the inverse transformation of the stick-breaking weights, we would obtain highly unbalanced clustering probabilities. To simulate approximately balanced clusters, the inverse transformation of the stick-breaking weights must be applied to $\{\psi\true_{ik}-\log(K-k)\}_{i=1,\dots,259;k = 1,\dots,K\true-1}$. Then, the model allocates the spatial areas into clusters, storing the labels into the matrix $\mathbf{Z}\true$.
Lastly, given the matrix of centroids $\Mu\true$ and the covariance matrix of diseases $\bSigma\true$, the model generates the observed data in the form of the logarithm of mortality rates of $d$ diseases (named also death causes) as
\begin{equation}
    \label{formula:modelsimulation2}
\mathbf{y}_{i.}\sim\mathcal{N}_d\left\{\left(\mathbf{Z}\true_{i.}\right)^T\Mu\true,\bSigma\true\right\}.
\end{equation}

In \review{all} the proposed simulation experiments, we take our conclusions using 20 replicates of the data generated under the same experimental conditions. 

\subsection{Simulation 1}
\label{subsec:simulation1}
The first experiment serves a dual purpose. On one hand, it compares \perla~with some competing clustering models, such as mixture models and SSS. 
On the other hand, it examines the gain brought by the use of shrinkage priors, such as those appearing in Formulas \eqref{formula:prior_mu} and \eqref{formula:prior_mu2}, in posterior inference compared to using non-informative priors when the number of death causes considered in the analysis is large. 

\begin{figure}[t]
    \centering
    \includegraphics[width=0.4\linewidth]{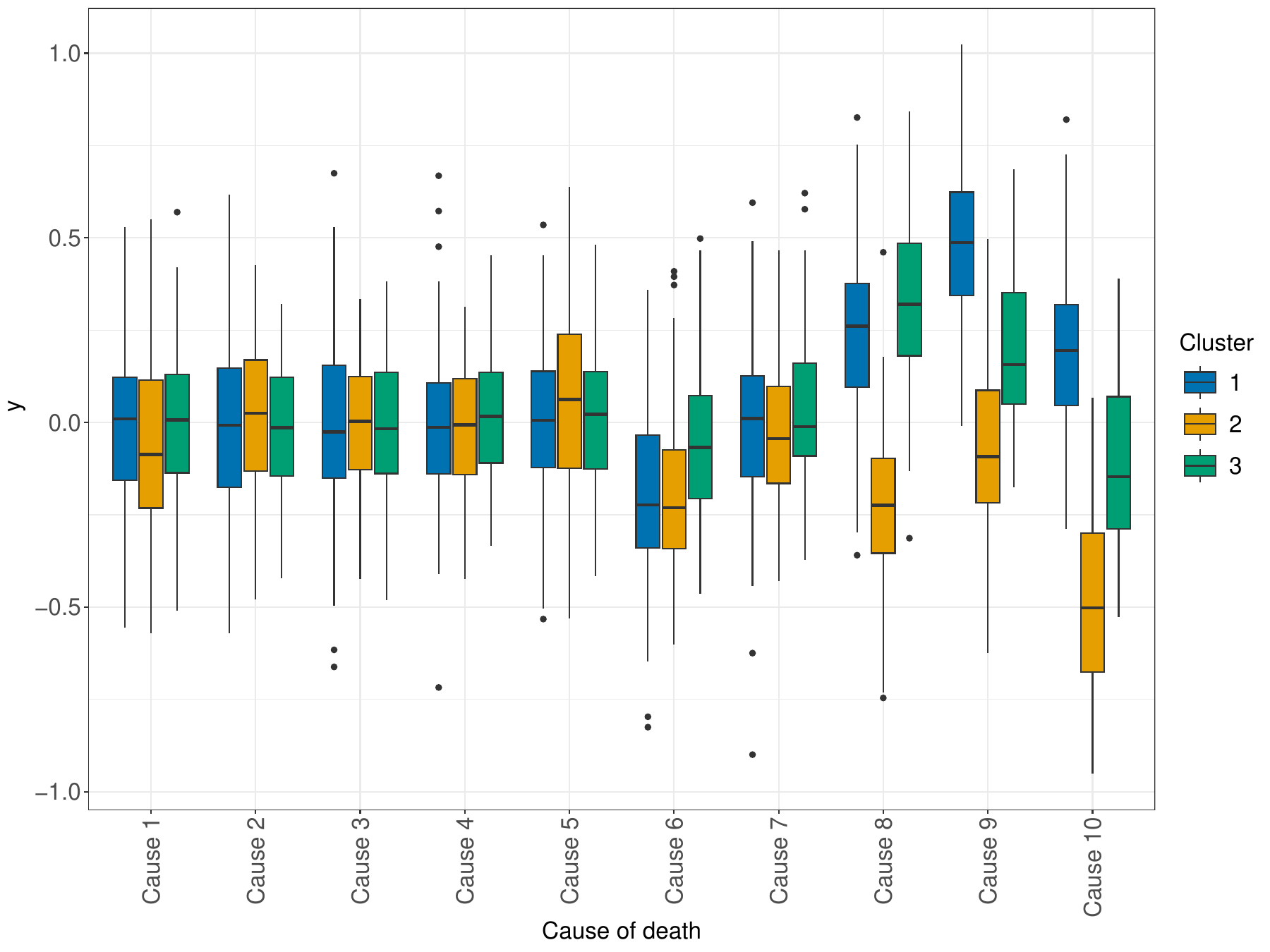}
    \includegraphics[width=0.4\linewidth]{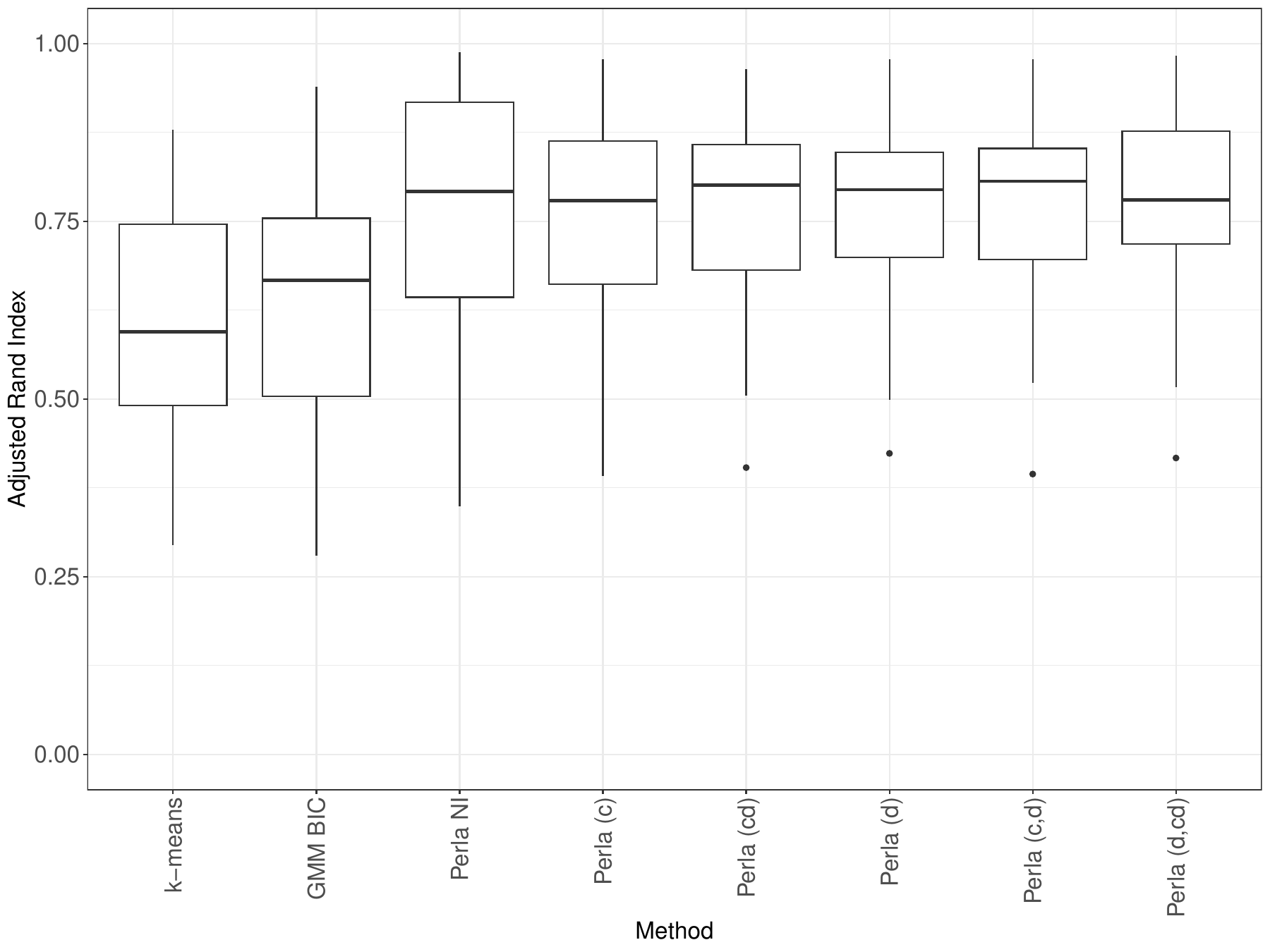}
    \caption{\review{Data and results from Simulation 1. Left: boxplots representing the distribution of $d = 10$ outcomes in  $K\true = 3$ clusters, as they appear in one of the 20 simulated maps.  Death causes 1-5 and 7 are not informative of the clustering structure. Right: boxplots distribution of the ARI obtained from several models on the 20 simulated datasets, using $K = 4$. `Perla NI' denotes the implementation of PERLA~that uses non-informative priors.  We do not report any results obtained with \texttt{Dclusterm} as it never identifies any cluster.
    }}
    \label{fig:sim1_rand}
\end{figure}

To accomplish this, we propose a scenario where the formation of mortality clusters is driven only by a subset of the $d$ diseases considered. We use $K\true = 3$ and $d = 10$. For each of the 20 simulated datasets, we induce large spatial correlation across clustering probabilities, taking ${\rho}\true_k\sim\mathcal{U}(0.8,1)$, for $k = 1,2$. Next, we randomly determine the diseases that do not contribute to cluster formation: for $j = 1,\dots,d$, we introduce a random variable $I_j\sim\mathcal{B}e(0.5)$, which takes value 1 if disease $j$ informs the clustering, and 0 if it does not. Then, we draw $\mu\true_{kj}|I_j = 1\sim\mathcal{U}(-0.5,0.5)$, while we set $\mu_{kj} = 0$ if $I_j = 0$, for all $k$.
To simulate the covariance matrix of the $d$ diseases, we draw from the uniform distribution $\mathcal{U}(-0.1,0.1)$ to generate the correlations values, which are used to fill the off-diagonal elements of an identity matrix of order $d$. Then, we check that the resulting matrix is semi-positive definite. Finally, we multiply all the elements of the correlation matrix by a factor of 0.05 to obtain $\bSigma\true$. The 20 synthetic datasets are generated following the simulation model described in Section \ref{subsec:simulation_model}. The left panel of Figure \ref{fig:sim1_rand} illustrates the distribution of the outcome variables in one of the 20 datasets, where $I_j = 0$ for $j = 1,\dots,5$ and $j = 7$.

For each of the 20 scenarios, we fit the following models:
\begin{itemize}
    \item \perla, with shrinkage factors $\phi\zeta$, $\phi\delta$, $\phi\gamma$, $\phi\zeta\delta$, and $\phi\zeta\gamma$. \review{To recall them, we use the notation reported in the left column of Table \ref{tab:list_of_shrinkage_parameters}};
    \item \perla, with the non-informative priors $\mu_{kj}\sim\mathcal{N}(0,10)$, for all $k$ and $j$; 
    \item a multivariate GMM, selected with the BIC criterion, using the \texttt{R} library \texttt{mclust} \citep{scrucca_etal.2016}; 
    \item the k-means algorithm, considered also by \cite{Aungkulanon_etal.2017};
    \item   the SSS of \cite{gomez-rubio_etal.2019}, implemented in the \texttt{R} library \texttt{Dclusterm}. This algorithm fits a series of generalised linear models using the observed death counts as outcomes, and the expected death counts as offsets. However, our simulation model produces continuous data. Therefore, we apply \texttt{Dclusterm} within a linear regression model, rather than a Poisson regression. In addition, this method does not handle the analysis of multiple causes of death simultaneously, so it needs to be applied to every disease separately.
\end{itemize}

To the best of our knowledge, a library software that implements the multivariate HPM is not available in the \texttt{R} computing environment. In addition, the current version of \texttt{bayesImageS} \citep{moores_etal.2020} does not fully support non-regular lattice data. This does not allow us to compare \perla~with HPM.

We apply \perla, GMM and k-means using $K=4$ to test how effectively each model can retrieve the clustering structure of the data when the number of clusters is incorrectly specified. 
The right panel of Figure \ref{fig:sim1_rand} displays the \review{adjusted Rand index (ARI, \citealt{Hubert_Arabie.1985})} distribution on the 20 datasets given by each model. 
We notice that all the configurations of \perla~lead to comparable results in terms of classification accuracy. This result can be explained by the presence of some $\mu^{\text{true}}_{kj}$ whose value is close to 0 for some $k$ even when $I_j = 1$. For example, in the left plot of Figure \ref{fig:sim1_rand} we see that the distribution of $\{y_{i6}: \mathrm{Z}\true_i = 3\}$, \textit{i.e.}, the death cause 6 in cluster 3, is much closer to zero than the distribution of the same death cause in the two other clusters. Therefore, in this simulation context, it is reasonable that different combinations of the shrinkage parameters can be equally efficient in retrieving the clustering structure.
As expected, all the configurations of \perla~globally perform better than mixture models not informed by the spatial data structure. 
We do not report the results obtained with the \texttt{Dclusterm} algorithm, as it failed to retrieve the clustering structure in all of the 20 datasets.

\begin{figure}[t]
    \centering
    \includegraphics[width=9.7cm,height = 6.5cm]{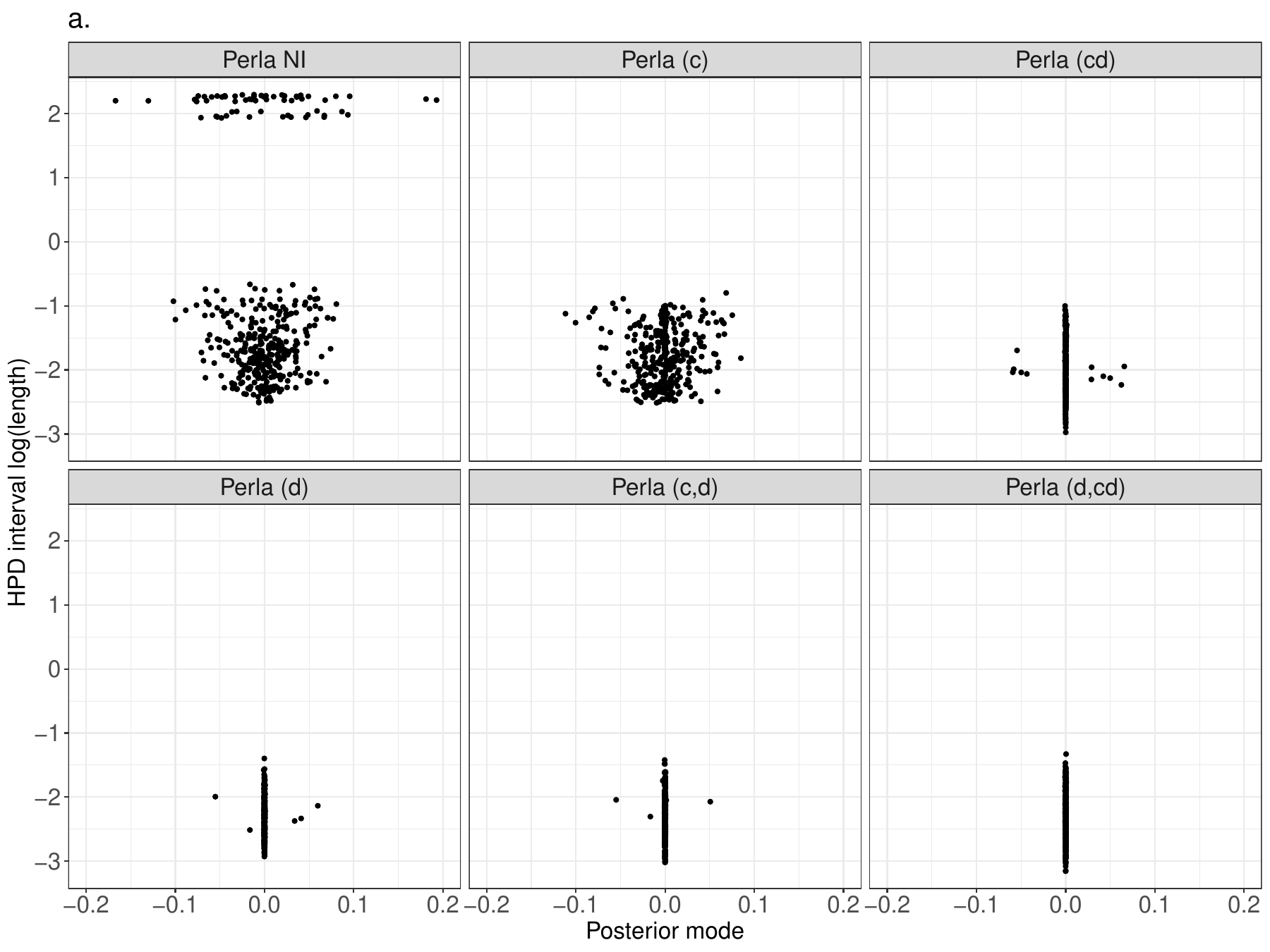}
    \includegraphics[width=10cm,height = 6cm]{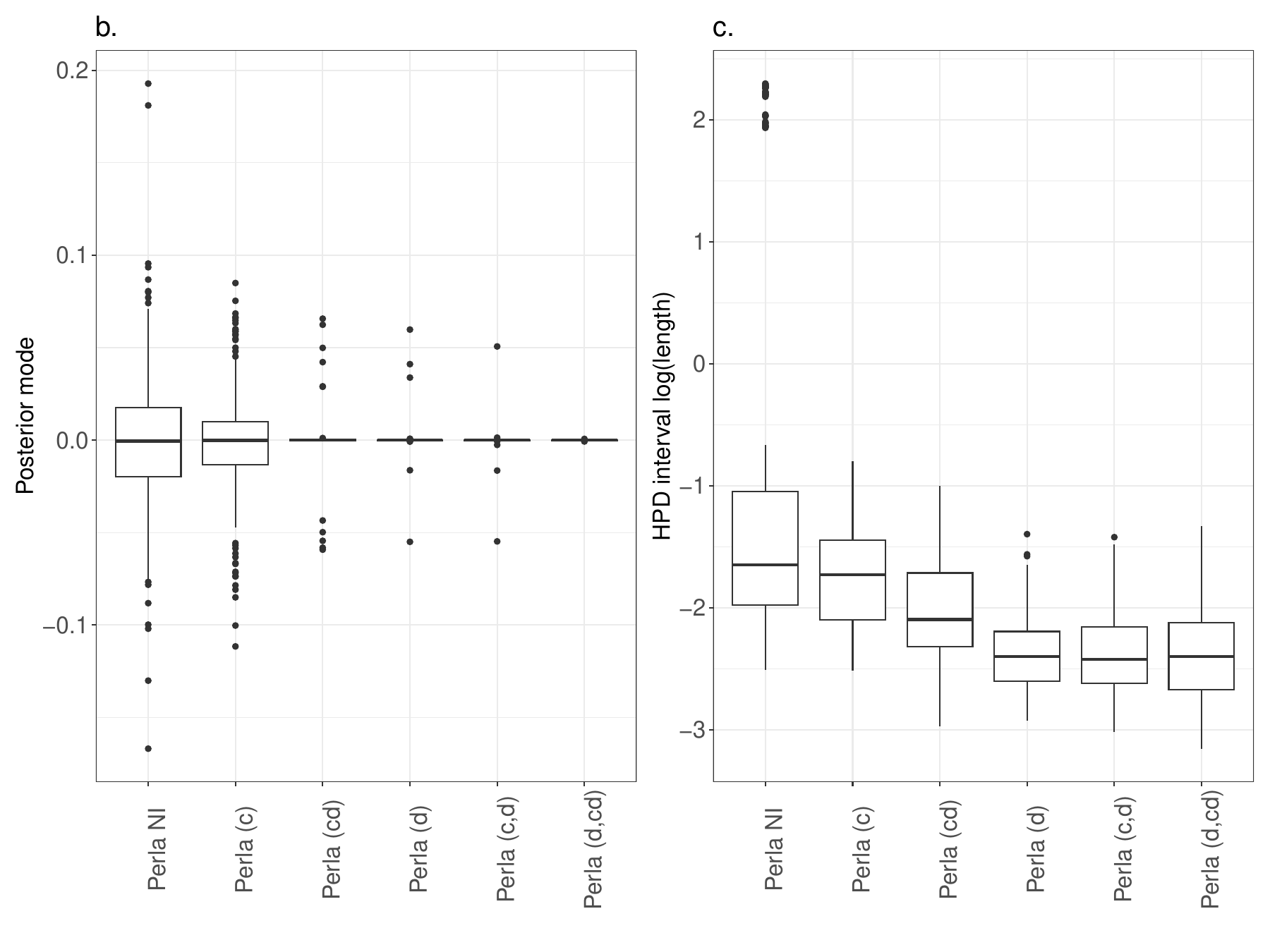}
    \caption{Results from Simulation 1 based on 20 simulated maps. The graphs display the posterior modes and of logarithm of the 95\% HPD interval lengths of the $\mu_{kj}$ for those $j$ such that $I_j = 0$, obtained with six versions of PERLA~under study. The top row displays the joint values, while the bottom row displays their marginal distributions using boxplots. `Perla NI' denotes the implementation of PERLA~that uses non-informative priors.}
    \label{fig:sim1_not_informative}
\end{figure}

\begin{figure}[t]
    \centering
    \includegraphics[width=9.7cm,height = 6.5cm]{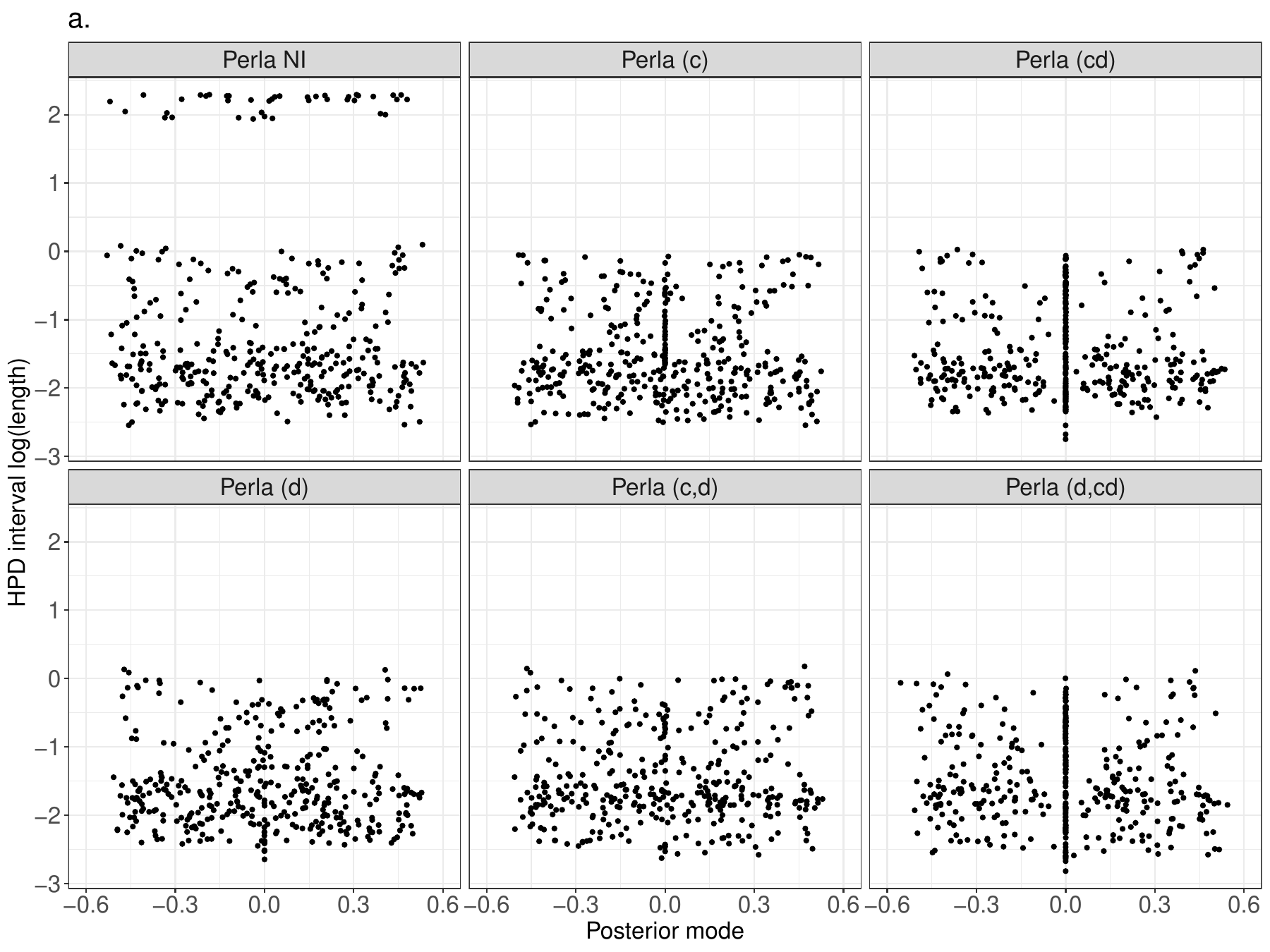}
    \includegraphics[width=10cm,height = 6cm]{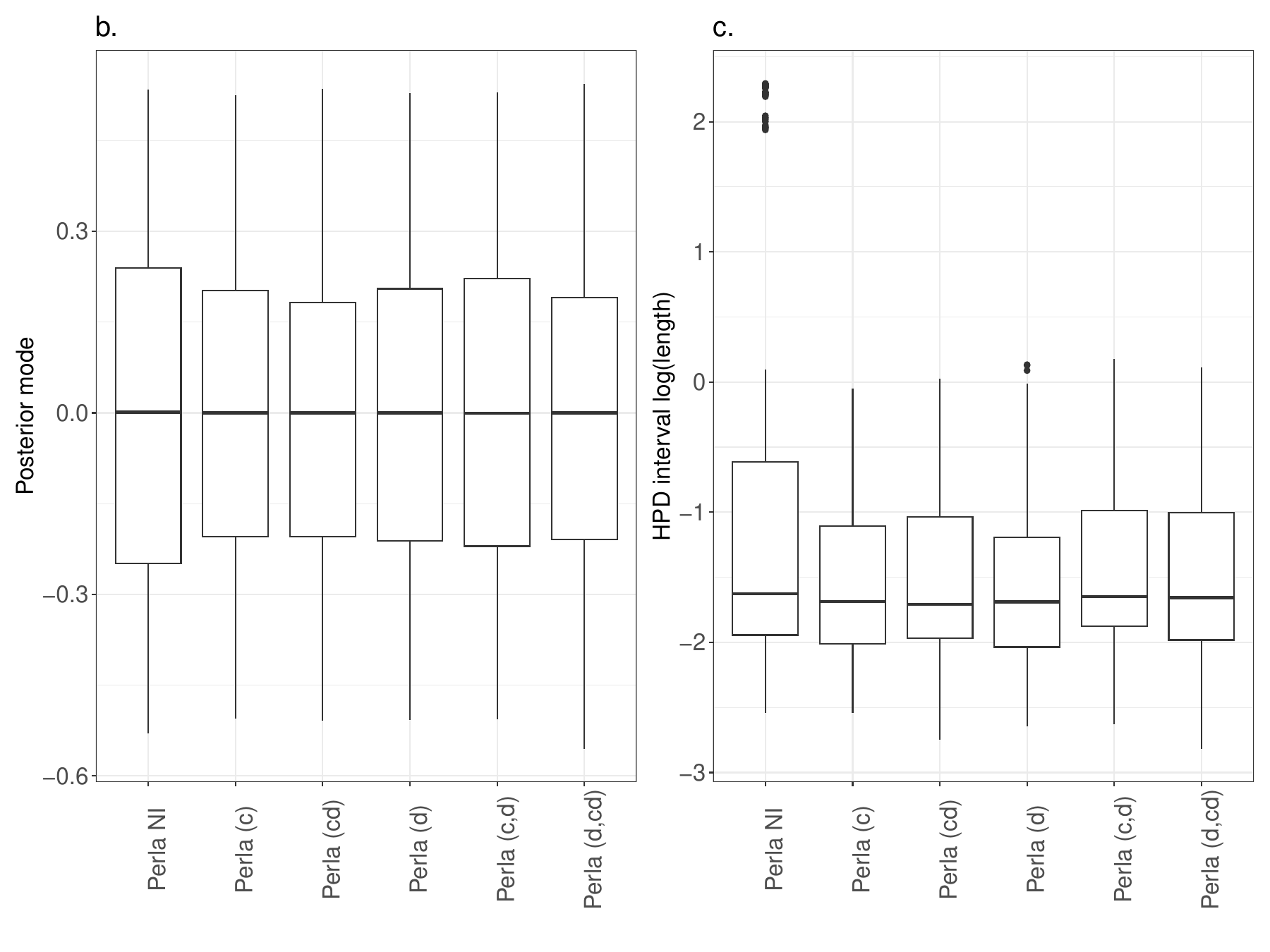}
    \caption{Results from Simulation 1 based on 20 simulated maps. The graphs display the posterior modes and of logarithm of the 95\% HPD interval lengths of the $\mu_{kj}$ for those $j$ such that $I_j = 1$, obtained with six versions of PERLA~under study. The top row displays the joint values, while the bottom row displays their marginal distributions using boxplots. `Perla NI' denotes the implementation of PERLA~that uses non-informative priors.}
    \label{fig:sim1_informative}
\end{figure}

\review{Despite the ARI values obtained with different \perla~setups are comparable, we now show that the contribute brought by different shrinkage priors extends beyond the overall clustering accuracy and consists of retrieving which diseases are truly responsible of the formation of mortality clusters. We display in Figure \ref{fig:sim1_not_informative} the posterior distributions of the  $\mu_{kj}$ parameters across the 20 simulated maps for which $I_j = 0$. We compare the results obtained from the five penalised versions of \perla, and \perla~with non-informative priors (denoted `Perla NI' in the figures).  Ideally, since the corresponding $\mu^{\text{true}}_{kj}$ are zero, the posterior distributions should be concentrated around zero. Figure \ref{fig:sim1_not_informative}\textit{a} displays the posterior modes and the logarithm of the 95\% HPD interval lengths. From top-left panel, it emerges that `Perla NI' provides highly variable posterior estimates of the $\mu_{kj}$, and large HPD intervals. This is also evident from the marginal distributions of the posterior modes and interval lengths, displayed in Figures \ref{fig:sim1_not_informative}\textit{b} and \ref{fig:sim1_not_informative}\textit{c}. We notice that, among the estimates provided by `Perla NI', some parameters show very large HPD interval sizes. These parameters are associated to the fourth mixture component, which does not find a correspondence with a real cluster. This effect is no longer present when using \perla~with only the cluster-specific shrinkage factors (setup \texttt{(c)}), as seen in the top-central panel of Figure \ref{fig:sim1_not_informative}\textit{a}, although the posterior modes remain generally variable. 
With the setups \texttt{(cd)}, \texttt{(d)}, and \texttt{(c,d)}, many posterior modes are shrunk toward zero, while a few of them remain insufficiently shrunk, and the HPD interval lengths decrease. 
In addition, note from Figure \ref{fig:sim1_not_informative}\textit{c} that the HPD intervals of \perla~\texttt{(cd)}, which uses the horseshoe prior, remain wider that the ones provided by the two other setups. Finally, our proposed prior \eqref{formula:prior_mu}, denoted  \texttt{(d,cd)}, effectively shrinks all posterior modes towards zero, with narrow HPD intervals.

Figure \ref{fig:sim1_informative} displays the posterior distributions of the $\mu_{kj}$ parameters for whichh $I_j = 1$, corresponding to diseases that effectively contribute to the formation of mortality clusters. Since $\mu^{\text{true}}_{kj}\mid I_j = 1 \sim \mathcal{U}(-0.5,0.5)$, some coefficients may be generated very close to zero, and we expect our shrinkage priors to estimate them as null. This is observed in Figure \ref{fig:sim1_informative}\textit{a}, where several posterior distributions have mode on zero, especially under the priors \texttt{(cd)} and \texttt{(d,cd)}. The reason is that these setups employ the factor $\gamma_{kj}$ to apply a targeted shrinkage on $\mu_{kj}$, without affecting the other elements $\Mu$. In contrast, the setup \texttt{(d)}  applies the same amount of shrinkage to all the coefficients of the same disease, while \texttt{(c,d)} applies to $\mu_{kj}$ a combination of the shrinkage factors $\delta_k$ and $\zeta_j$, which account for all the other elements in $\Mu_{k.}$ and $\Mu_{.j}$.  Finally, the version \texttt{(c)} effectively shrinks towards zero all the coefficients that do not match a real cluster, which is something that `Perla NI' fails to achieve. Overall, the marginal distributions of the posterior modes and interval lengths from the five penalised versions of \perla~appear similar (see Figures \ref{fig:sim1_informative}\textit{b} and \ref{fig:sim1_informative}\textit{c}).

We conclude that clustering accuracy is not the only metric to consider for evaluating the model performance. Although \perla~provides stable results in terms of ARI using different prior setups, our results demonstrate that our shrinkage priors improve interpretability by indicating which diseases do, and which do not, contribute to cluster formation. If disease $j$ does not affect the overall clustering, a prior setup that uses just its specific shrinkage parameter $\zeta_j$ is sufficient to shrink the means $\Mu_{.j}$ towards zero while recovering the clustering using information from informative diseases. Differently, if disease $j$ is informative for only a subset of the $K\true$ clusters, then the model requires the use of additional cluster-disease shrinkage parameters $\gamma_{kj}$. Our results also indicate that combining disease-specific shrinkage factors $\zeta_j$ with cluster-disease shrinkage factors $\gamma_{kj}$ yields better performance in distinguishing between informative and uninformative diseases, compared to using cluster-disease shrinkage factors alone. Finally, the cluster-specific factors $\delta_k$ efficiently shrink additional mean parameters, a feature that is particularly useful when the number of mixture components is taken larger than the number of clusters in the data.

Model selection can be guided by the $\mathrm{DIC}_3$ criterion, as discussed in Section \ref{subsec:postprocessing_modelselection}. In our practical applications presented in Section \ref{sec:application}, we select the model based on both the $\mathrm{DIC}_3$ and an examination of the posterior distributions of the $\mu_{kj}$, evaluating which models produce well-separated mixture components that can be used to describe the underlying epidemiological phenomena.
}

\subsection{Simulation 2}
\label{subsec:simulation2}
The second simulation experiment aims to study the performance of \perla~when the clustering probabilities have different levels of spatial correlation. This setting reflects a scenario where some clusters exhibit strong spatial connection, resulting in geographically concentrated and interconnected areas, while others may show weak spatial connection, with areas scattered across the whole territory. \review{Since \perla~accounts for the spatial structure of the data through the CAR priors in \eqref{formula:CAR}, an additional scope of this experiment is investigating the effects of different a priori configurations for the parameters $\tau_k$ and $\rho_k$.}

To accomplish this, we generate 20 simulated maps of mortality rates, considering $d = 3$ death causes and $K\true = 4$ clusters. To induce different levels of spatial correlation among clustering probabilities, we consider three equispaced spatial correlation values $\boldsymbol{\rho}^{\text{true}} = (0.01, 0.455, 0.9)$, that remain fixed across the 20 replicates. To induce marginal correlation across causes of death, \review{we generate $\bSigma\true$ as described in Section \ref{subsec:simulation1}: correlation values are drawn from a uniform distribution $\mathcal{U}(-0.2, 0.2)$. After ensuring the semi-positive definiteness of the correlation matrix, we scale all its elements by a factor of 0.07 to obtain the covariance matrix}. We generate the cluster- and disease-specific intercepts $\mu_{kj}$, for $k = 1,\dots,K\true$ and $j = 1,\dots,d$, from $\mathcal{U}(-1,1)$. Lastly, each dataset is drawn from the simulation model detailed in Section \ref{subsec:simulation_model}. Supplementary Figure 5 displays, on the left panel, the spatial clusters in one of the 20 simulated datasets and, in the right panel, the distribution of the three outcomes within each cluster appearing in the left panel. The map shows that counties in cluster 1 are spread throughout the territory, while those in clusters 2 and 3 are more geographically concentrated due to the larger spatial correlation values employed. 
From the spatial shape of the clusters, we highlight the importance of using a model that leverages spatial information without strictly enforcing clusters to be made solely of contiguous areas. 
For example, cluster 3 (shown in green in the left panel) consists of separate agglomerates of contiguous areas that are not directly connected. It is essential that the clustering model recognises these separate agglomerates as part of a unique cluster, since they share the same epidemiological characteristics.

First, we study the effects of different a priori assumptions on $\rho_k$, keeping $\tau_k = 1$ for all $k$.
\review{For this experiment,} we consider \review{three prior setups: \textit{i.)} $\rho_k = 0.99$ $\forall k$, that is a proper CAR model with large spatial correlation; \textit{ii.)} $\rho_k = 1$ $\forall k$, that is an improper CAR; and \textit{iii.)} the prior given in Formula  \eqref{formula:prior_rho_SS}} (denoted as `S\&S' in the figures). \review{We evaluate} whether \review{allowing $\rho_k$ to be learnt from the data} 
offers advantages in clustering accuracy and model fitting compared to assuming a constant value, especially in a framework \review{where clusters are characterised by} different levels of spatial correlation. \review{In addition, we check whether assuming a proper or an improper CAR model yields meaningful differences in performance. We fit only \perla~\texttt{(d,cd)}, as the evaluation of different prior models for $\mu_{kj}$ is already treated in Section \ref{subsec:simulation1}}.

\begin{figure}[t]
\includegraphics[width=0.47\linewidth]{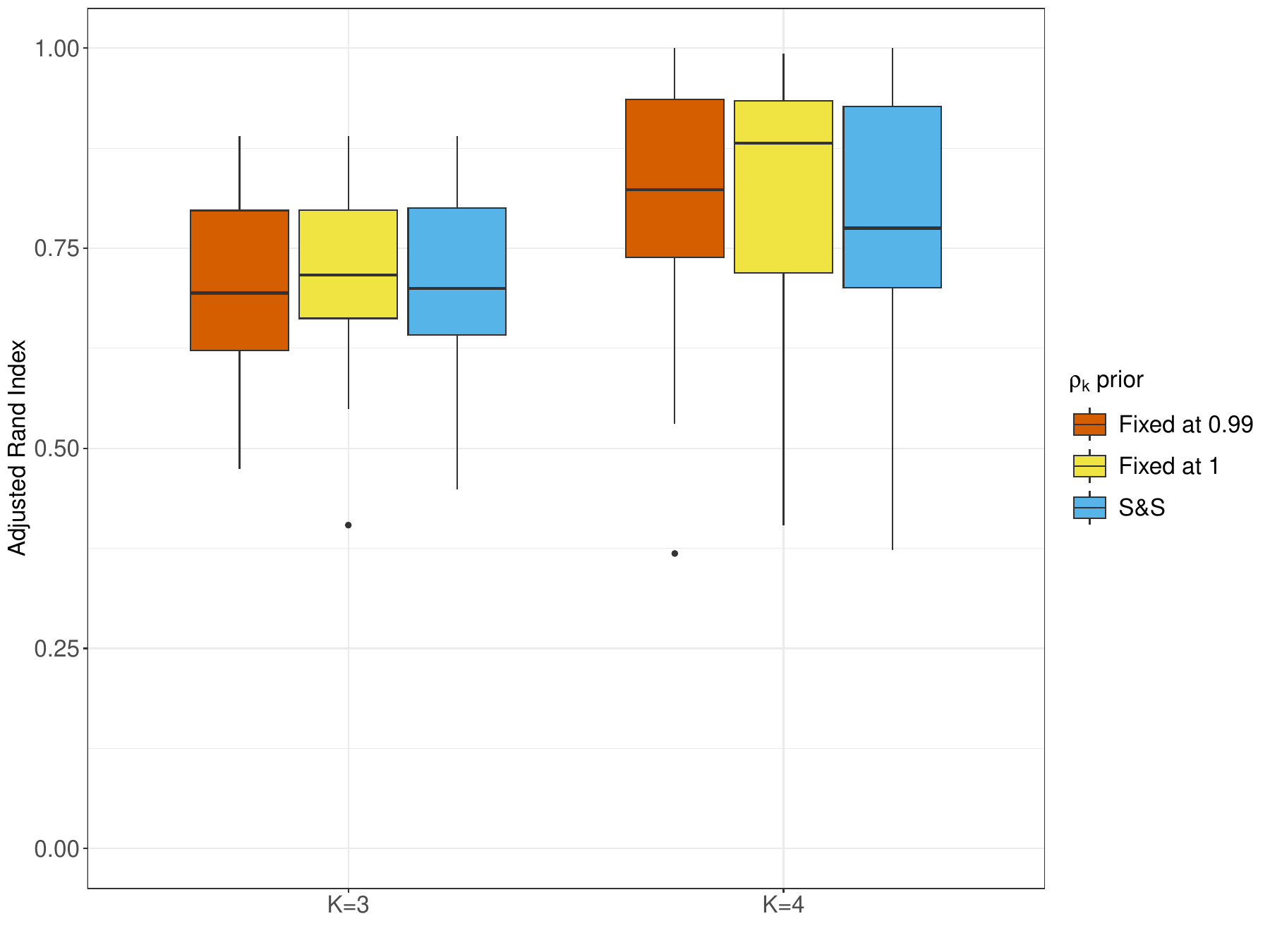}
\includegraphics[width=0.47\linewidth]{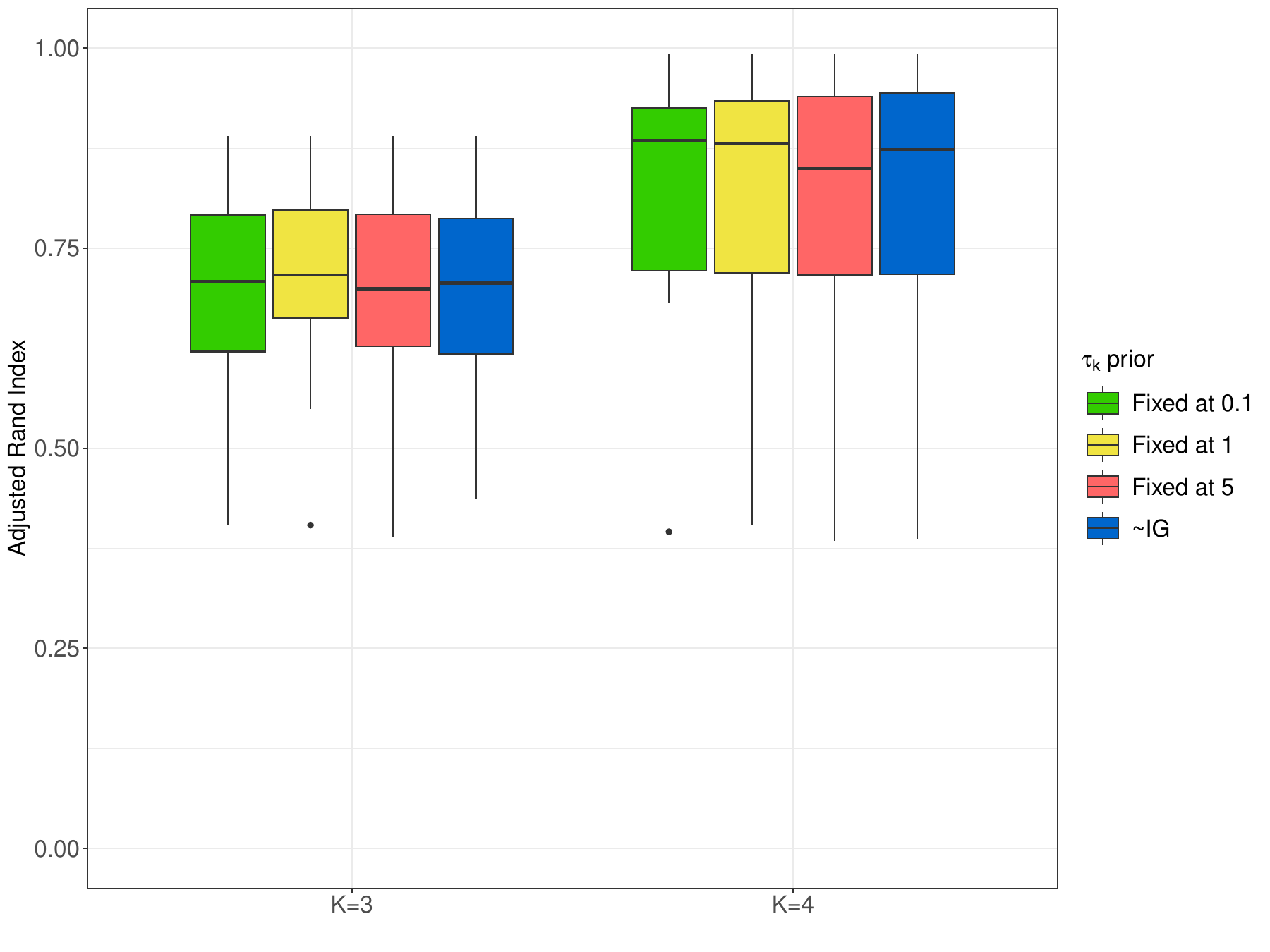}
\centering
\caption{\review{Results from Simulation 2 based on 20 simulated datasets. The two panels display the ARI distributions obtained with different a priori configurations for $\rho_k$, keeping $\tau_k = 1$ $\forall k$ (left panel) and with different a priori configurations for $\tau_k$, keeping $\rho_k = 1$ $\forall k$ (right panel). The model is estimated with $K = 3$ and $K = 4$ clusters.}}
\label{fig:simulation2_boxplots}
\end{figure}

\begin{figure}[t]
\includegraphics[width = 0.8\linewidth]{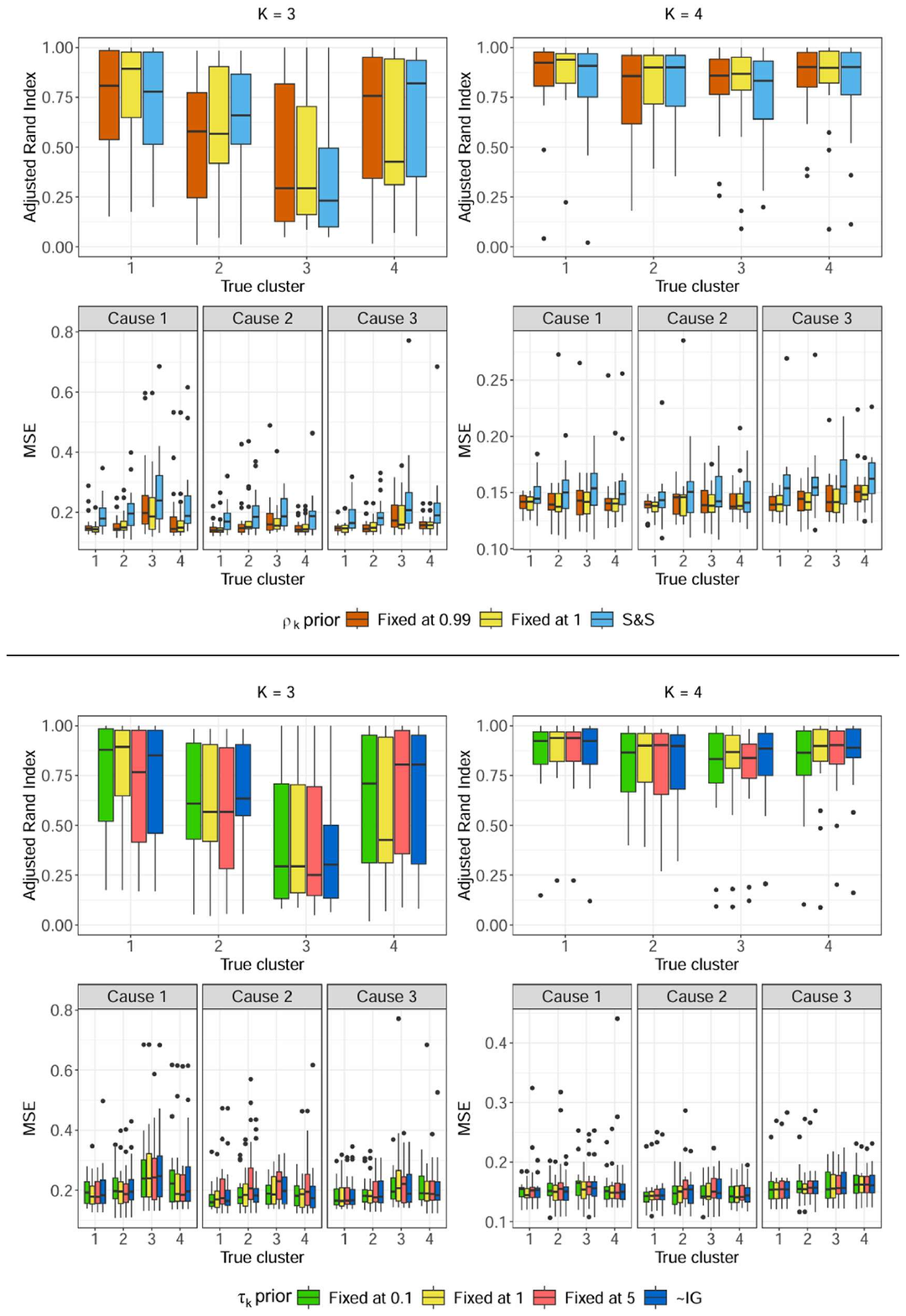}
\centering
\caption{\review{Results from Simulation 2 based on 20 simulated datasets. The first two rows give the results obtained with different prior configurations for $\rho_k$, keeping $\tau_k = 1$ $\forall k$, and the last two rows with different prior configurations for $\tau_k$, keeping $\rho_k = 1$ $\forall k$. The first and the third lines show the ARI distribution of each of the four true clusters. The second and the fourth lines show the MSE distribution within each of the four true clusters and for the three causes of death. The left column displays the results obtained using $K = 3$ estimated clusters, and the right column using $K = 4$ clusters.}}
\label{fig:simulation2_varyingRhoTau}
\end{figure}

The \review{left} panel of Figure \ref{fig:simulation2_boxplots} displays the \review{ARI} index distribution based on 20 simulated datasets. We observe that, when the number of clusters is correctly specified, \perla~accurately recovers the underlying clustering structure (median \review{ARI} $> 0.9$). As expected, the performance is generally worse when the number of clusters is misspecified (median \review{ARI} $\in(0.85,0.9)$). \review{All the} prior setups perform similarly in terms of clustering accuracy for $K = 3$ and $K = 4$. Since clusters are characterised by different levels of spatial correlation, we also measure the accuracy in recovering each cluster separately. To do so, let $\mathbf{Z}^{\mathrm{post}}$ be the posterior estimate of the matrix $\mathbf{Z}$, and $h(k)=\arg\max_{\ell = 1,\dots,K}\mathrm{ARI}(\Z\true_{.k}, \Z^{\mathrm{post}}_{.\ell})$ be the index of the estimated cluster that is most associated with the $k$-th real cluster. \review{The top line of} Figure \ref{fig:simulation2_varyingRhoTau} displays the ARI distribution between $\Z\true_{.k}$ and $\Z^{\mathrm{post}}_{.h(k)}$, for each true cluster $k = 1,\dots,K\true$. These graphs help to visualise which clusters are more challenging to recover. When $K = 3$, the third cluster remains the most difficult to retrieve, and the model employing the spike-and-slab-type prior `S\&S' performs worse than the models with constant $\rho_k$. When $K = 4$, we do not observe significant differences across clusters or models.

\review{Last, we test if using different prior setups for $\rho_k$ has an impact or not on model fitting. To do so, we display in the second row of} Figure \ref{fig:simulation2_varyingRhoTau} the mean squared error (MSE) distribution of each real cluster, \review{that is computed
as $\sum_{i: \mathrm{Z}\true_{ik} = 1} (y_{ij}-\mathbf{Z}_{i.}^{\mathrm{post}}\Mu^{\mathrm{post}}_{.j})^2$, where $\Mu^{\mathrm{post}}_{.j}$ denotes the posterior estimate of $\Mu_{.j}$.} 
\review{When the number of clusters is misspecified, all models struggle to fit the data in cluster 3. However, the model denoted `S\&S' consistently performs poorly in all clusters and causes of death compared to the CAR priors with fixed $\rho_k$ values, which perform similarly. This trend is also observed when $K = 4$.}

\review{The second step involves studying the effects of different prior models for $\tau_k$, keeping $\rho_k = 1$ $\forall k$. For this experiment, we assume $\tau_k = \tau$ $\forall k$, and we investigate the following options: \textit{i.)} fixing $\tau$ respectively to 0.1,  1, and 5, and \textit{ii.)} allowing $\tau$ to be learned from the data, assuming the conjugate prior $\mathcal{IG}(2.1, 3.1)$, which implies $\mathcal{M}(\tau) = 1$ and a 95\% HPD interval that ranges approximately between 0.3 and 7.85.

The right panel of Figure \ref{fig:simulation2_boxplots} displays the distribution of the ARI across the 20 simulated datasets. We observe that any choice of $\tau$ leads essentially to equivalent results in terms of the overall classification accuracy. We display also in the third row of Figure \ref{fig:simulation2_varyingRhoTau} the distribution of the ARI between $\Z\true_{.k}$ and $\Z^{\mathrm{post}}_{.h(k)}$, for each true cluster $k = 1,\dots,K\true$. When $K = 3$, the third cluster remains the most difficult to recover, and the model assuming $\tau\sim\mathcal{IG}$ performs slightly worse in recovering it than the models with fixed $\tau$. When $K = 4$, we do not observe significant differences across clusters or across models.

Finally, the last row of Figure \ref{fig:simulation2_varyingRhoTau} displays the distribution of the mean squared error (MSE) within each true cluster. The third cluster remains the most difficult to recover when $K = 3$, while the quality of the model fitting remains stable across every cluster when $K = 4$. 
}

We conclude that, when the data-generating mechanism presents several clusters with different levels of spatial correlation, \perla~performs well in recovering the clustering structure when the number of clusters is correctly specified. In addition, it appears not to be influenced by the varying levels of spatial correlation. On the contrary, when the assumed number of clusters is lower than the true number, both the clustering accuracy and the model fitting are more affected in areas with high spatial correlation. 
\review{This experiment has served also to frame the impact of the hyperparameters values of the CAR prior on the model performance. Our simulations show that specifying a prior distribution on $\rho_k$ or fixing it at a constant value yields equivalent results in terms of clustering accuracy, while the  model fitting is poorer when $\rho_k$ is not taken as constant. In addition, we have observed that assuming an improper CAR ($\rho_k = 1$) is practically equivalent to assuming a proper CAR with a sufficiently large $\rho_k$. Finally, our model appears to be robust to different choices of $\tau_k$, producing practically equivalent results whether $\tau_k$ is assumed variable or fixed in advance.}

\review{
\subsection{Simulation 3}
\label{subsec:simulation3}
In epidemiological studies, data are typically collected as counts. For instance, mortality studies rely on the number of deaths attributed to various diseases within each area of the analysed territory. However, these raw counts are not meaningful \textit{per se} unless compared with an appropriate reference measure. For example, the observed number of deaths can be assessed against the expected number in the territory, which accounts for population stratification by age and allows to identify areas with mortality excess or deficit.  

From a statistical modelling perspective, Poisson regression is a natural choice for count data, with the chosen reference measure incorporated as an offset. In practice, however, while raw counts are often available, suitable comparison measures are frequently inaccessible due to privacy restrictions. This limitation motivates us to develop a model that works with continuous outcome variables, such as rates, which are both readily available and unaffected by privacy constraints. Nonetheless, we believe it is essential to assess our model’s performance in scenarios where both observed counts and a valid comparison measure are available.

In this section, we propose an experiment that stresses \perla~by simulating data using a generating mechanism that substantially deviates from our model in three aspects: first, it uses Formula \eqref{formula:modelsimulation2} to simulate the log-risks, incorporating also three covariates into the generative mechanism; then, it simulates the observed deaths from a Poisson model; and last, it assumes that different diseases yield different spatial clustering patterns. 
We consider $d = 5$ causes of death. The first two diseases yield two clusters, obtained by dividing the map along a line at latitude 40°10'33.888'' N. This line lies at the midpoint of the latitude range, so one cluster occupies the northern part of the map and the other the southern part. A representation of these two clusters is provided in the left panel of {Supplementary Figure 6}. We store the clustering information into the $n\times 2$ matrix ${\mathbf{Z}}^{(1:2)\true}$, and the mean levels of the two diseases observed in the two clusters into the $2\times 2$ matrix ${\Mu^{(1:2)}}\true$. The next two diseases yield two clusters, obtained by dividing the map along a line at longitude 117°8'2.4'' W. As this line lies at the midpoint of the longitude range, one cluster occupies the western part of the territory and the other the eastern part. A representation of these two clusters is shown in the central panel of {Supplementary Figure 6}. We store the clustering information into the $n\times 2$ matrix ${\mathbf{Z}}^{(3:4)\true}$, and the mean levels into the $2\times 2$ matrix ${\Mu^{(3:4)}}\true$. The last disease does not show any clustering pattern.

The proposed data-generating scheme simulates the log-risks as
$$
\log(\boldsymbol{\theta}\true_{i.})\sim\mathcal{N}_d\left\{\left[\left({\mathbf{Z}^{(1:2)}_{i.}}\true\right)^T{\Mu^{(1:2)}}\true,\left({\mathbf{Z}^{(3:4)}_{i.}}\true\right)^T{\Mu^{(3:4)}}\true,0\right]^T+\mathbf{x}^T_{i.}\boldsymbol{\beta}\true,\bSigma\true\right\}.
$$
The independent variables are generated as $x_{i\ell} \sim \mathcal{N}(-0.2,\, 0.5^2)$ and the regression coefficients as $\beta_{\ell j} \sim \mathcal{N}(0.02,\, 0.05^2)$, for $\ell = 1,2,3$. These values were manually tuned to avoid unrealistically high mortality risks. The covariance matrix $\boldsymbol{\Sigma}^{\text{true}}$ is assumed to be block diagonal, with the blocks $\boldsymbol{\Sigma}^{\text{true}}_{1:2,1:2}$ and $\boldsymbol{\Sigma}^{\text{true}}_{3:4,3:4}$ generated as in Simulation 1, and with $\Sigma^{\text{true}}_{55} = 0.05$.

To generate the observed deaths, we consider the expected ones as offset. These are obtained from $E_{ij} \sim \mathcal{G}(1,0.03)$. This distribution was chosen because it assumes that the average number of expected deaths is 30 and approximately the 95\% of the areas in the territory have at most 100 deaths, while the remaining 5\% can have much higher values, representing a few areas with large population sizes (an consequently more deaths). The observed counts in the $i$-th area of the map and for the $j$-th cause of death are simulated from $O_{ij} \sim {\mathcal{P}oi}(\theta^{\text{true}}_{ij} E_{ij})$. Finally, to analyse these data using \perla, we compute the logarithm of the \textit{standardised mortality rations} (SMR) as $y_{ij} = \log(O_{ij} / E_{ij})$ or, in case $O_{ij}=0$, $y_{ij} = \log\{(O_{ij} +1/2)/ E_{ij}\}$, as suggested by \cite{Bivand_etal.2013}. 

We generate 20 replicates of the experiment, and we apply \perla~\texttt{(d,cd)} to ensure flexibility, incorporating the model covariates. For comparison, we apply the SSS of \cite{gomez-rubio_etal.2019}, already considered in Section \ref{subsec:simulation1}, that must be applied to each outcome separately and works directly with counts data~$O_{ij}$. In order to obtain a final clustering, we combine the labels obtained on the five outcomes into a unique clustering configuration.
In addition, we consider also GMM that assumes $\mathbf{y}_{i\cdot} \mid \Z_{ik}=1 \sim \mathcal{N}_d\Bigl(\boldsymbol{\mu}_{k\cdot} + \mathbf{x}_{i\cdot}^T \boldsymbol{\beta}, \, \boldsymbol{\Sigma}_k\Bigr)$, with $\Pr(\Z_{ik} = 1) = \pi_k$ and the covariance structure $\boldsymbol{\Sigma}_k$ selected using the BIC criterion. Similarly to \perla, this model assumes that the covariates are not directly related to the clustering. The estimation is performed via maximum likelihood. 

Due to the complex clustering structure used to generate these data, evaluating the clustering accuracy of the models is not immediate. By combining ${\mathbf{Z}^{(1:2)}}^{\text{true}}$ and ${\mathbf{Z}^{(3:4)}}^{\text{true}}$, we obtain a partition of the territory into four clusters (see the right panel of {Supplementary Figure 6}) that we use to evaluate the models.  
 We fit \perla~and GMM using $K = 2, 3, 4$. The ARI distributions in the left panel of Figure \ref{fig:simulation3} show that the highest clustering quality is achieved by \perla~with $K = 4$. This result is particularly remarkable considering that the SSS works directly with raw counts data.
 We notice also that both \perla~and GMM perform better when the number of clusters is enlarged. This is due to the nested clustering structure that characterises this simulation. \reviewm{The improvement is particularly evident when moving from 2 to 3 clusters, as shown by the gain in ARI level, and, separately, by the substantial decrease in $\mathrm{DIC}_3$ (see the central panel of Figure~\ref{fig:simulation3}). } 
 We conclude that, when different clusters vary with the outcome variables, an effective strategy is to increase the number of clusters assumed by the model. In addition, we observe that transforming the outcomes into continuous variables rather that using directly the counts data does not lead to any loss in clustering accuracy.

In addition, we report in the right panel of Figure \ref{fig:simulation3} the distribution of the computational cost of \perla~as a function of the number of clusters. Each model estimation involves running four Markov chains of length 10,000 sequentially using the algorithm detailed in Section \ref{subsec:mcmc_algorithm} and combining the results with the post-processing algorithm described in Section \ref{subsec:postprocessing_modelselection}. Consequently, the runtime for a single Markov chain of length 10,000 can be roughly deduced by dividing the values in the graph by four.

}

\section{Case studies}
\label{sec:application}

In this section, we investigate the existence of mortality clusters in the case studies discussed in Section \ref{subsec:motivation}. 
In both scenarios, we apply \perla~to male and female populations separately. We do not include sex as a covariate in the model because, as outlined in Section 2.1, \perla~assumes that exogenous variables influence the outcomes but do not regulate the formation of clusters. However, we are interested ing investigating whether the mortality clusters based on the three causes of death considered show significant differences between males and females in both maps. This requires us to conduct the analysis separately for each group. We consider all the combinations of shrinkage factors shown in Table \ref{tab:list_of_shrinkage_parameters}, assuming \review{$\tau_k = 1$ and} $\rho_k = 0.99$, $\forall k$. \reviewm{The only setup that we do not consider is the one that applies only a global shrinkage to the $\Mu$ through $\phi$.} 
Every model is fitted running the MCMC algorithm four times separately, each for 10,000 iterations, discarding the first half of each run, and merging the remaining draws into a unique chain of length 20,000. Lastly, the ECR algorithm is applied to resolve eventual label switching. 

\subsection{Mortality in the Padua province}
\label{subsec:padua}
We investigate the presence of mortality clusters in the Padua province using age-adjusted standardised mortality ratios $SMR_{ij}=O_{ij}/E_{ij}$, derived as the ratio between observed death counts ($O_{ij}$) and expected death counts ($E_{ij}$) in the $i = 1,\dots,106$ areas and for the $j = 1,2,3$ diseases considered. The expected number of deaths represents the number of deaths per age class that would occur if the mortality rates per age class of the entire province were applied. Thus, assuming $T$ age classes, the expected number of deaths due to disease $j$ in the $i$-th area is $E_{ij} = \sum_{t = 1}^T n_{it} r_{jt}$, where $r_{jt}$ is the provincial mortality rate of the $j$-th disease within the $t$-th age class, and $n_{it}$ is the population size within the $i$-th area and the $t$-th age class. For our analysis, we consider
$$
y_{ij} = \log\left(
SMR_{ij}
\right).
$$
The logarithm is applied so that $y_{ij} = 0$ represents the case of  concordance between observed and expected deaths, $y_{ij} > 0$ denotes a mortality excess compared to the province level, and $y_{ij} < 0$ denotes a mortality deficit compared to the province level.  In cases where $O_{ij} = 0$ for some $i$ and $j$, one can apply the alternative formula $SMR_{ij} = (O_{ij} + 1/2)/E_{ij}$ to enable the computation of the logarithm \reviewm{(\citealt[Section 5]{Bivand_etal.2013}}). 
\reviewm{As discussed by \cite{Moller_Ahrenfeldt.2021}, this approach is preferable to adding 1 to each $O_{ij}$ equal to zero, as it reduces distortion of the sample size; however, it does not allow a clear probabilistic interpretation, since the resulting counts are not integers}. \reviewm{Nonetheless}, it is important to clarify that, although we know how the standardised mortality ratios are computed, the raw data $O_{ij}$ and $E_{ij}$ are not provided by ULSS6 \textit{Euganea} due to privacy considerations; only the $SMR_{ij}$ values are available for the analysis.

We consider two variables that are representative of the socioeconomic status of individuals in an area to be used as covariates: 
\begin{itemize}
\item \emph{Deprivation index}, quantified through the \textit{Caranci's index} \citep{caranci.2009}.
It quantifies the deprivation, which represents socio-economic disadvantages, by considering various factors affecting residents in the territory.
It is calculated as the standardised sum of five indicators, with higher scores indicating higher deprivation \citep{rosano2020}.
These five indicators include the level of education among individuals aged 15-60, the unemployment status, the condition of single-parent family with underage children, house renting, and housing density per 100 m$^2$.
The index typically ranges between $-5$ and 40, with most cases between $-$2 and 2.
The data used to compute the index, sourced from the Italian 2011 census, are available at census section level and were aggregated to obtain data at municipality level.
\item \emph{Sale price of real estate per squared meter}, which serves as an indicator of the general economic status of municipalities.
These price data were scraped from \url{www.immobiliare.it}, a web portal that collects property sales listings in Italy, using the average of the prices in euro/m$^2$ on January 1$^{\makebox{st}}$ of 2017, 2018, and 2019 \citep{immobiliare.it}.
The specific time frame was chosen as it is characterised by stable price trends,  without notable increases or decreases.
\end{itemize}

On male data, we fit \perla~considering $K = 2$ and $K = 3$, mainly motivated by the relatively small number of areas in the map. Results of the model selection are shown in \reviewm{Supplementary Figure 7}. 
The $\mathrm{DIC}_3$ criterion clearly supports $K = 3$. 
\review{In fact,} the plot of the posterior distribution of $\Mu$ under $K = 2$  (top panel of Supplementary Figure 8) displays multiple modes, which is a clear indication that a model with an additional component should be considered. 
We thus select \perla~with $K = 3$ and shrinkage factor \review{\texttt{(c,d)}, which is more parsimonious than \texttt{(d,cd)} while providing in practice the same value of $\mathrm{DIC}_3$}. The final classification of the map is reported in the top-left plot of Figure \ref{fig:map_posterior_Padua}. While cluster 1 extends over the whole territory, particularly on the northern area of the province, clusters 2 and 3 are \review{prevalently} located in the south-west.  The posterior densities of $\{\exp(\mu_{kj})\}_{k=1,2,3;j=1,2,3}$ are displayed in the top-right panel of Figure \ref{fig:map_posterior_Padua}, showing that cluster 1 is characterised by an excess mortality due to respiratory diseases ($\mathcal{M}(\mu_{12}|\mathrm{data}) = 1.216$, $\Pr\{\exp(\mu_{12})> 1|\mathrm{data}\} = 0.997$), and \review{cluster 2} by a mortality excess due to circulatory diseases ($\mathcal{M}(\mu_{21}|\mathrm{data}) = 1.166$, $\Pr\{\exp(\mu_{21})> 1|\mathrm{data}\} = 0.966$) and by a mortality deficit due to respiratory diseases ($\mathcal{M}(\mu_{22}|\mathrm{data}) = 0.318$, $\Pr\{\exp(\mu_{22})> 1|\mathrm{data}\} = 0$). Lastly, \review{cluster 3} does not show evidence of neither excess nor deficit of mortality due to the three causes of death considered. \review{This is visible from the posterior distributions of the elements in $\{\exp(\mu_{3j})\}_{j = 1,2,3}$, which are shrank toward one thanks to the cluster-specific shrinkage factor $\delta_3$.} 
In Table \ref{tab:males_females_padua_mortality}, we report all the posterior SMR estimates and the probabilities of observing an excess mortality. Notice that, according to our model, deaths for tumours appear not to impact on the formation of spatial mortality clusters. \review{This is  highlighted by the disease-specific shrinkage factor $\zeta_3$, which collapses the posterior estimates of $\{\exp(\mu_{k3})\}_{k=1,2,3}$ toward one.} We report also the trace plots of the cluster centroids $\Mu$ in the top panel of {Supplementary Figure 9}. Chains appear stable, demonstrating consistency across different parallel runs of the MCMC algorithm.

Figure~\ref{fig:shrink_Padova} displays the posterior estimates of the shrinkage factors used in the analysis of male data. The plot consists of a $K \times d$ matrix, coloured according to the posterior mode of the normalised disease-specific shrinkage factors, $\{\zeta_j / \sum_{l=1}^d \zeta_l\}_{j=1,\dots,d}$. This normalisation allows this quantity to be interpreted as the percentage contribution of disease~$j$ to the clustering structure. Since $\zeta_j$ either amplifies or shrinks the role of disease~$j$ in formation of clusters, in the figure every $\zeta_j / \sum_{l=1}^d \zeta_l$ remains consistent across all clusters. 
The results indicate that cluster formation is driven primarily by respiratory diseases, accounting for roughly 90\% of the observed structure. This is further visible from the second column of Table~\ref{tab:males_females_padua_mortality}, where 
\reviewm{the variability of risk posterior estimates across the three clusters in respiratory diseases is higher than in the other two causes of death.}
Thus, while mortality from respiratory diseases varies substantially across the territory, leading to distinct spatial clusters, the other two diseases show more stable geographical patterns. The shrinkage factors $\{\delta_k\}_{k=1,\dots,K}$ quantify the variability in mortality levels across different causes of death within each cluster, with larger values indicating greater heterogeneity. However, from an epidemiological perspective, we find these parameters less interpretable than ${\zeta}_j$, and therefore we do not report their graphical representation.

On female data, \review{all the model setups, except \texttt{(d)}, are comparable in terms of $\mathrm{DIC}_3$ (see the right panel of \reviewm{Supplementary Figure 7}
). Therefore, 
we apply \perla~with $K = 4$ and shrinkage factor \texttt{(c)}, which is the most parsimonious. In alternative, we could have considered the model with three clusters, following the conclusions taken from male data. However, the posterior distribution of $\Mu$ under $K = 3$ displays multiple modes, as shown in the bottom panel of Supplementary Figure 8.}
The final classification of the map is displayed in the bottom-left plot of Figure \ref{fig:map_posterior_Padua}. Some differences emerge with respect to male data. Cluster 3 is geographically located in the same area of cluster 2 identified on male data, but its composition is rather different; for instance, it does not include any municipality in the northern province. Additionally, the model detects a fourth cluster consisting of only two separate municipalities. The posterior density of every $\exp(\mu_{kj})$ are displayed in the bottom-right plot of Figure \ref{fig:map_posterior_Padua}, and detailed estimates are reported in the bottom part of Table \ref{tab:males_females_padua_mortality}. We detect an excess mortality in cluster 1, that corresponds to the northern province, due to respiratory diseases, and a deficit of mortality for the same cause in clusters 3 and 4. In addition, cluster 4 is characterised by a non-negligible deficit of mortality due to tumours. Cluster 3 shows also a probability of excess mortality due to circulatory diseases of almost 0.9. 
\review{We report also the trace plots of the cluster centroids $\Mu$ in the bottom panel of {Supplementary Figure 9}.}

Although cluster 4 comprises only two municipalities, there is strong evidence of its separation from cluster 3. Supplementary Figure 10 illustrates the percentage of times each pair of municipalities in clusters 3 and 4 was assigned to the same cluster across the MCMC iterations. The estimated probability that the two municipalities finally assigned to cluster 4 (\textit{Barbona} and \textit{Arquà Petrarca}) belong to the same cluster exceeds 0.9. In contrast, the estimated probability of either municipality being placed in the same cluster with any of the nine municipalities finally assigned to cluster 3 ranges from 0 to 0.05.

We also notice that the best partition selected by the ECR algorithm does not include cluster 2. Observations assigned to cluster 2 in some MCMC iterations are all part of the group of 95 observations denoted as cluster 1 in the final estimate of the data partition. We conclude that, although cluster 2 does not have a practical epidemiological interpretation, its inclusion in some MCMC iterations help achieve unimodal posterior distributions, facilitating the inference on the model parameters and the computation of some quantities of interest, such as the probability of mortality excess. Notice also that the posterior distributions of each element in $\{\exp(\mu_{2j})\}_{j = 1,2,3}$ is shrank toward one thanks to the cluster-specific shrinkage factor $\delta_2$. 
Lastly, it is worth reporting that the data partition shown in the bottom-left panel of Figure \ref{fig:map_posterior_Padua} is equivalent to the partition returned by the SALSO algorithm, which confirms the robustness of the results discussed.

Both male and female mortality data do not show substantial evidence of marginal correlation across the three causes of death considered, as shown in Supplementary Table 1. 

\subsection{Mortality in the U.S. counties}
\label{subsec:USA}
We now investigate the presence of mortality clusters in the U.S. from 2016 to 2019. Due to the high diversity of the U.S. territory already mentioned in Section \ref{subsec:motivation}, we restrict our attention to the north-east and central-east coasts, selecting the U.S. counties whose centroids are located east of the 80\textdegree~W longitude meridian. This territory consists of 388 counties, which are displayed in the right plot of Supplementary Figure 1. 
We examine the relative \review{age-adjusted} mortality risks  \review{${RR}_{ij}=MR_{ij}/TMR_{j}$}, derived as the ratio between the age-adjusted mortality rate ($MR_{ij}$) and  the total age-adjusted mortality rate \review{($TMR_{j}$)} in the $i=1,\dots,388$ counties and for the $j = 1,2,3$ diseases considered. 
These rates are obtained from the Centres for Disease Control and Prevention (CDC) WONDER Online Databases~\cite{WONDER}. Assuming $T$ age classes, the age-adjusted mortality rates in county $i$ are computed as $MR_{ij} = \sum_{t = 1}^T MR_{ijt}P_{t}/P$, where $P$ \review{is the size of} the standard population (\textit{i.e.}, the U.S. population in 2000) $P_{t}$ is \review{the size of} the standard population in the $t$-th age class, and $MR_{ijt}$ is the age-specific mortality rate in the $t$-th age class. Similarly, \review{$TMR_{j} = \sum_{t = 1}^T TMR_{jt}P_{t}/P$, where $TMR_{jt}$ is the total} age-specific mortality rate in the $t$-th age class over the entire U.S. territory. \review{Both $MR_{ij}$ and $TMR_{j}$ are directly available in the WONDER Online Databases; however, the specific information of age classes (\textit{e.g.}, the $MR_{ijt}$) is not.}

In our analysis, we apply
$$
y_{ij} = \log\left( RR_{ij} \right),
$$
so that $y_{ij} = 0$, $y_{ij} > 0$ and $y_{ij} < 0$ represent respectively the cases where the mortality in county $i$ due to disease $j$ is equal, larger or smaller than the mortality across the entire U.S. territory. To enable the computation of the logarithm of the relative age-adjusted mortality risk when $MR_{ij} = 0$ for some $i$ and $j$, one can adopt, for example, the approach described by \cite{Moller_Ahrenfeldt.2021}.
\reviewm{For counties with at least 10 but fewer than 20 deaths due to a specific cause, the WONDER database does not report the mortality rate directly, but provides a confidence interval instead. In these cases, we impute the mortality rate using the \textit{Byar's method} (\citealt{breslow1987volume}, Formula 2.13). If the number of deaths is fewer than 10, neither the mortality rate nor its confidence interval is reported.
In the map considered in this section,} the mortality level due to respiratory causes is not available for the male population in \textit{Tyrrell County} (NC) and \textit{Grand Isle County} (VT), and it is unavailable for both male and female populations in \textit{Highland County} (VA).
In such cases, we impute the missing data using the average mortality rates for the same death cause from neighbouring counties. 
\review{It is worth underlying that, although the WONDER Online Databases also provide the number of deaths for every county, information of the population size per age class is not available. This prevents us from computing a possible measure of comparison of the observed deaths that accounts for the age population structure, such as the expected deaths. The use of the crude death rate -- the ratio between death counts and population size in each county -- does not account for the age population structure. This motivates the use of the age-adjusted mortality rates $MR_{ij}$ (and, consequently, the relative age-adjusted mortality risks $RR_{ij}$) in our analysis.}

On male data, we choose the model with $K = 3$ clusters \review{that exploits cluster}-specific penalisation factors, \review{denoted} \texttt{(c)}, \review{as it is the model that returns the smallest $\mathrm{DIC}_3$ value (see the left panel of Supplementary Figure 11)}. Results are shown in the top row of Figure \ref{fig:map_posterior_east_USA}. A predominant cluster that spans the entire territory is characterised by elevated mortality rates for all causes of death considered. The second cluster comprises primarily large urban areas or their neighbouring counties (including Boston, New York, Long Island, Philadelphia, Baltimore, and the urban area around Washington D.C.), along with some isolated regions. Mortality levels in these areas are uniformly below 1, indicating lower mortality compared to the entire U.S. territory. Lastly, the third cluster consists of isolated counties, showing a mortality deficit for respiratory diseases, and a mortality excess for circulatory diseases and tumours. Further details are provided in Table \ref{tab:males_females_EST_mortality}. It also appears that the marginal correlation across causes of death is not negligible. The posterior estimates and the corresponding 95\% HPD intervals of the marginal correlations are: 0.439 (0.319,0.544) between circulatory and respiratory causes, 0.424 (0.307,0.508) between circulatory causes and tumours, and 0.363 (0.250,0.478) between respiratory causes and tumours. 

On female data, we select the model \texttt{(d,cd)}, \review{as it returns the minimum value of $\mathrm{DIC}_3$} \review{(see the right panel of Supplementary Figure 11)}. The results displayed in the bottom row of Figure \ref{fig:map_posterior_east_USA} reveal a spatial distribution of the clusters that mirrors the patterns observed in the male data. Once again, cluster 1 spans the entire territory, cluster 2 captures prevalently large urban areas, and cluster 3 consists of isolated counties. Additionally, cluster 4 comprises only two counties. The posterior distributions in the bottom-right plot illustrate the variations in mortality levels across clusters. The mortality rates for circulatory and respiratory diseases observed in cluster 4 are similar to those in cluster 2, while mortality rate for tumours is substantially lower. Unlike the male data, there is not a mortality excess due to circulatory diseases in cluster 3. Further details are provided in the bottom section of Table \ref{tab:males_females_EST_mortality}. The marginal correlation of both diseases with tumours appear to be lower in female data than in male data. The posterior estimates and corresponding 95\% HPD intervals are: 0.301 (0.168, 0.432) between circulatory and respiratory causes, 0.357 (0.239, 0.453) between circulatory causes and tumours, and 0.37 (0.264, 0.486) between respiratory causes and tumours.

Figure \ref{fig:shrink_USA} displays the posterior estimates of the shrinkage factors employed by the models used for the analysis of \review{female} data. Each figure represents a $ K \times d $ matrix, coloured according to the posterior mode of the shrinkage factors. The larger the value in the $ kj $-th cell, the more significant the contribution of disease $ j $ to the formation of the $ k $-th cluster. The results for $\zeta_j$ are displayed as the normalised version $\zeta_j/\sum_{l = 1}^d \zeta_l$, which can be interpreted as the percentage contribution of disease $ j $ to the whole clustering structure. 
\review{The left panel shows that territorial mortality clusters are predominantly shaped by respiratory diseases ($\sim 50\%$) and tumours ($\sim40\%$).} 
The second panel shows the joint shrinkage factors $\gamma_{kj}$. Although this quantity cannot be normalised as $\zeta_j$, it indicates that tumours are particularly responsible for the formation of cluster 4, and respiratory diseases significantly contribute to the formation of cluster 3.

The two clusters denominated `cluster 1' in both male and female data encompass the majority of counties on the map and exhibit higher mortality rates compared to the entire U.S. territory. The absence of spatial clusters \review{with average mortality levels in keeping with} national mortality levels might raise concerns about potential systematic biases in the data. Therefore, we conduct an additional analysis in a geographically distant area. Specifically, we apply \perla~to female data from the 259 counties on the West side map of the U.S., which is the territory used for the simulation experiments in Section \ref{sec:simulation_experiments}. The prior setup considered is the same used for the East coast female data. In this region, 27 areas have at least one missing data point and their value are imputed similarly to the missing data on the East coast territory. The analysis reveals the presence of a predominant cluster characterised by lower mortality rates for circulatory diseases and tumours compared to the national average. The remaining two clusters do not correspond to major urban areas as observed in the East coast data; instead, they include regions with low population densities. Detailed plots are provided in \review{Supplementary Figure 12}. We conclude that the U.S. East coast exhibits higher mortality levels for major causes of death compared to the rest of the nation, while the West coast demonstrates generally lower mortality levels.

\section{Discussion}
\label{section:discussion}
The analysis of the mortality distribution due to multiple death causes provides an overview of the health conditions and disparities within a territory. In particular, detecting mortality clusters means identifying areas with similar levels of mortality and investigating the presence of unknown risk factors. 
In this work, we have proposed \perla, a multivariate Bayesian model that addresses the dual problem of detecting spatial mortality clusters simultaneously considering multiple death causes, and identifying the diseases responsible for cluster formation. \perla~integrates the spatial structure of the data directly into the clustering probabilities, enabling tractable posterior distributions. Additionally, it accounts for the presence of external covariates, if available.

We have proposed different setups for \perla~considering different types of global-local shrinkage priors. These setups help the model \review{avoid the identification of} spurious clusters, and \reviewm{can be used to quantify which diseases are responsible of the overall clustering structure underlying the territory, as well as for the formation of specific mortality clusters.}

Through two different case-studies, we have demonstrated the various insights achievable with our methodology: it partitions the areas of a territory into clusters, it estimates the probability of risk excess in each cluster, and it quantifies the marginal correlation across mortality levels due to different death causes. 

Although this article has introduced a comprehensive solution to address some key questions in the analysis of multivariate areal data,  we believe that there is space for further extensions.
While we have assumed that the data $\mathbf{Y}$ are realisations of  continuous and zero-centred variables, mortality data often come in the form of counts, as discussed in Section \ref{subsec:simulation3}.
Consequently, following the approach of \cite{Green_Richardson.2002}, the regression specified in Formula \eqref{formula:model} could be extended to the generalised linear model framework, and considering a prior distribution on $\boldsymbol{\beta}$ that induces variable selection when $p$ grows substantially.  This approach should be performed taking into consideration the $d$ regressions jointly, as suggested by \cite{Deshpande_etal.2019}. 

\review{Our simulation experiments have shown that \perla~can recover the clustering structure of the data even when the clusters are characterised by different levels of spatial correlation. Nonetheless, in view of potential applications in other frameworks, \perla~could be extended by incorporating an unstructured random component both at the data level in Formula \eqref{formula:model}, mirroring the structure of the BYM model \citep{Besag_etal.1991}, and in the clustering probabilities. Such an extension, however, complicates the posterior sampling since some steps of the simulation scheme would require the use of the Metropolis-Hastings algorithm.}

In the case-studies treated in this work, we have performed the model selection looking both at the $\mathrm{DIC}_3$ information criterion and the quality of the MCMC samples, preferring models with unimodal posterior distributions. However, this approach necessitates running \perla~multiple times.
Alternatively, a simulation algorithm such as the reversible jump Markov chain Monte Carlo (RJMCMC, \citealt{Green.1995}) could streamline this process.  RJMCMC would automatically test multiple model setups and provide a probabilistic assessment of the number of clusters and the optimal combination of penalisation parameters that best fit the data. \review{Bayesian nonparametric methods are also valuable for selecting the number of mixture components \citep{Rodriguez_Muller.2013}.
In our case studies, we have observed that a small number of clusters is enough to describe the data. However, in other applications, this step may be necessary. In such cases, \perla~could be easily extended to an infinite mixture model by allowing the stick-breaking process in \eqref{formula:stick_breaking} to continue without stopping at a fixed number of components, thereby generating progressively smaller probability sticks (see, for instance, \citealt{Jo_etal.2017}).}

 Finally, as we previously mentioned, inference on the parameters can be performed easily via MCMC methods, thanks to multiple data augmentation strategies that allow for expressing most conditional posterior distributions in closed form.  These computational characteristics make it possible to explore potentially more efficient estimation procedures, such as variational Bayes methods, which are particularly appealing for the analysis of massive datasets.

\section{Supplementary material}

Contains details about the derivation of the $\mathrm{DIC}_3$ information criterion, a discussion on the role of the CAR parameters in the variation of the clustering probabilities, and  additional figures and tables related to both the simulation experiments and case studies.

\section{Software}

The methodology presented in this work is implemented in the \texttt{R} package \texttt{perla}, which can be accessed online at \url{https://github.com/andreasottosanti/perla}.
This package allows to reproduce the analyses conducted on both simulated data and the publicly available mortality U.S. data.
Data from the Italian health care system are not accessible due to privacy concerns.

\section{Competing interests}

No competing interest is declared.




\section{Funding}
This work was supported by Next Generation EU, in the context of the National Recovery and Resilience Plan, Investment PE8 – Project Age-It: \virgolette{Ageing Well in an Ageing Society}, CUP C93C22005240007
 [DM 1557 11.10.2022]. The views and opinions expressed are only those of the authors and do not necessarily reflect those of the European Union or the European Commission. Neither the European Union nor the European Commission can be held responsible for them. Additionally, GB and PB acknowledge funding from PRIN SOcial and health Frailty as determinants of Inequality in Aging. (SOFIA), project n. 020KHSSKE, CUP C93C22000270001, directorial decree ministry of university n. 222 dated February 18, 2022.

\section{Acknowledgments}
The authors are grateful to the editorial board and the three anonymous referees for their constructive comments and valuable suggestions, which greatly improved the quality of this work. In addition, the authors extend their gratitude to C. Castiglione (Bocconi University) and P. Onorati (University of Padua) for the helpful discussions about global-local shrinkage priors, to F. Denti (University of Padua), for precious suggestions after reading the first draft of the manuscript, and to B. Bezzon (University of Padua) for linguistic support.
Authors also acknowledge ULSS6 \textit{Euganea}, particularly L. Benacchio, for providing the aggregated SMR used in this study, within the framework of the STHEP (State of Health in Padua) collaboration between ULSS6 \textit{Euganea} and the Department of Statistical Sciences of the University of Padua.
\color{black}

\bibliographystyle{abbrvnat}
\bibliography{reference}

\begin{thebibliography}{73}
\providecommand{\natexlab}[1]{#1}
\providecommand{\url}[1]{\texttt{#1}}
\expandafter\ifx\csname urlstyle\endcsname\relax
  \providecommand{\doi}[1]{doi: #1}\else
  \providecommand{\doi}{doi: \begingroup \urlstyle{rm}\Url}\fi

\bibitem[Ahmed and Genin(2020)]{ahmed_genin.2020}
M.-S. Ahmed and M.~Genin.
\newblock A functional-model-adjusted spatial scan statistic.
\newblock \emph{Stat Med}, 39\penalty0 (8):\penalty0 1025--1040, 2020.
\newblock ISSN 1097-0258.
\newblock \doi{10.1002/sim.8459}.

\bibitem[Amini et~al.(2023)Amini, Azizmohammad~Looha, Rahimi~Pordanjani,
  Asadzadeh~Aghdaei, and Pourhoseingholi]{Amini_etal.2023}
M.~Amini, M.~Azizmohammad~Looha, S.~Rahimi~Pordanjani, H.~Asadzadeh~Aghdaei,
  and M.~A. Pourhoseingholi.
\newblock Global long-term trends and spatial cluster analysis of pancreatic
  cancer incidence and mortality over a 30-year period using the global burden
  of disease study 2019 data.
\newblock \emph{PloS One}, 18\penalty0 (7), 2023.

\bibitem[Assunção and Krainski(2009)]{Assuncao_Krainski.2009}
R.~Assunção and E.~Krainski.
\newblock Neighborhood {dependence} in {bayesian} {spatial} {models}.
\newblock \emph{Biom J}, 51\penalty0 (5):\penalty0 851--869, 2009.
\newblock ISSN 1521-4036.
\newblock \doi{10.1002/bimj.200900056}.

\bibitem[Aungkulanon et~al.(2017)Aungkulanon, Tangcharoensathien, Shibuya,
  Bundhamcharoen, and Chongsuvivatwong]{Aungkulanon_etal.2017}
S.~Aungkulanon, V.~Tangcharoensathien, K.~Shibuya, K.~Bundhamcharoen, and
  V.~Chongsuvivatwong.
\newblock Area-level socioeconomic deprivation and mortality differentials in
  {Thailand}: results from principal component analysis and cluster analysis.
\newblock \emph{Int J Equity Health}, 16:\penalty0 1--12, 2017.

\bibitem[Banerjee et~al.(2015)Banerjee, Carlin, and
  Gelfand]{Banerjee_etal.2015}
S.~Banerjee, B.~P. Carlin, and A.~E. Gelfand.
\newblock \emph{Hierarchical modeling and analysis for spatial data}, volume
  135 of \emph{Monographs on {Statistics} and {Applied} {Probability}}.
\newblock CRC Press, Boca Raton, FL, second edition, 2015.
\newblock ISBN 978-1-4398-1917-3.

\bibitem[Besag(1974)]{Besag.1974}
J.~Besag.
\newblock Spatial {interaction} and the {statistical} {analysis} of {lattice}
  {systems}.
\newblock \emph{J R Stat Soc Ser B Stat Methodol}, 36\penalty0 (2):\penalty0
  192--236, 1974.
\newblock ISSN 0035-9246.

\bibitem[Besag et~al.(1991)Besag, York, and Mollié]{Besag_etal.1991}
J.~Besag, J.~York, and A.~Mollié.
\newblock Bayesian image restoration, with two applications in spatial
  statistics.
\newblock \emph{Ann Inst Stat Math}, 43\penalty0 (1):\penalty0 1--20, 1991.
\newblock ISSN 1572-9052.
\newblock \doi{10.1007/BF00116466}.

\bibitem[Bhadra et~al.(2019)Bhadra, Datta, Polson, and
  Willard]{Bhadra_etal.2019}
A.~Bhadra, J.~Datta, N.~G. Polson, and B.~Willard.
\newblock Lasso {meets} {horseshoe}: {a} {survey}.
\newblock \emph{Stat Sci}, 34\penalty0 (3):\penalty0 405--427, 2019.
\newblock ISSN 0883-4237.

\bibitem[Bivand et~al.(2013)Bivand, Pebesma, and
  Gómez-Rubio]{Bivand_etal.2013}
R.~S. Bivand, E.~Pebesma, and V.~Gómez-Rubio.
\newblock \emph{Applied {spatial} {data} {analysis} with {R}}.
\newblock Springer, New York, NY, 2013.
\newblock ISBN 978-1-4614-7617-7 978-1-4614-7618-4.

\bibitem[Bovo et~al.(2023)Bovo, Belloni, Sottosanti, and
  Boccuzzo]{Bovo_etal.2023}
E.~Bovo, P.~Belloni, A.~Sottosanti, and G.~Boccuzzo.
\newblock Territorial clusters of mortality and role of social and
  environmental factors: the case of {ULSS 6} {Euganea} ({Italy}).
\newblock In \emph{2023 IEEE Conference on Computational Intelligence in
  Bioinformatics and Computational Biology (CIBCB)}, pages 1--5. IEEE, 2023.

\bibitem[Breslow and Day(1987)]{breslow1987volume}
N.~Breslow and N.~Day.
\newblock \emph{Statistical methods in cancer research. Volume II-The design
  and analysis of cohort studies}.
\newblock International agency for research on cancer, 1987.

\bibitem[Caranci and Costa(2009)]{caranci.2009}
N.~Caranci and G.~Costa.
\newblock Un indice di deprivazione a livello aggregato da utilizzare su scala
  nazionale: giustificazioni e composizione.
\newblock \emph{Salute Soc. Fascicolo 1, 2009}, pages 1000--1021, 2009.

\bibitem[Carlin and Banerjee(2003)]{Carlin_Banerjee.2003}
B.~P. Carlin and S.~Banerjee.
\newblock Hierarchical {multivarite} {CAR} {models} for {spatio}-{temporally}
  {correlated} {survival} {data}.
\newblock \emph{Bayesian Statistics}, 7\penalty0 (7):\penalty0 45--63, 2003.

\bibitem[Carvalho et~al.(2010)Carvalho, Polson, and Scott]{Carvalho_etal.2010}
C.~M. Carvalho, N.~G. Polson, and J.~G. Scott.
\newblock The horseshoe estimator for sparse signals.
\newblock \emph{Biometrika}, 97\penalty0 (2):\penalty0 465--480, 2010.
\newblock ISSN 0006-3444.

\bibitem[CDC(2021)]{WONDER}
CDC.
\newblock National vital statistics system, mortality 1999-2020 on {CDC}
  {WONDER} online database, 2021.
\newblock Accessed on: Apr 11, 2024.
  \url{http://wonder.cdc.gov/mcd-icd10.html}.

\bibitem[Celeux and Govaert(1995)]{celeux_govaert.1995}
G.~Celeux and G.~Govaert.
\newblock Gaussian parsimonious clustering models.
\newblock \emph{Pattern Recognit}, 28\penalty0 (5):\penalty0 781--793, 1995.
\newblock ISSN 0031-3203.
\newblock \doi{10.1016/0031-3203(94)00125-6}.

\bibitem[Celeux et~al.(2006)Celeux, Forbes, Robert, and
  Titterington]{Celeux_etal.2006}
G.~Celeux, F.~Forbes, C.~P. Robert, and D.~M. Titterington.
\newblock Deviance information criteria for missing data models.
\newblock \emph{Bayesian Anal}, 1\penalty0 (4):\penalty0 651--673, 2006.
\newblock ISSN 1936-0975, 1931-6690.
\newblock \doi{10.1214/06-BA122}.

\bibitem[Coker et~al.(2023)Coker, Molitor, Liverani, Martin, Maranzano,
  Pontarollo, and Vergalli]{Coker_etal.2023}
E.~S. Coker, J.~Molitor, S.~Liverani, J.~Martin, P.~Maranzano, N.~Pontarollo,
  and S.~Vergalli.
\newblock Bayesian profile regression to study the ecologic associations of
  correlated environmental exposures with excess mortality risk during the
  first year of the {Covid-19} epidemic in {Lombardy}, {Italy}.
\newblock \emph{Environ Res}, 216, 2023.

\bibitem[Cucala(2016)]{Cucala.2016}
L.~Cucala.
\newblock A {Mann}–{Whitney} scan statistic for continuous data.
\newblock \emph{Commun Stat Theory Methods}, 45\penalty0 (2):\penalty0
  321--329, Jan. 2016.
\newblock ISSN 0361-0926.
\newblock \doi{10.1080/03610926.2013.806667}.

\bibitem[Cucala et~al.(2017)Cucala, Genin, Lanier, and
  Occelli]{Cucala_etal.2017}
L.~Cucala, M.~Genin, C.~Lanier, and F.~Occelli.
\newblock A multivariate {Gaussian} scan statistic for spatial data.
\newblock \emph{Spat Stat}, 21:\penalty0 66--74, Aug. 2017.
\newblock ISSN 2211-6753.
\newblock \doi{10.1016/j.spasta.2017.06.001}.

\bibitem[Cucala et~al.(2019)Cucala, Genin, Occelli, and
  Soula]{Cucala_etal.2019}
L.~Cucala, M.~Genin, F.~Occelli, and J.~Soula.
\newblock A multivariate nonparametric scan statistic for spatial data.
\newblock \emph{Spat Stat}, 29:\penalty0 1--14, Mar. 2019.
\newblock ISSN 2211-6753.
\newblock \doi{10.1016/j.spasta.2018.10.002}.

\bibitem[Dahl et~al.(2022)Dahl, Johnson, and Müller]{Dahl_etal.2022}
D.~B. Dahl, D.~J. Johnson, and P.~Müller.
\newblock Search {algorithms} and {loss} {functions} for {Bayesian}
  {clustering}.
\newblock \emph{J Comput Graph Stat}, 31\penalty0 (4):\penalty0 1189--1201,
  2022.
\newblock ISSN 1061-8600.
\newblock \doi{10.1080/10618600.2022.2069779}.

\bibitem[Datta et~al.(2019)Datta, Banerjee, Hodges, and Gao]{Datta_etal.2019}
A.~Datta, S.~Banerjee, J.~S. Hodges, and L.~Gao.
\newblock Spatial {disease} {mapping} {using} {directed} {acyclic} {graph}
  {auto}-{regressive} ({DAGAR}) {models}.
\newblock \emph{Bayesian Anal}, 14\penalty0 (4), 2019.
\newblock ISSN 1936-0975.
\newblock \doi{10.1214/19-BA1177}.

\bibitem[Denti et~al.(2023)Denti, Azevedo, Lo, Wheeler, Gandhi, Guindani, and
  Shahbaba]{Denti_etal.2023}
F.~Denti, R.~Azevedo, C.~Lo, D.~G. Wheeler, S.~P. Gandhi, M.~Guindani, and
  B.~Shahbaba.
\newblock A horseshoe mixture model for {Bayesian} screening with an
  application to light sheet fluorescence microscopy in brain imaging.
\newblock \emph{Ann Appl Stat}, 17\penalty0 (3):\penalty0 2639--2658, Sept.
  2023.
\newblock ISSN 1932-6157, 1941-7330.
\newblock \doi{10.1214/23-AOAS1736}.

\bibitem[Deshpande et~al.(2019)Deshpande, Ročková, and
  George]{Deshpande_etal.2019}
S.~K. Deshpande, V.~Ročková, and E.~I. George.
\newblock Simultaneous {variable} and {covariance} {selection} {with} the
  {multivariate} {spike}-and-{slab} {LASSO}.
\newblock \emph{J Comput Graph Stat}, 28\penalty0 (4):\penalty0 921--931, 2019.
\newblock ISSN 1061-8600.
\newblock \doi{10.1080/10618600.2019.1593179}.

\bibitem[Fedeli et~al.(2021)Fedeli, Schievano, Avossa, Baruffa, Rampado, Lucia,
  Veronese, Gennaro, Pellizzari, Pinato, Ferroni, Basso, Netti, Cestari, Paoli,
  Dotto, Pierobon, Casotto, Costa, Braggion, Lamattina, and
  Zabeo]{fedeli_etal.2021}
U.~Fedeli, E.~Schievano, F.~Avossa, T.~Baruffa, A.~Rampado, A.~Lucia,
  M.~Veronese, N.~Gennaro, M.~Pellizzari, E.~Pinato, E.~Ferroni, C.~Basso,
  S.~T. Netti, L.~Cestari, A.~D. Paoli, M.~Dotto, S.~Pierobon, V.~Casotto,
  M.~Costa, M.~Braggion, M.~R. Lamattina, and V.~Zabeo.
\newblock La mortalità nella {Regione} del {Veneto}. {Periodo} 2016-2019,
  2021.

\bibitem[Frévent et~al.(2023)Frévent, Ahmed, Dabo-Niang, and
  Genin]{Frevent_etal.2023}
C.~Frévent, M.-S. Ahmed, S.~Dabo-Niang, and M.~Genin.
\newblock Investigating spatial scan statistics for multivariate functional
  data.
\newblock \emph{J R Stat Soc Ser C Appl Stat}, 72\penalty0 (2):\penalty0
  450--475, May 2023.
\newblock ISSN 0035-9254.
\newblock \doi{10.1093/jrsssc/qlad017}.

\bibitem[Green(1995)]{Green.1995}
P.~J. Green.
\newblock Reversible jump {Markov} chain {Monte} {Carlo} computation and
  {Bayesian} model determination.
\newblock \emph{Biometrika}, 82\penalty0 (4):\penalty0 711--732, 1995.
\newblock ISSN 0006-3444.
\newblock \doi{10.1093/biomet/82.4.711}.

\bibitem[Green and Richardson(2002)]{Green_Richardson.2002}
P.~J. Green and S.~Richardson.
\newblock Hidden {Markov} {models} and {disease} {mapping}.
\newblock \emph{J Am Stat Assoc}, 97\penalty0 (460):\penalty0 1055--1070, 2002.
\newblock ISSN 0162-1459.

\bibitem[Griffith(2023)]{griffith.2023}
D.~A. Griffith.
\newblock Spatial autocorrelation mixtures in geospatial disease data: an
  important global epidemiologic/public health assessment ingredient?
\newblock \emph{Trans GIS}, 27\penalty0 (3):\penalty0 730--751, 2023.

\bibitem[Gómez-Rubio et~al.(2019)Gómez-Rubio, Moraga, Molitor, and
  Rowlingson]{gomez-rubio_etal.2019}
V.~Gómez-Rubio, P.~Moraga, J.~Molitor, and B.~Rowlingson.
\newblock {DClusterm}: {model}-{based} {detection} of {disease} {clusters}.
\newblock \emph{J Stat Softw}, 90:\penalty0 1--26, Aug. 2019.
\newblock ISSN 1548-7660.
\newblock \doi{10.18637/jss.v090.i14}.

\bibitem[Hubert and Arabie(1985)]{Hubert_Arabie.1985}
L.~Hubert and P.~Arabie.
\newblock Comparing partitions.
\newblock \emph{J Classif}, 2\penalty0 (1):\penalty0 193--218, Dec. 1985.
\newblock ISSN 1432-1343.
\newblock \doi{10.1007/BF01908075}.

\bibitem[immobiliare.it(2022)]{immobiliare.it}
immobiliare.it.
\newblock Real estate price data in {Italy}, 2022.
\newblock Accessed on: Dec 19, 2022.
  \url{https://www.immobiliare.it/mercato-immobiliare}.

\bibitem[Jo et~al.(2017)Jo, Lee, Müller, Quintana, and Trippa]{Jo_etal.2017}
S.~Jo, J.~Lee, P.~Müller, F.~A. Quintana, and L.~Trippa.
\newblock Dependent {species} {sampling} {models} for {spatial} {density}
  {estimation}.
\newblock \emph{Bayesian Anal}, 12\penalty0 (2), June 2017.
\newblock ISSN 1936-0975.
\newblock \doi{10.1214/16-BA1006}.

\bibitem[Kim(2021)]{Kim.2021}
C.~Kim.
\newblock Deviance information criteria for mixtures of distributions.
\newblock \emph{Commun Stat Simul Comput}, 50\penalty0 (10):\penalty0
  2935--2948, 2021.
\newblock ISSN 0361-0918.
\newblock \doi{10.1080/03610918.2019.1617878}.

\bibitem[Kulldorff(1997)]{kulldorff.1997}
M.~Kulldorff.
\newblock A spatial scan statistic.
\newblock \emph{Commun Stat Theory Methods}, 26\penalty0 (6):\penalty0
  1481--1496, 1997.

\bibitem[Kulldorff et~al.(2005)Kulldorff, Heffernan, Hartman, Assun{\c{c}}ao,
  and Mostashari]{Kulldorff_etal.2005}
M.~Kulldorff, R.~Heffernan, J.~Hartman, R.~Assun{\c{c}}ao, and F.~Mostashari.
\newblock A space--time permutation scan statistic for disease outbreak
  detection.
\newblock \emph{PLoS Med}, 2\penalty0 (3):\penalty0 e59, 2005.

\bibitem[Künzli et~al.(2000)Künzli, Kaiser, Medina, Studnicka, Chanel,
  Filliger, Herry, Horak, Puybonnieux-Texier, Quénel, Schneider, Seethaler,
  Vergnaud, and Sommer]{Kunzli_etal.2000}
N.~Künzli, R.~Kaiser, S.~Medina, M.~Studnicka, O.~Chanel, P.~Filliger,
  M.~Herry, F.~Horak, V.~Puybonnieux-Texier, P.~Quénel, J.~Schneider,
  R.~Seethaler, J.-C. Vergnaud, and H.~Sommer.
\newblock Public-health impact of outdoor and traffic-related air pollution: a
  {European} assessment.
\newblock \emph{Lancet}, 356\penalty0 (9232):\penalty0 795--801, Sept. 2000.
\newblock ISSN 0140-6736.
\newblock \doi{10.1016/S0140-6736(00)02653-2}.

\bibitem[Lafferty and Blei(2005)]{Lafferty_Blei.2005}
J.~Lafferty and D.~Blei.
\newblock Correlated {topic} {models}.
\newblock In \emph{Advances in {Neural} {Information} {Processing} {Systems}},
  volume~18. MIT Press, 2005.

\bibitem[Lawson(2013)]{lawson.2013}
A.~B. Lawson.
\newblock \emph{Statistical methods in spatial epidemiology}.
\newblock John Wiley \& Sons, 2013.

\bibitem[Lee et~al.(2009)Lee, Ferguson, and Mitchell]{Lee_etal.2009}
D.~Lee, C.~Ferguson, and R.~Mitchell.
\newblock Air pollution and health in {Scotland}: a multicity study.
\newblock \emph{Biostatistics}, 10\penalty0 (3):\penalty0 409--423, 2009.

\bibitem[Li et~al.(2012)Li, Best, Hansell, Ahmed, and Richardson]{Li_etal.2012}
G.~Li, N.~Best, A.~L. Hansell, I.~Ahmed, and S.~Richardson.
\newblock {BaySTDetect: detecting unusual temporal patterns in small area data
  via Bayesian model choice}.
\newblock \emph{Biostatistics}, 13\penalty0 (4):\penalty0 695--710, 2012.
\newblock ISSN 1465-4644.
\newblock \doi{10.1093/biostatistics/kxs005}.

\bibitem[Li et~al.(2021)Li, Nguyen, Banerjee, Rhee, Kalantar-Zadeh, Kürüm,
  and Şentürk]{Li_etal.2021}
Y.~Li, D.~V. Nguyen, S.~Banerjee, C.~M. Rhee, K.~Kalantar-Zadeh, E.~Kürüm,
  and D.~Şentürk.
\newblock Multilevel {modeling} of {spatially} {nested} {functional} {data}:
  {spatiotemporal} {patterns} of {hospitalization} {rates} in the {U}.{S}.
  {dialysis} {population}.
\newblock \emph{Stat Med}, 40\penalty0 (17):\penalty0 3937--3952, July 2021.
\newblock ISSN 0277-6715.
\newblock \doi{10.1002/sim.9007}.

\bibitem[Linderman et~al.(2015)Linderman, Johnson, and
  Adams]{Linderman_etal.2015}
S.~W. Linderman, M.~J. Johnson, and R.~P. Adams.
\newblock Dependent multinomial models made easy: stick breaking with the
  {Pólya}-gamma augmentation.
\newblock In \emph{Proceedings of the 28th {International} {Conference} on
  {Neural} {Information} {Processing} {Systems} - {Volume} 2}, {NIPS}'15, pages
  3456--3464, Cambridge, MA, USA, 2015. MIT Press.

\bibitem[Liu et~al.(2018)Liu, Liu, and Zhang]{Liu_etal.2018}
Y.~Liu, Y.~Liu, and T.~Zhang.
\newblock Wald-based spatial scan statistics for cluster detection.
\newblock \emph{Comp Stat Data Anal}, 127:\penalty0 298--310, Nov. 2018.
\newblock ISSN 0167-9473.
\newblock \doi{10.1016/j.csda.2018.06.002}.

\bibitem[Makalic and Schmidt(2016)]{Makalic_Schmidt.2016}
E.~Makalic and D.~F. Schmidt.
\newblock A {simple} {sampler} for the {horseshoe} {estimator}.
\newblock \emph{IEEE Signal Processing Letters}, 23\penalty0 (1):\penalty0
  179--182, 2016.
\newblock ISSN 1558-2361.
\newblock \doi{10.1109/LSP.2015.2503725}.

\bibitem[Marin and Robert(2007)]{Marin_Robert.2007}
J.-M. Marin and C.~P. Robert.
\newblock Regression and {variable} {selection}.
\newblock In \emph{Bayesian {Core}: {A} {Practical} {Approach} to
  {Computational} {Bayesian} {Statistics}}, pages 47--84. Springer, New York,
  NY, 2007.
\newblock ISBN 978-0-387-38983-7.

\bibitem[Moores et~al.(2020{\natexlab{a}})Moores, Nicholls, Pettitt, and
  Mengersen]{moores_etal.2020}
M.~Moores, G.~Nicholls, A.~Pettitt, and K.~Mengersen.
\newblock Scalable {Bayesian} {inference} for the {inverse} {temperature} of a
  {hidden} {Potts} {model}.
\newblock \emph{Bayesian Anal}, 15\penalty0 (1):\penalty0 1--27,
  2020{\natexlab{a}}.
\newblock ISSN 1936-0975, 1931-6690.
\newblock \doi{10.1214/18-BA1130}.

\bibitem[Moores et~al.(2015)Moores, Drovandi, Mengersen, and
  Robert]{Moores_etal.2015}
M.~T. Moores, C.~C. Drovandi, K.~Mengersen, and C.~P. Robert.
\newblock Pre-processing for approximate {Bayesian} computation in image
  analysis.
\newblock \emph{Stat and Comput}, 25\penalty0 (1):\penalty0 23--33, 2015.
\newblock ISSN 1573-1375.
\newblock \doi{10.1007/s11222-014-9525-6}.

\bibitem[Moores et~al.(2020{\natexlab{b}})Moores, Pettitt, and
  Mengersen]{Moores_etal.2018}
M.~T. Moores, A.~N. Pettitt, and K.~L. Mengersen.
\newblock Bayesian {computation} with {intractable} {likelihoods}.
\newblock In K.~L. Mengersen, P.~Pudlo, and C.~P. Robert, editors, \emph{Case
  {Studies} in {Applied} {Bayesian} {Data} {Science}: {CIRM} {Jean}-{Morlet}
  {Chair}, {Fall} 2018}, pages 137--151. Springer International Publishing,
  Cham, 2020{\natexlab{b}}.
\newblock ISBN 978-3-030-42553-1.

\bibitem[Möller and Ahrenfeldt(2021)]{Moller_Ahrenfeldt.2021}
S.~Möller and L.~J. Ahrenfeldt.
\newblock Estimating {relative} {risk} {when} {observing} {zero}
  {events}—{frequentist} {inference} and {Bayesian} {credibility}
  {intervals}.
\newblock \emph{Int J Environ Res Public Health}, 18\penalty0 (11):\penalty0
  5527, May 2021.
\newblock ISSN 1661-7827.
\newblock \doi{10.3390/ijerph18115527}.

\bibitem[Papastamoulis(2016)]{Papastamoulis.2016}
P.~Papastamoulis.
\newblock label.switching: {an} {R} {package} for {dealing} with the {label}
  {switching} {problem} in {MCMC} {outputs}.
\newblock \emph{J Stat Softw}, 69:\penalty0 1--24, 2016.
\newblock ISSN 1548-7660.
\newblock \doi{10.18637/jss.v069.c01}.

\bibitem[Papastamoulis and Iliopoulos(2013)]{Papastamoulis_Iliopoulos.2010}
P.~Papastamoulis and G.~Iliopoulos.
\newblock On the {convergence} {rate} of {random} {permutation} {sampler} and
  {ECR} {algorithm} in {missing} {data} {models}.
\newblock \emph{Methodol Comput Appl Probab}, 15\penalty0 (2):\penalty0
  293--304, 2013.
\newblock ISSN 1573-7713.
\newblock \doi{10.1007/s11009-011-9238-7}.

\bibitem[Piironen and Vehtari(2017)]{Piironen_Vehtari.2017}
J.~Piironen and A.~Vehtari.
\newblock Sparsity information and regularization in the horseshoe and other
  shrinkage priors.
\newblock \emph{Electron J Stat}, 11\penalty0 (2):\penalty0 5018--5051, Jan.
  2017.
\newblock ISSN 1935-7524, 1935-7524.
\newblock \doi{10.1214/17-EJS1337SI}.

\bibitem[Polson and Scott(2011)]{polson_scott.2011}
N.~G. Polson and J.~G. Scott.
\newblock Shrink {globally}, {act} {locally}: {sparse} {Bayesian}
  {regularization} and {prediction}.
\newblock In J.~M. Bernardo, M.~J. Bayarri, J.~O. Berger, A.~P. Dawid,
  D.~Heckerman, A.~F.~M. Smith, and M.~West, editors, \emph{Bayesian
  {Statistics} 9}, page~0. Oxford University Press, Oct. 2011.
\newblock ISBN 978-0-19-969458-7.

\bibitem[Potts(1952)]{Potts.1952}
R.~B. Potts.
\newblock Some generalized order-disorder transformations.
\newblock \emph{Math Proc Camb Philos Soc}, 48\penalty0 (1):\penalty0 106--109,
  1952.
\newblock ISSN 1469-8064, 0305-0041.
\newblock \doi{10.1017/S0305004100027419}.

\bibitem[Qian et~al.(2023)Qian, Nguyen, Telesca, Kurum, Rhee, Banerjee, Li, and
  Senturk]{Qian_etal.2023}
Q.~Qian, D.~V. Nguyen, D.~Telesca, E.~Kurum, C.~M. Rhee, S.~Banerjee, Y.~Li,
  and D.~Senturk.
\newblock Multivariate spatiotemporal functional principal component analysis
  for modeling hospitalization and mortality rates in the dialysis population.
\newblock \emph{Biostatistics}, 2023.

\bibitem[{R Core Team}(2024)]{Rlibrary}
{R Core Team}.
\newblock \emph{R: A language and environment for statistical computing}.
\newblock R Foundation for Statistical Computing, Vienna, Austria, 2024.
\newblock URL \url{https://www.R-project.org/}.

\bibitem[Richardson et~al.(2004)Richardson, Thomson, Best, and
  Elliott]{Richardson_etal.2004}
S.~Richardson, A.~Thomson, N.~Best, and P.~Elliott.
\newblock Interpreting {posterior} {relative} {risk} {estimates} in
  {disease}-{mapping} {studies}.
\newblock \emph{Environ Health Perspect}, 112\penalty0 (9):\penalty0
  1016--1025, 2004.
\newblock \doi{10.1289/ehp.6740}.

\bibitem[Robertson et~al.(2010)Robertson, Nelson, MacNab, and
  Lawson]{Robertson_etal.2010}
C.~Robertson, T.~A. Nelson, Y.~C. MacNab, and A.~B. Lawson.
\newblock Review of methods for space–time disease surveillance.
\newblock \emph{Spat Spatiotemporal Epidemiol}, 1\penalty0 (2):\penalty0
  105--116, July 2010.
\newblock ISSN 1877-5845.
\newblock \doi{10.1016/j.sste.2009.12.001}.

\bibitem[Rodriguez and Müller(2013)]{Rodriguez_Muller.2013}
A.~Rodriguez and P.~Müller.
\newblock Nonparametric {Bayesian} {inference}.
\newblock \emph{NSF-CBMS Reg Conf Ser Probab Stat}, 9:\penalty0 i--110, 2013.
\newblock ISSN 1935-5920.

\bibitem[Rosano et~al.(2020)Rosano, Pacelli, Zengarini, Costa, Cislaghi, and
  Caranci]{rosano2020}
A.~Rosano, B.~Pacelli, N.~Zengarini, G.~Costa, C.~Cislaghi, and N.~Caranci.
\newblock Aggiornamento e revisione dell’indice di deprivazione italiano 2011
  a livello di sezione di censimento.
\newblock \emph{Epidemiol Prev}, 44\penalty0 (2-3):\penalty0 162--70, 2020.

\bibitem[Ročková(2018)]{rockova.2018}
V.~Ročková.
\newblock Bayesian estimation of sparse signals with a continuous
  spike-and-slab prior.
\newblock \emph{Ann Stat}, 46\penalty0 (1):\penalty0 401--437, Feb. 2018.
\newblock ISSN 0090-5364, 2168-8966.
\newblock \doi{10.1214/17-AOS1554}.

\bibitem[Russo et~al.(2022)Russo, Singer, and Dunson]{Russo_etal.2022}
M.~Russo, B.~H. Singer, and D.~B. Dunson.
\newblock Multivariate mixed membership modeling: {inferring} domain-specific
  risk profiles.
\newblock \emph{Ann Appl Stat}, 16\penalty0 (1):\penalty0 391--413, 2022.
\newblock ISSN 1932-6157.
\newblock \doi{10.1214/21-aoas1496}.

\bibitem[Scrucca et~al.(2016)Scrucca, Fop, Murphy, and
  Raftery]{scrucca_etal.2016}
L.~Scrucca, M.~Fop, T.~B. Murphy, and A.~E. Raftery.
\newblock mclust 5: {clustering}, {classification} and {density} {estimation}
  {using} {Gaussian} {finite} {mixture} {models}.
\newblock \emph{R J}, 8\penalty0 (1):\penalty0 289--317, 2016.
\newblock ISSN 2073-4859.

\bibitem[Tango(2021)]{tango.2021}
T.~Tango.
\newblock Spatial scan statistics can be dangerous.
\newblock \emph{Stat Methods Med Res}, 30\penalty0 (1):\penalty0 75--86, 2021.

\bibitem[Taylor et~al.(2015)Taylor, Wilkinson, Davies, Armstrong, Chalabi,
  Mavrogianni, Symonds, Oikonomou, and Bohnenstengel]{Taylor_etal.2015}
J.~Taylor, P.~Wilkinson, M.~Davies, B.~Armstrong, Z.~Chalabi, A.~Mavrogianni,
  P.~Symonds, E.~Oikonomou, and S.~I. Bohnenstengel.
\newblock Mapping the effects of urban heat island, housing, and age on excess
  heat-related mortality in {London}.
\newblock \emph{Urban Clim}, 14:\penalty0 517--528, 2015.

\bibitem[Tesema et~al.(2023)Tesema, Tessema, Heritier, Stirling, and
  Earnest]{tesema.2023}
G.~A. Tesema, Z.~T. Tessema, S.~Heritier, R.~G. Stirling, and A.~Earnest.
\newblock A systematic review of joint spatial and spatiotemporal models in
  health research.
\newblock \emph{Int J Environ Res Public Health}, 20\penalty0 (7):\penalty0
  5295, 2023.

\bibitem[van Erp et~al.(2019)van Erp, Oberski, and Mulder]{vanErp_etal.2019}
S.~van Erp, D.~L. Oberski, and J.~Mulder.
\newblock Shrinkage priors for {Bayesian} penalized regression.
\newblock \emph{J Math Psychol}, 89:\penalty0 31--50, Apr. 2019.
\newblock ISSN 0022-2496.
\newblock \doi{10.1016/j.jmp.2018.12.004}.

\bibitem[Ver~Hoef et~al.(2018)Ver~Hoef, Peterson, Hooten, Hanks, and
  Fortin]{VerHoef_etal.2018}
J.~M. Ver~Hoef, E.~E. Peterson, M.~B. Hooten, E.~M. Hanks, and M.-J. Fortin.
\newblock Spatial autoregressive models for statistical inference from
  ecological data.
\newblock \emph{Ecol Monogr}, 88\penalty0 (1):\penalty0 36--59, 2018.
\newblock ISSN 1557-7015.
\newblock \doi{10.1002/ecm.1283}.

\bibitem[Waller and Gotway(2004)]{waller.2004}
L.~A. Waller and C.~A. Gotway.
\newblock \emph{Applied spatial statistics for public health data}.
\newblock John Wiley \& Sons, 2004.

\bibitem[Wong et~al.(2008)Wong, Vichit-Vadakan, Kan, Qian, and {the PAPA
  Project Teams}]{Wong_etal.2008}
C.-M. Wong, N.~Vichit-Vadakan, H.~Kan, Z.~Qian, and {the PAPA Project Teams}.
\newblock Public {health} and {air} {pollution} in {Asia} ({PAPA}): {a}
  {multicity} {study} of {short}-{term} {effects} of {air} {pollution} on
  {mortality}.
\newblock \emph{Environ Health Perspect}, 116\penalty0 (9):\penalty0
  1195--1202, Sept. 2008.
\newblock \doi{10.1289/ehp.11257}.

\bibitem[Zhao et~al.(2021)Zhao, Stone, Ren, Guenthoer, Smythe, Pulliam,
  Williams, Uytingco, Taylor, Nghiem, Bielas, and Gottardo]{zhao_etal.2021}
E.~Zhao, M.~R. Stone, X.~Ren, J.~Guenthoer, K.~S. Smythe, T.~Pulliam, S.~R.
  Williams, C.~R. Uytingco, S.~E.~B. Taylor, P.~Nghiem, J.~H. Bielas, and
  R.~Gottardo.
\newblock Spatial transcriptomics at subspot resolution with {BayesSpace}.
\newblock \emph{Nat Biotechnol}, 39\penalty0 (11):\penalty0 1375--1384, 2021.
\newblock ISSN 1546-1696.
\newblock \doi{10.1038/s41587-021-00935-2}.

\end{thebibliography}



\newpage

\begin{figure}[t]
\vspace{1cm}
\includegraphics[width= 1\linewidth, height = 5cm]{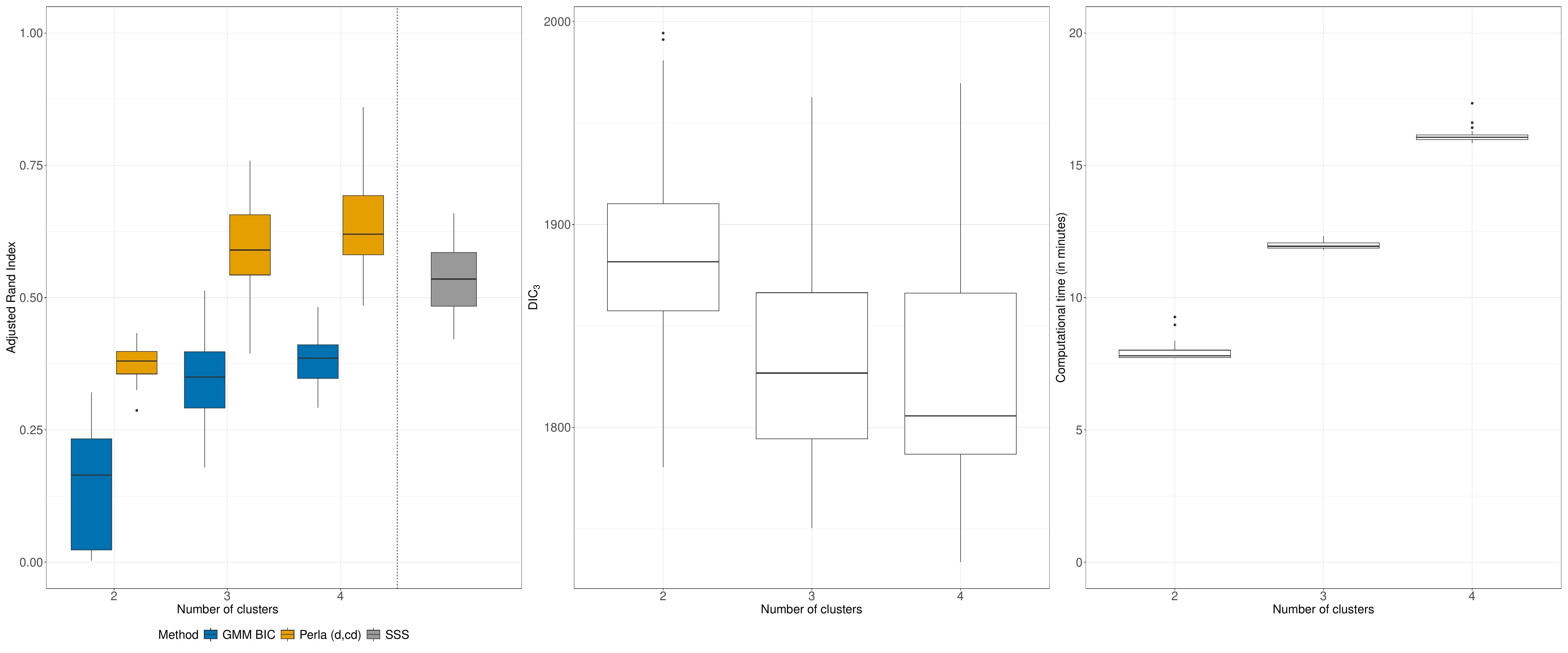}
\centering
\caption{\review{Results from Simulation 3 based on 20 simulated datasets. Left: comparison of the ARI distribution obtained using PERLA with prior setup \texttt{(d,cd)} with covariates, GMM with covariates, and the SSS of \cite{gomez-rubio_etal.2019}. \reviewm{Centre: distribution of the $\mathrm{DIC}_3$ information criterion for PERLA with $K = 2, 3, 4$ clusters.} Right: distribution of computational time (in minutes) required to run four Markov chains of length 10,000 sequentially using our MCMC algorithm using PERLA with setup (d,cd).
}}
\label{fig:simulation3}
\end{figure}



\begin{figure}[t]
    \centering
        \includegraphics[width=0.48\linewidth]{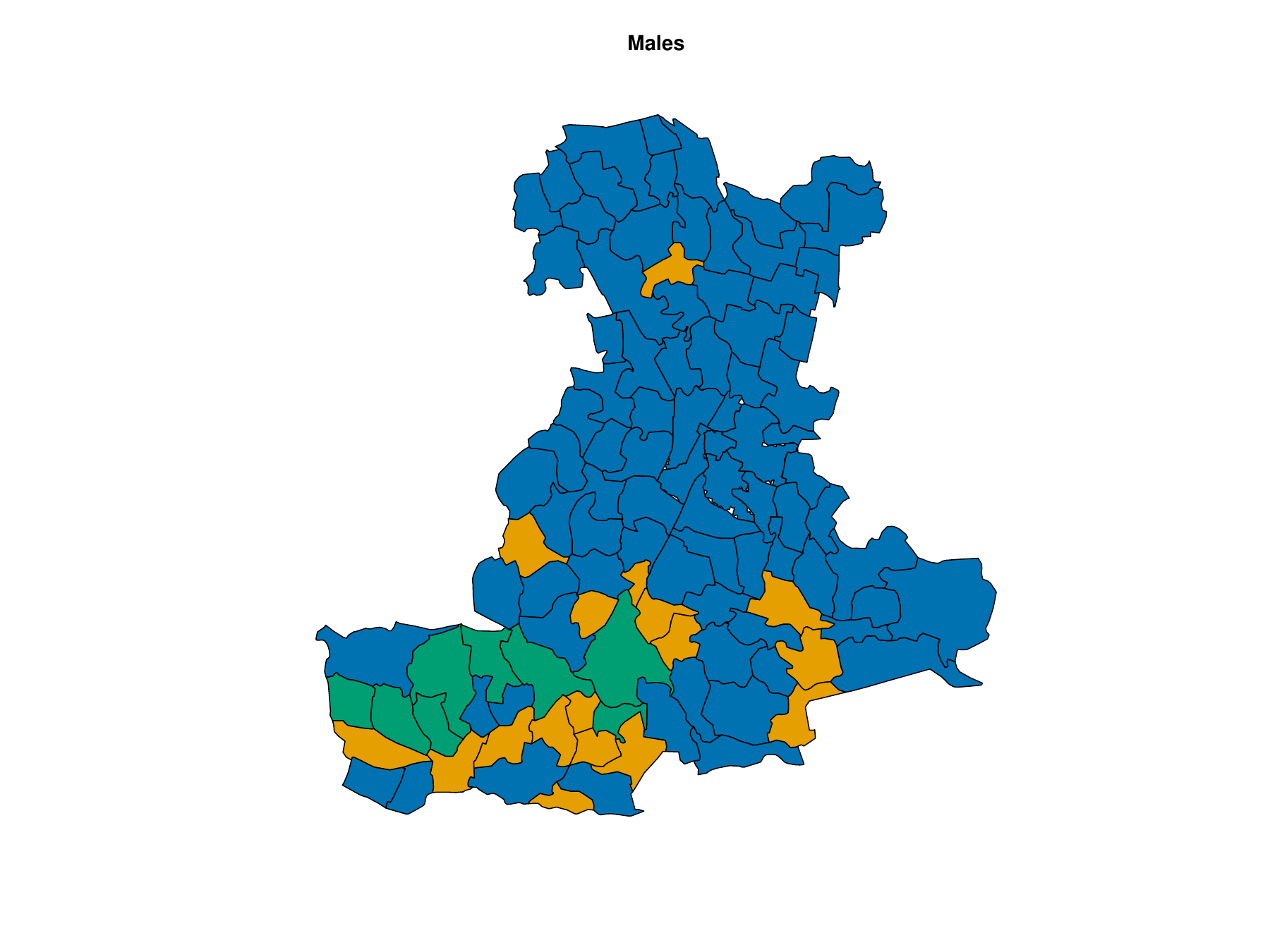}
 \includegraphics[width=7.8cm, height = 5.5cm]{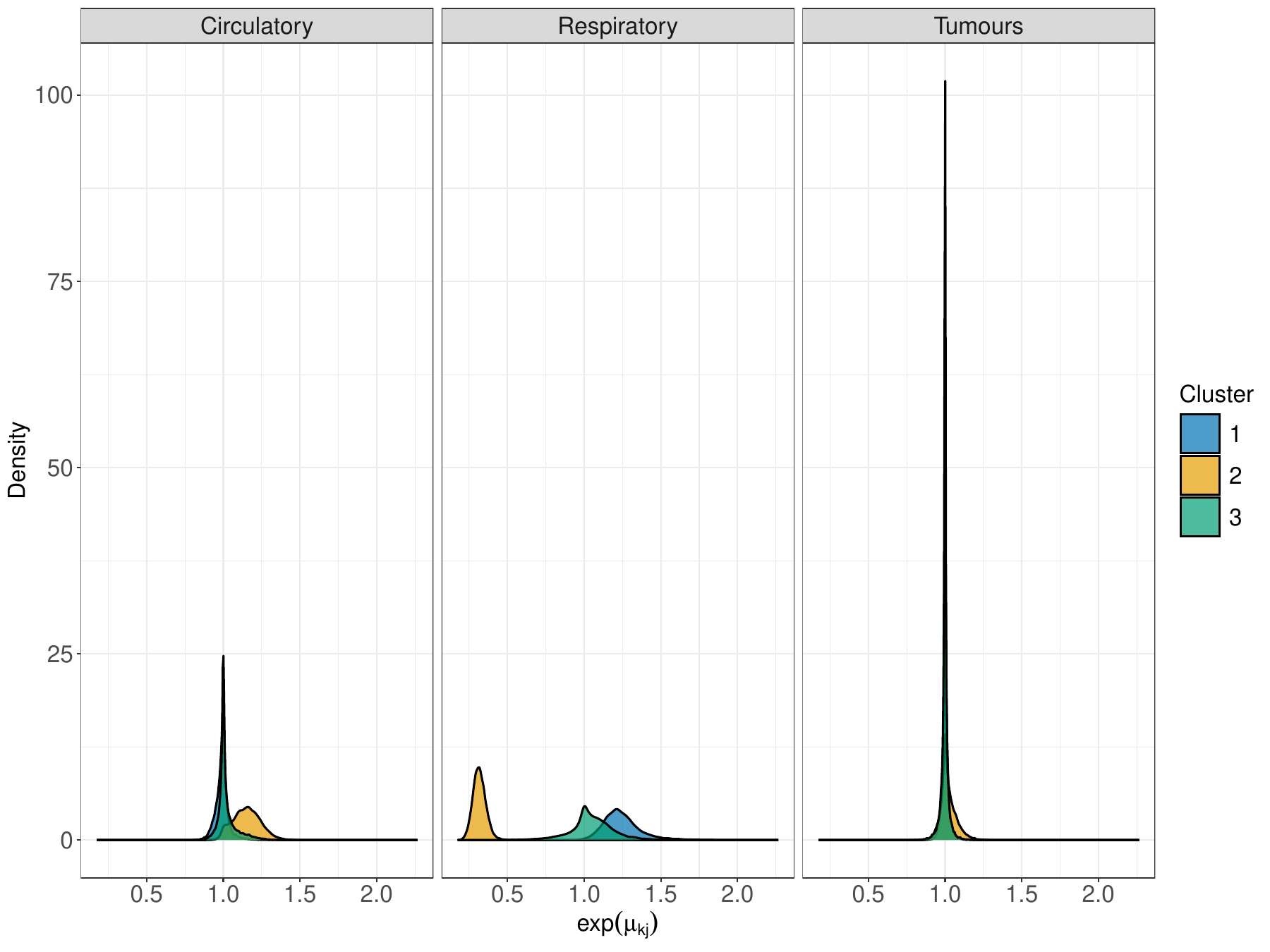}\\
 \includegraphics[width=0.48\linewidth]{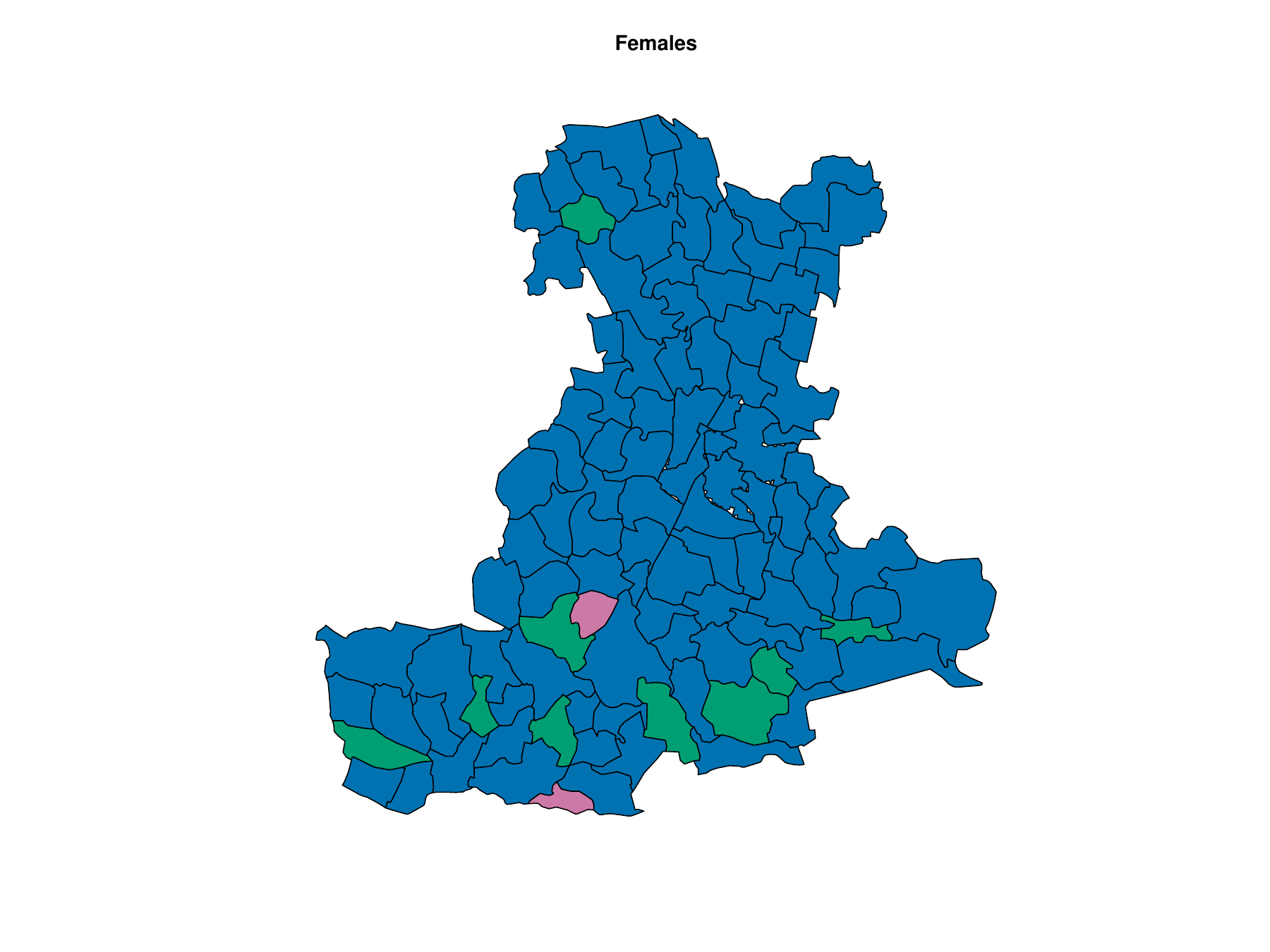}
 \includegraphics[width=7.8cm, height = 5.5cm]{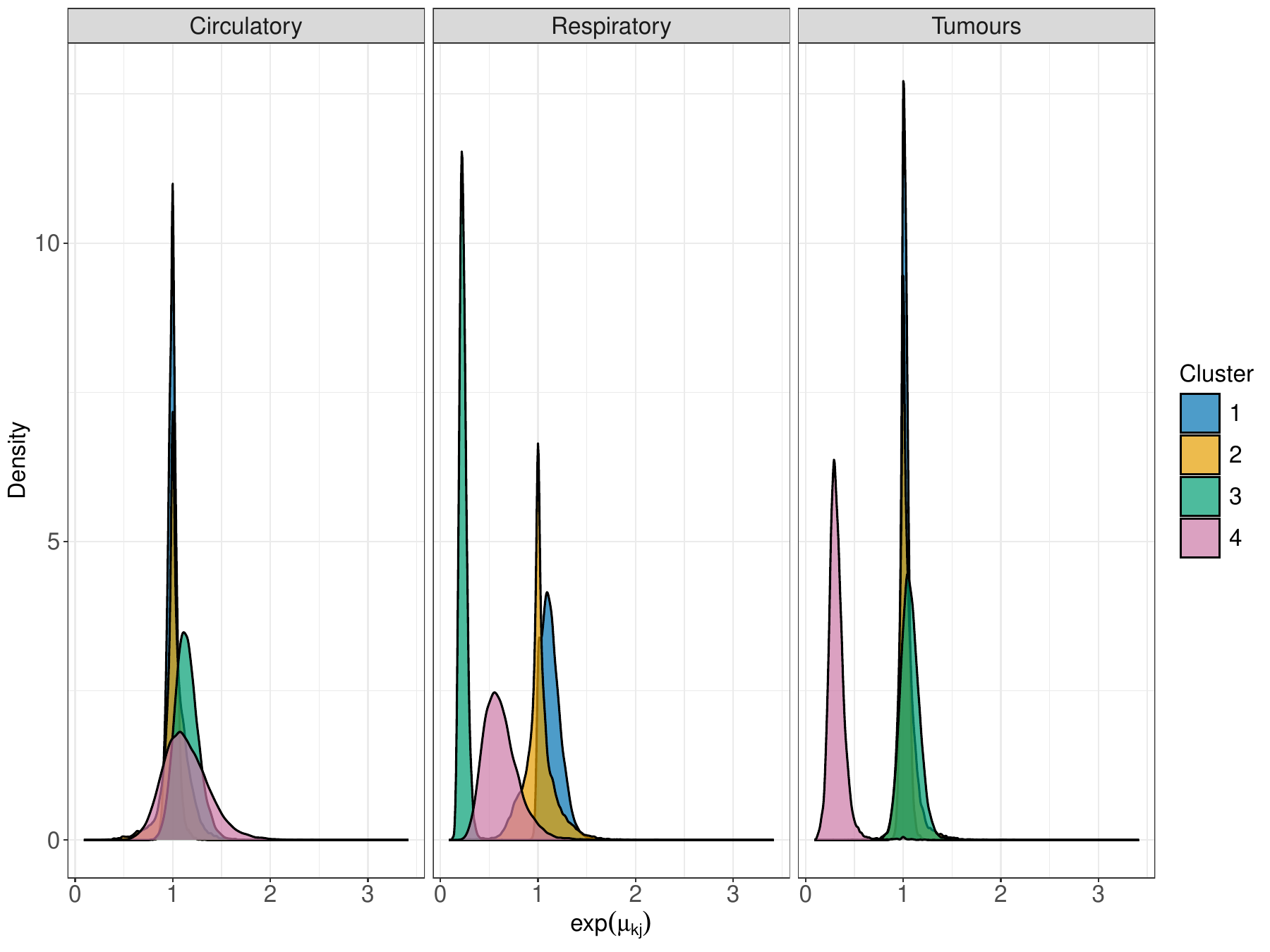}\\
    \caption{Results on the Padua province data, divided for males (top line) and females (bottom line). Left plots display the 106 areas coloured according to the best partition returned by the ECR relabelling algorithm. Right plots display the posterior densities of the $\{\exp(\mu_{kj})\}$, based on 20,000 posterior draws, obtained from the model with $K = 3$ (males) and $K = 4$ (females).}
    \label{fig:map_posterior_Padua}
\end{figure}

\begin{figure}[t]
    \centering
    \includegraphics[width=.48\linewidth]{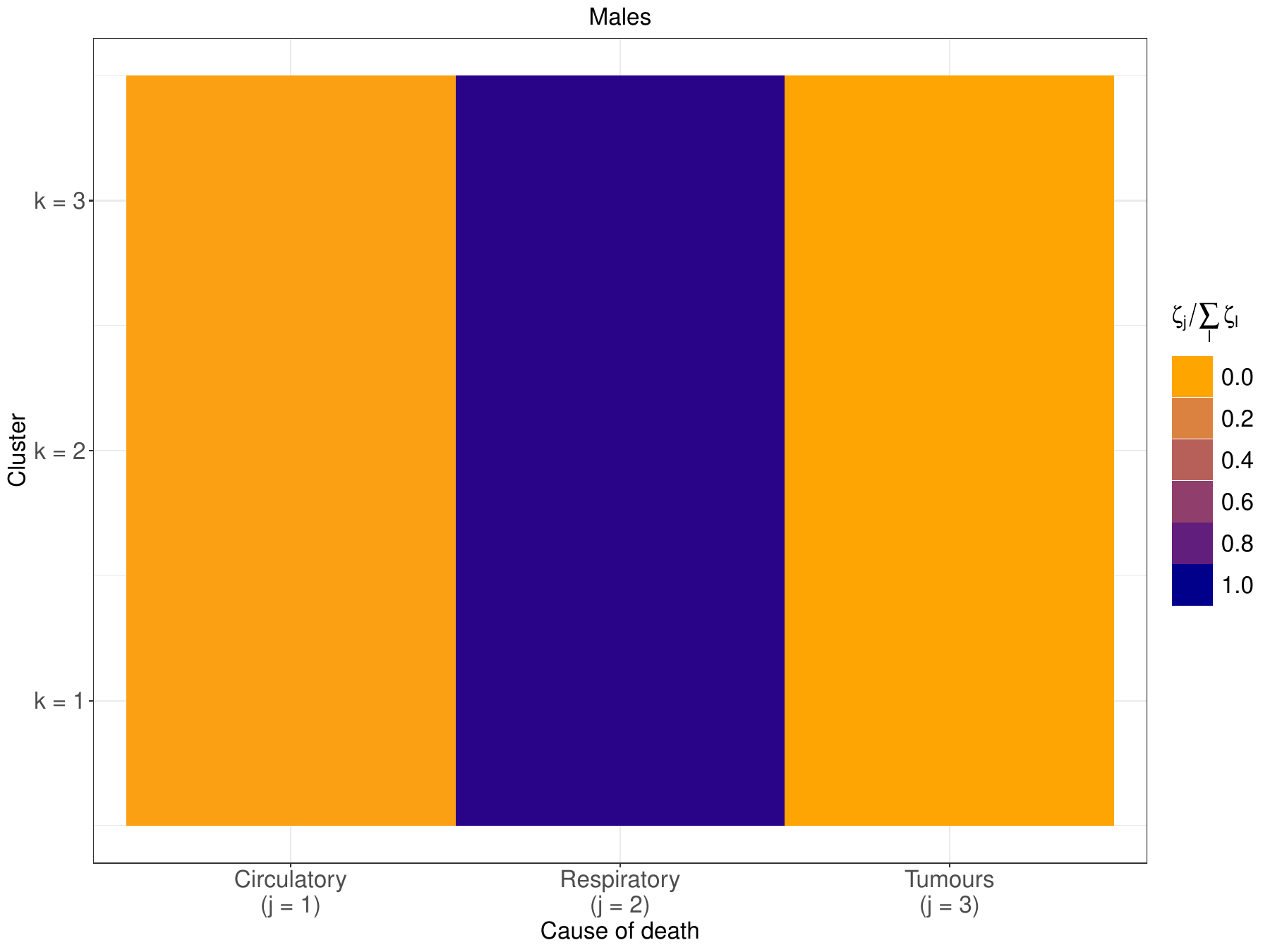}
    \caption{\reviewm{Posterior estimates of the shrinkage factors applied to the intercepts $\mu_{kj}$ for  each cluster $k = 1,\dots,K$ and each disease $j = 1,\dots,d$, obtained from the Padua male data. 
    The graph is coloured according to  $\zeta_j/\sum_{l = 1}^d \zeta_l$, the normalised version of the disease shrinkage factors $\zeta_j$. The $j$-th factor applies the same amount of shrinkage to every element of $\Mu_{.j}$.}}
    \label{fig:shrink_Padova}
\end{figure}


\begin{figure}[t]
    \centering
        \includegraphics[width=0.32\linewidth]{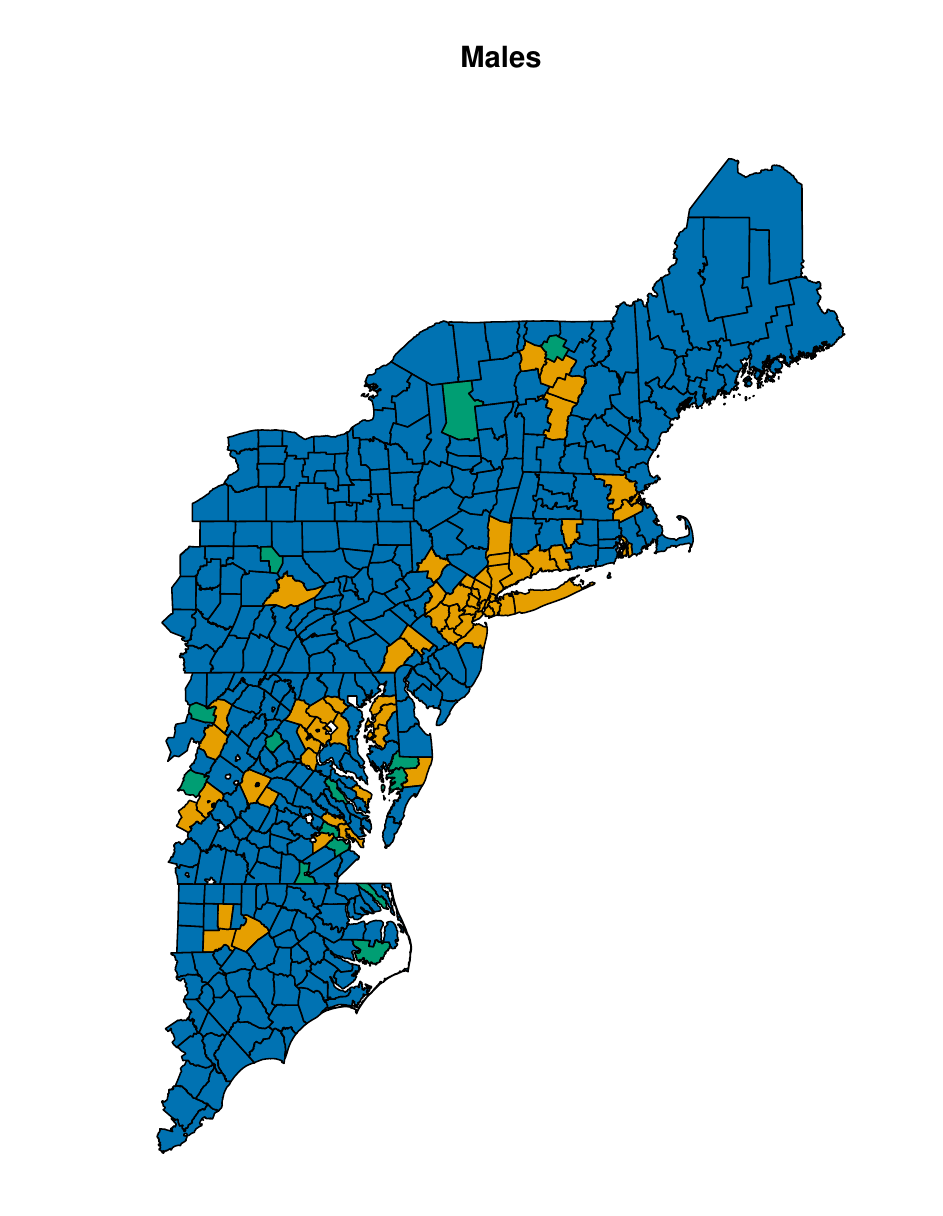}
 \includegraphics[width=7.8cm, height = 5.5cm]{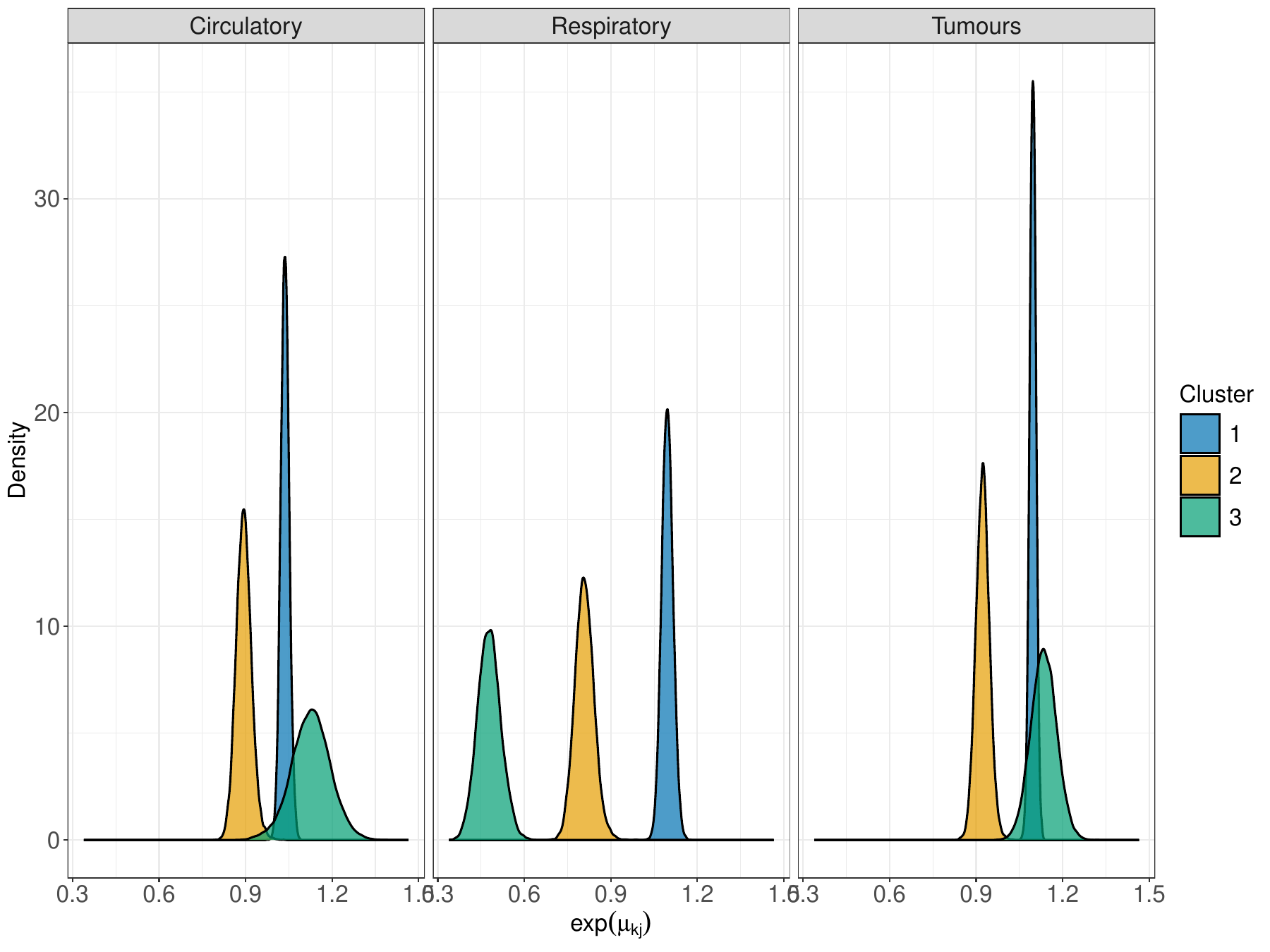}\\
 \includegraphics[width=0.32\linewidth]{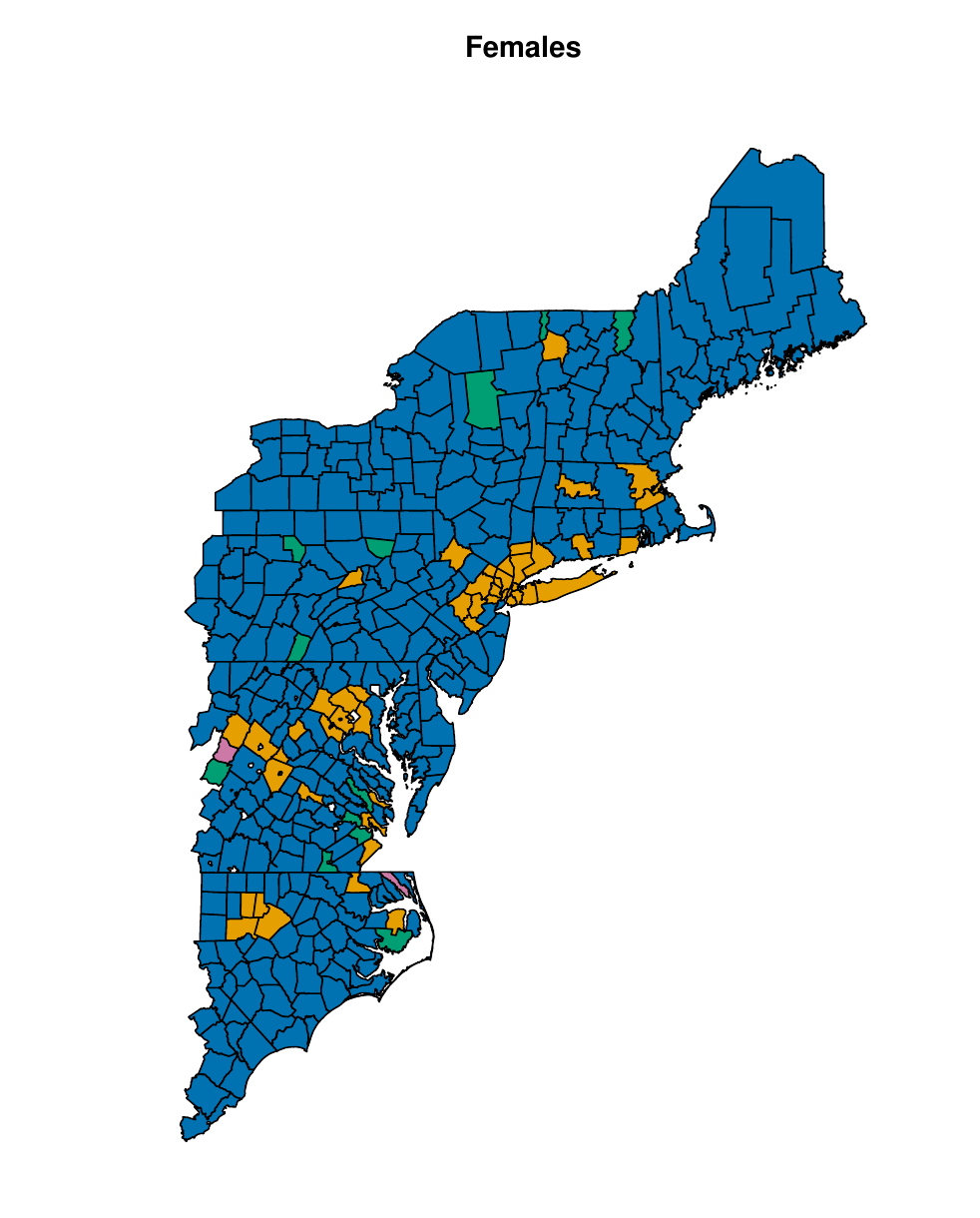}
 \includegraphics[width=7.8cm, height = 5.5cm]{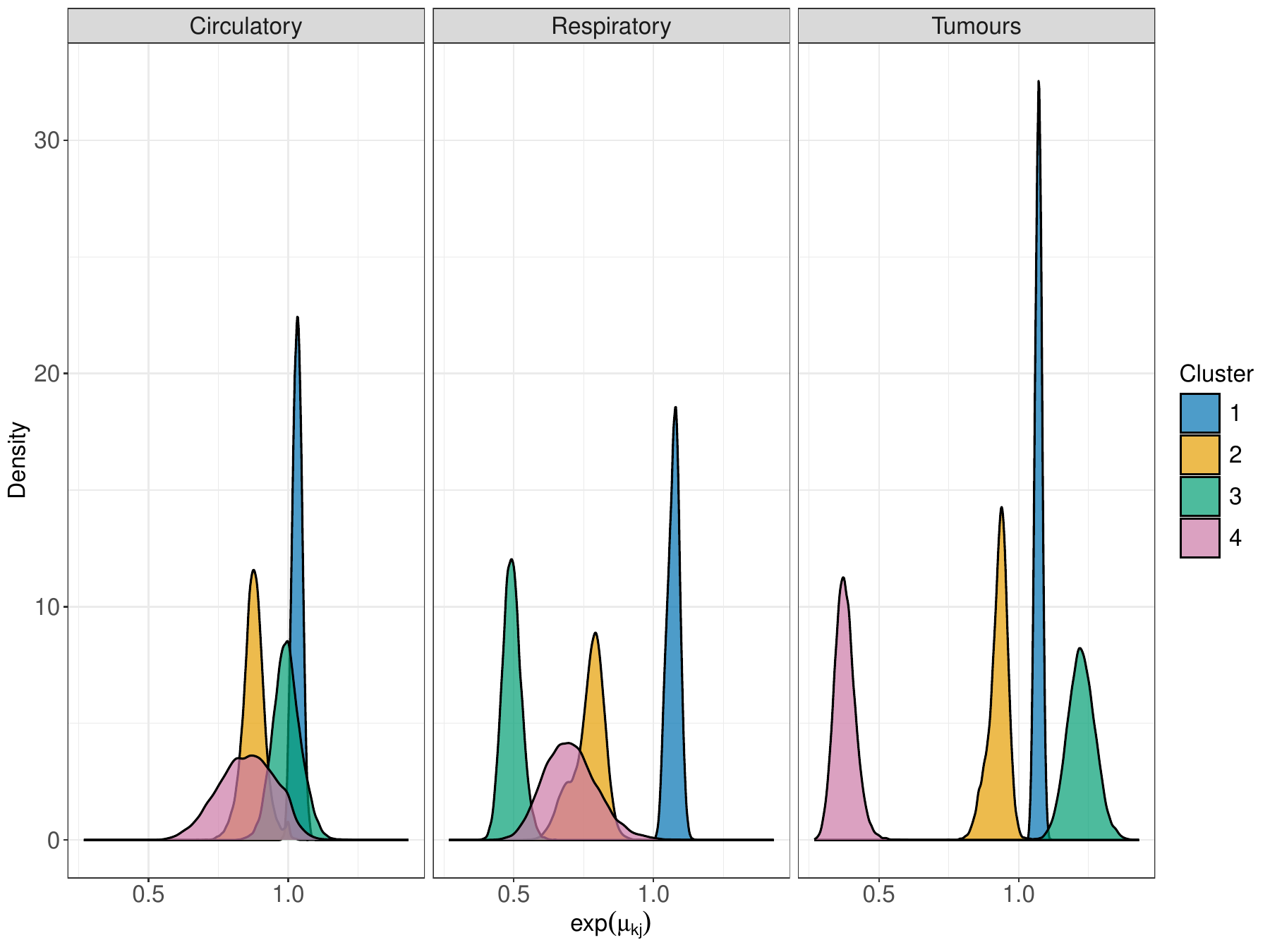}\\
    \caption{Results on the U.S. East coast data, divided for males (top line) and females (bottom line). Left plots display the 388 counties coloured according to the best partition returned by the ECR relabelling algorithm. Right plots display the posterior densities of the $\{\exp(\mu_{kj})\}$, based on 20,000 posterior draws, obtained from the model with $K = 3$ (males) and $K = 4$ (females).}
    \label{fig:map_posterior_east_USA}
\end{figure}

\clearpage
\begin{figure}[t]
    \centering
    \includegraphics[width=\linewidth]{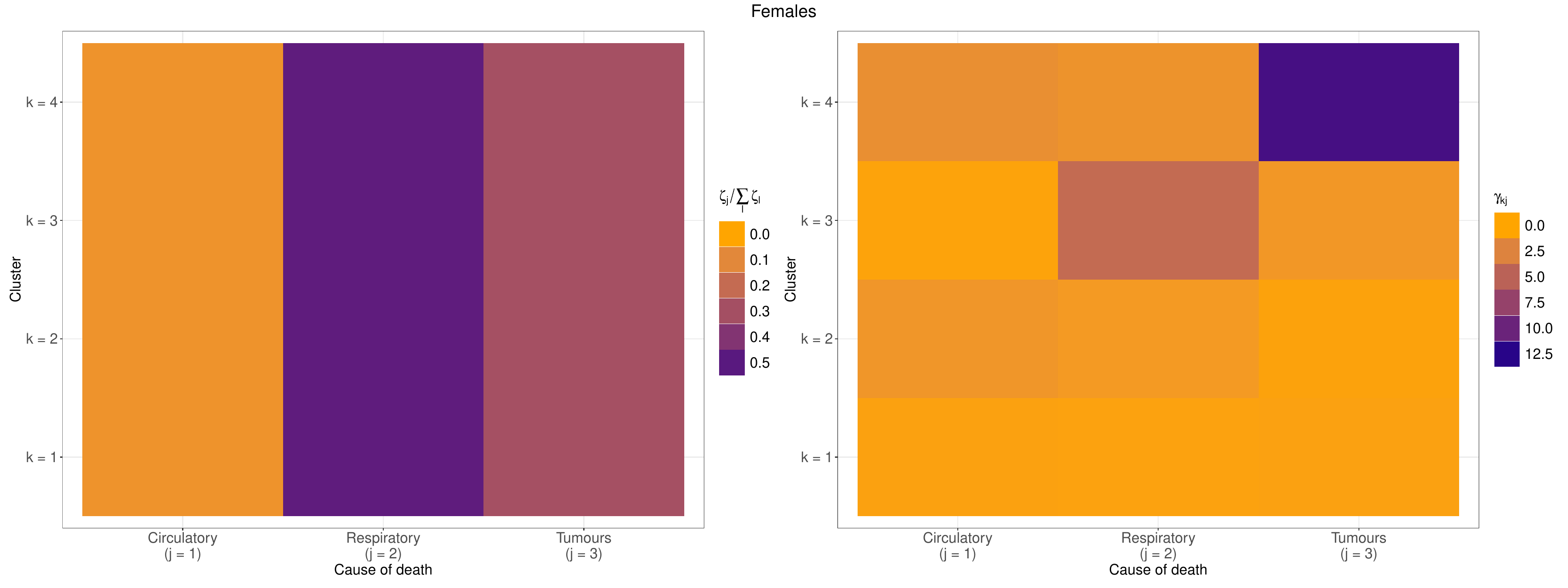}
    \caption{\reviewm{Posterior estimates of the shrinkage factors applied to the intercepts $\mu_{kj}$ for  each cluster $k = 1,\dots,K$ and each disease $j = 1,\dots,d$, obtained from the U.S. female data. 
    The left panel is coloured according to  $\zeta_j/\sum_{l = 1}^d \zeta_l$, the normalised version of the disease shrinkage factors $\zeta_j$. The $j$-th factor applies the same amount of shrinkage to every element of $\Mu_{.j}$. The right panel is coloured according to the cluster-disease shrinkage factors $\gamma_{kj}$, for $k = 1,\dots,K$ and $j = 1,\dots,d$, so the shrinkage level varies across both clusters and diseases.}}
    \label{fig:shrink_USA}
\end{figure}


\begin{table}
\caption{\review{
List of the combinations of shrinkage parameters considered for the prior models for $\Mu$. The first column reports the \texttt{R} code used to denote the different prior setups in our library \texttt{perla}, the second lists the model parameters, and the third describes the shrinkage effects.}}
    \centering
    	\begin{tabular}{ccc}
    \hline 
    \multirow{2}{*}{Code in \texttt{perla}} & Shrinkage & \multirow{2}{*}{Description} \\ 
    & parameters & \\
    \hline 
    \texttt{(1)} & $\phi$ & global shrinkage \\ 
    \hline
    \multirow{2}{*}{\texttt{(c)}} & $\phi$ & global shrinkage; \\
                                  & $\boldsymbol{\delta}$ & cluster shrinkage \\
    \hline 
    \multirow{2}{*}{\texttt{(d)}} & $\phi$ & global shrinkage; \\
                                  & $\boldsymbol{\zeta}$ & disease shrinkage \\
    \hline 
    \multirow{2}{*}{\texttt{(cd)}} & $\phi$ & global shrinkage; \\
                                   & $\boldsymbol{\gamma}$ & cluster-disease shrinkage \\
    \hline 
    \multirow{3}{*}{\texttt{(c,d)}} & $\phi$ & global shrinkage; \\
                                    & $\boldsymbol{\delta}$ & cluster shrinkage; \\
                                    & $\boldsymbol{\zeta}$ & disease shrinkage \\
    \hline 
    \multirow{3}{*}{\texttt{(d,cd)}} & $\phi$ & global shrinkage; \\
                                     & $\boldsymbol{\zeta}$ & disease shrinkage; \\
                                     & $\boldsymbol{\gamma}$ & cluster-disease shrinkage \\
    \hline 
\end{tabular}
    \label{tab:list_of_shrinkage_parameters}
\end{table}

\begin{table}
\caption{Results on the Padua province, divided for males and females. Each cell of the table displays the posterior modal value of $\exp(\mu_{kj})$ and the probability of observing a mortality excess $\Pr\{\exp(\mu_{kj})>1|\mathrm{data}\}$. Values for which the probability of mortality excess is larger than 0.95 or smaller than 0.05 are reported in bold.}
\centering
\begin{tabular}{cr|ccc}
  \hline
  \hline
& & Circulatory ($j = 1$) & Respiratory ($j = 2$) & Tumours ($j = 3$)\\ 
  \hline
  \hline
\multirow{3}{*}{Males} & $k = 1$& 1.000 (0.393) & \textbf{1.216 (0.997)} & 1.000 (0.511) \\ 
&  $k = 2$ & \textbf{1.166 (0.966)} & \textbf{0.318 (0.000)} & 1.000 (0.638) \\ 
&  $k = 3$ & 1.000 (0.580) & 0.999 (0.677) & 1.000 (0.476) \\ 
   \hline
\multirow{4}{*}{Females} &   $k = 1$ & 1.001 (0.461) & \textbf{1.098 (0.953)} & 1.001 (0.69) \\ 
&  $k = 2$ & 1.000 (0.588) & 1.002 (0.523) & 0.999 (0.637) \\ 
&  $k = 3$ & 1.154 (0.898) & \textbf{0.229 (0.000)} & 1.042 (0.768) \\ 
&  $k = 4$ & 1.061 (0.704) & \textbf{0.614 (0.034)} & \textbf{0.295 (0.003)} \\ 
   \hline
   \vspace{.2cm}
\end{tabular}
\label{tab:males_females_padua_mortality}
\end{table}

\begin{table}
\caption{Results on the U.S. East coast data, divided for males and females. Each cell of the table displays the posterior modal value of $\exp(\mu_{kj})$ and the probability of observing a mortality excess $\Pr\{\exp(\mu_{kj})>1|\mathrm{data}\}$. Values for which the probability of mortality excess is larger than 0.95 or smaller than 0.05 are reported in bold.}
\centering
\begin{tabular}{cr|ccc}
  \hline
  \hline
& & Circulatory ($j = 1$) & Respiratory ($j = 2$) & Tumours ($j = 3$)\\ 
  \hline
  \hline
\multirow{3}{*}{Males} & $k = 1$& \textbf{1.033 (0.995)} & \textbf{1.098 (1.000)} & \textbf{1.095 (1.000) }\\ 
&  $k = 2$ & \textbf{0.887 (0.001)} & \textbf{0.802 (0.000)} & \textbf{0.919 (0.002)} \\
&  $k = 3$ & \textbf{1.128 (0.967)} & \textbf{0.483 (0.000)} & \textbf{1.135 (0.999)} \\ 
   \hline
\multirow{4}{*}{Females} &   $k = 1$ & \textbf{1.034 (0.975)} & \textbf{1.079 (1.000)} & \textbf{1.072 (1.000)} \\ 
 & $k = 2$ & \textbf{0.878 (0.007)} & \textbf{0.793 (0.000)} & \textbf{0.939 (0.006)} \\  
 & $k = 3$ & 0.998 (0.445) & \textbf{0.494 (0.000)} & \textbf{1.22 (1.000)} \\  
 & $k = 4$ & 0.883 (0.067) & \textbf{0.707 (0.004)} & \textbf{0.374 (0.000)} \\
   \hline
   \vspace{.2cm}
\end{tabular}
\label{tab:males_females_EST_mortality}
\end{table}

\end{document}